%% file: main.tex
\documentclass[journal]{IEEEtran}
\usepackage{amsmath,amssymb,amsfonts}
\usepackage{graphicx}
\usepackage{svg}
\usepackage{cite}
\usepackage{url}
\usepackage{verbatim}
\usepackage{textcomp}
\usepackage{xcolor}
\usepackage{balance}

\usepackage[algo2e,ruled,linesnumbered,lined,boxed,commentsnumbered]{algorithm2e}
\usepackage{array}
\usepackage{booktabs}
\usepackage{multirow}
\usepackage{tabularx}
\usepackage{threeparttable}
\usepackage{stfloats}

\usepackage{subcaption}
\usepackage{float}
\usepackage{placeins}

\usepackage{amsthm}

\usepackage[
    colorlinks=true,
    linkcolor=blue,
    citecolor=green,
    urlcolor=magenta
]{hyperref}

\usepackage{tikz}
\usetikzlibrary{shapes,arrows}

\newcommand*\emptycirc[1][0.7ex]{\tikz\draw (0,0) circle (#1);} 
\newcommand*\fullcirc[1][0.7ex]{\tikz\fill (0,0) circle (#1);} 
\newcommand*\halfcirc[1][0.7ex]{%
  \begin{tikzpicture}
    \draw[fill] (0,0)-- (90:#1) arc (90:270:#1) -- cycle ;
    \draw (0,0) circle (#1);
  \end{tikzpicture}}

\usepackage{caption}
\begin{document}

\title{A Tutorial on Movable Antenna-Enabled ISAC Systems: Fundamentals, Parameter Estimation, and Security Issues}

\author{Zhendong Li, Zhou Su, Jianle Ba, Linchu Chen, Yan Yang, Weichun Zhao, Jinyuan Huang, Tom H. Luan, Wen Chen, Qingqing Wu, Lipeng Zhu, Zhenyu Xiao, Weidong Mei, Nan Cheng, Ruijin Sun, Lin Chen, Ying Wang
\thanks{Z. Li, L. Chen, W. Zhao, and J. Huang are with the School of Information and Communication Engineering, Xi’an Jiaotong University, Xi’an 710049, China (email: lizhendong@xjtu.edu.cn; chenlinchu@stu.xjtu.edu.cn; zwc2856669274@stu.xjtu.edu.cn; 2246112066@stu.xjtu.edu.cn). Z. Su, J. Ba, and T. H. Luan are with the School of Cyber Science and Engineering, Xi'an Jiaotong University, Xi'an 710049, China (email: zhousu@ieee.org; bajianle2024@163.com; tom.luan@xjtu.edu.cn). Y. Yang is with the Rocket Force University of Engineering, Xi’an 710025, China, and also with the School of Information and Communication Engineering, Xi’an Jiaotong University, Xi’an 710049, China (e-mail: yangyanyy2022@163.com). W. Chen and Q. Wu are with the School of Integrated Circuits, Shanghai Jiao Tong University, Shanghai 200240, China (e-mail: wenchen@sjtu.edu.cn; qingqingwu@sjtu.edu.cn). L. Zhu is with the School of Interdisciplinary Science, Beijing Institute of Technology, Beijing 100081, China (e-mail: zhulp@bit.edu.cn). Z. Xiao is with the School of Electronic and Information Engineering, Beihang University, Beijing 100191, China (e-mail: xiaozy@buaa.edu.cn). W. Mei is the National Key Laboratory of Wireless Communications, University of Electronic Science and Technology of China, Chengdu 611731, China (email: wmei@uestc.edu.cn). N. Cheng and R. Sun are with the State Key Laboratory of Integrated Services Networks and School of Telecommunications Engineering, Xidian University, Xi'an 710126, China (e-mail: nancheng@xidian.edu.cn; sunruijin@xidian.edu.cn). L. Chen is with the Department of Electrical and Computer Engineering, Stevens Institute of Technology, Hoboken, NJ 07030, USA (e-mail: lchen53@stevens.edu). Y. Wang is with the State Key Laboratory of Networking and Switching Technology, Beijing University of Posts and Telecommunications, Beijing 100876, China (e-mail: wangying@bupt.edu.cn). }
\thanks{(Corresponding author: Zhou Su)}
\vspace{-1.5em}
}

% The paper headers
%\markboth{Journal of \LaTeX\ Class Files,~Vol.~14, No.~8, August~2021}%
%{Shell \MakeLowercase{\textit{et al.}}: A Sample Article Using IEEEtran.cls for IEEE Journals}
%
%\IEEEpubid{0000--0000/00\$00.00~\copyright~2021 IEEE}
% Remember, if you use this you must call \IEEEpubidadjcol in the second
% column for its text to clear the IEEEpubid mark.

\maketitle
\begin{abstract}
Movable antenna (MA) technology has recently emerged as a promising paradigm for enhancing the flexibility and performance of wireless systems by allowing antennas to dynamically adjust their spatial positions within a confined region. Meanwhile, integrated sensing and communication (ISAC) has attracted significant attention as a key technology for next-generation wireless networks, aiming to integrate communication and sensing within a shared hardware and spectral framework. By leveraging the additional spatial degrees of freedom offered by MA, MA-enabled ISAC systems provide new opportunities to enhance both communication performance and sensing accuracy. However, the integration of MA into ISAC also poses new challenges in parameter estimation and security. In this tutorial, we provide a comprehensive overview of MA-enabled ISAC systems, with a particular focus on their fundamentals, parameter estimation, and security issues. First, we introduce the basic principles of MA and ISAC, and review representative MA-enabled ISAC systems and their potential applications. Then, we establish unified mathematical models for communication and sensing, and systematically review representative channel and sensing parameter estimation methods tailored to MA-enabled ISAC systems. Numerical comparisons are provided to illustrate the characteristics and performance of different estimation methods. Furthermore, we investigate the security issues of MA-enabled ISAC systems, introduce the fundamentals of physical-layer security (PLS), and summarize how the spatial reconfigurability of MAs can be exploited to enhance the security of ISAC systems. Representative case studies are presented to demonstrate the performance gains of MA-enabled secure ISAC systems. Finally, we identify several open issues and future research directions toward practical MA-enabled ISAC systems. This tutorial aims to provide a unified perspective on the fundamentals, parameter estimation, and security of MA-enabled ISAC, and to facilitate further research in this emerging area.
\end{abstract}

\begin{IEEEkeywords}
Movable antenna, integrated sensing and communication, parameter estimation, physical layer security.
\end{IEEEkeywords}

\section{Introduction}
\subsection{Background}
\begin{figure}
    \centering
    \includegraphics[width=\linewidth]{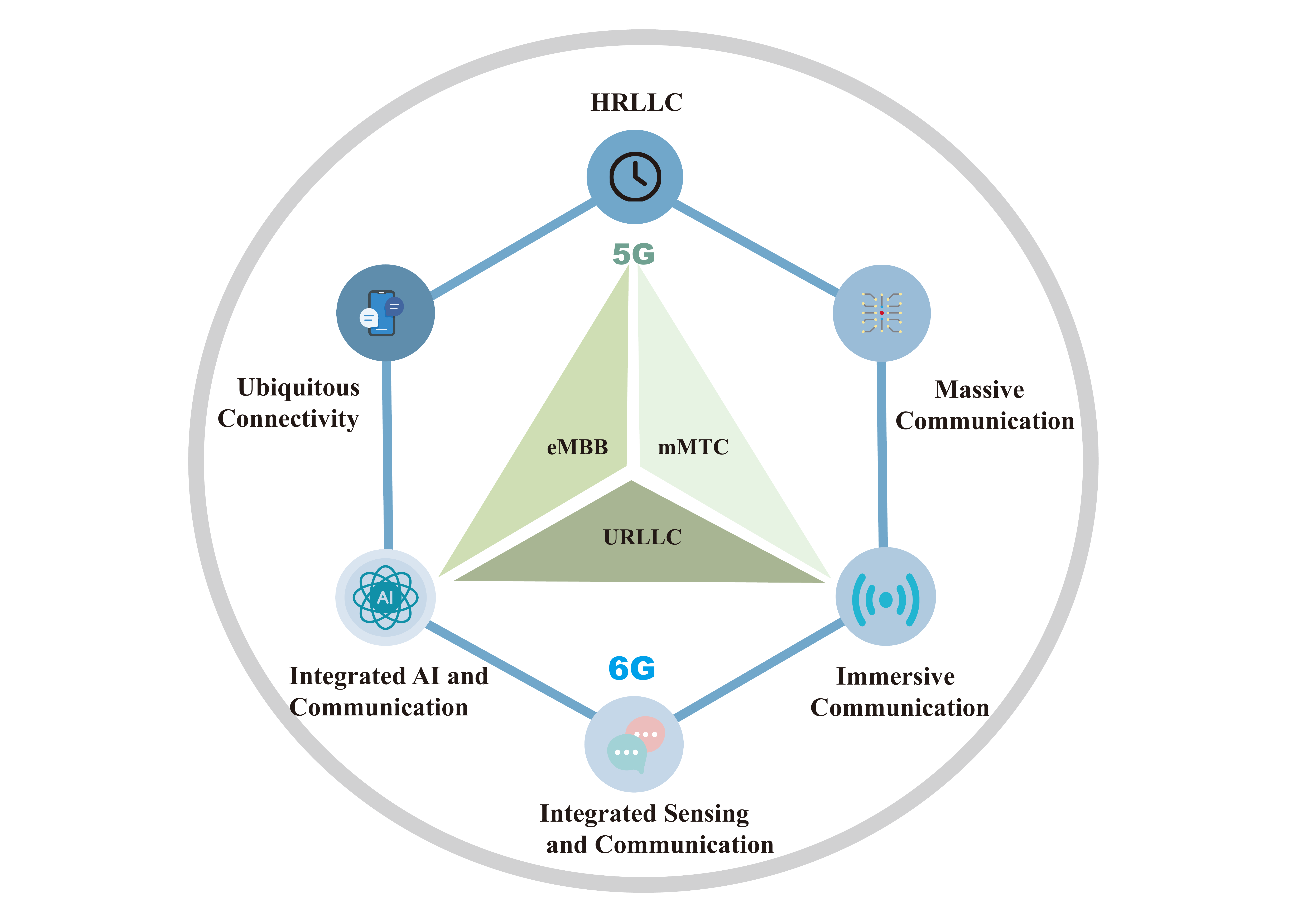}
    \caption{Six typical application scenarios of 6G.}
    \label{fig:placeholder}
    \vspace{-1em}
\end{figure}

With the evolution of mobile communication technologies, sixth generation (6G) networks are widely envisioned as key enablers for future digital and intelligent societies. Building on the three fifth generation (5G) service scenarios, namely enhanced mobile broadband (eMBB), massive machine type communications (mMTC), and ultra reliable and low latency communications (URLLC) \cite{8879484,8314688,9318506}, 6G further extends network capabilities and introduces integrated service paradigms such as artificial intelligence (AI)-empowered communication. Prior studies have identified six typical 6G scenarios, including immersive communication, massive communication, hyper reliable and low latency communication (HRLLC), ubiquitous connectivity, integrated AI and communication, and integrated sensing and communication (ISAC) \cite{10054381,9795904}. As shown in Fig.~\ref{fig:placeholder}, these scenarios expand the service space of 5G and support diverse requirements from enhanced user experiences to industrial digital transformation.

Immersive communication represents a major evolution beyond the 5G eMBB paradigm \cite{11247926,11193869}. It exploits extremely large bandwidth, low latency, accurate positioning, and tight synchronization, targeting peak rates for each user from 10~Gbps to 1~Tbps, rates experienced by users from 1~Gbps to 10~Gbps, air interface latency below 1~ms, positioning accuracy better than 1~cm, and synchronization errors under $10~\mu\mathrm{s}$. Supported by extended reality and holographic technologies, it enables seamless interaction between physical and virtual spaces. Massive communication extends the 5G mMTC scenario toward extremely high connection density \cite{11192777,11203888}, supporting massive access in smart cities, industrial Internet, and smart homes with diverse data rate requirements. HRLLC further advances the 5G URLLC scenario \cite{11251017,11218921}, targeting latency from 0.1~ms to 1~ms and network jitter below $1~\mu\mathrm{s}$ for critical applications such as industrial control, autonomous driving, and remote procedures. It emphasizes deterministic latency, high reliability, and coordinated resource management across communication, computing, and networking domains.

Ubiquitous connectivity aims to overcome the coverage limitations of terrestrial networks through an integrated architecture across space, air, and ground \cite{Xiao2024SAGIN6G}. By integrating satellites, high-altitude platforms (HAPs), unmanned aerial vehicles (UAVs), and terrestrial networks with ultra-dense deployment, it targets global coverage above 99\%, downlink rates of at least 10~Mbps in remote areas, and end-to-end latency below 50~ms in high mobility scenarios \cite{11175010}. Low Earth orbit (LEO) satellites provide broad remote coverage \cite{9371230}, while HAPs and UAVs offer near stationary, temporary, or supplementary aerial coverage \cite{Abir2023SDUAV}. Terrestrial ultra-dense networks further enhance capacity and link diversity in areas with heavy traffic. This paradigm supports maritime communications \cite{11222777}, broadband connectivity during flight, data exchange between aircraft and ground control \cite{9780352,10124871}, and communication recovery in regions affected by disasters \cite{Matracia2021PostDisasterUAV}.

Integrated AI and communication is regarded as a defining 6G paradigm, forming a reciprocal framework between AI for communication and communication for AI \cite{9200330}. In AI for communication, deep learning supports channel estimation, signal reconstruction, and physical layer optimization in complex wireless environments \cite{10490142}. In millimeter wave systems, it improves channel state information (CSI) acquisition under multipath propagation, blockages, and channel dynamics, thereby enhancing link reliability and bit error rate (BER) performance \cite{8752012}. Reinforcement learning (RL) further enables adaptive resource scheduling by learning from network states, link quality, and quality of service requirements \cite{8936987}. In communication for AI, 6G provides high bandwidth, low latency, and ubiquitous coverage for distributed model training and deployment. With edge computing, AI models can be trained and inferred at terminals or edge nodes, reducing latency and privacy risks caused by cloud data offloading \cite{9453402}. In UAV swarms, federated learning enables collaborative model updates without sharing raw data \cite{Zhang2020FLMultiUAV}. Privacy preserving federated learning and learning based scheduling can further coordinate model training, resource allocation, and communication links in UAV supported networks \cite{Yang2021PrivacyFLUAV}. Benefiting from the high data rate and very low latency of 6G, UAVs can accelerate model synchronization for intelligent collaboration and swarm decision making \cite{10113154}.

ISAC represents a novel 6G application paradigm that integrates communication and sensing within a unified wireless infrastructure \cite{10845869}. By sharing hardware platforms and  radio frequency signals, it enables simultaneous data transmission and environmental sensing. Its key targets include range resolution better than 1~cm, velocity resolution better than 0.1~m/s, a sensing refresh rate of tens of frames per second, and sensing latency below 10~ms \cite{9737357}. ISAC supports positioning, tracking, imaging, and reconstruction. High accuracy positioning can be achieved through round trip delay, signal strength, and angle of arrival (AoA) estimation \cite{10883658}. Environmental reconstruction relies on echo reflections \cite{10129074}, while gesture recognition exploits wireless signal variations caused by human movements \cite{Zhang2022WiFiSensingGesture}. ISAC further enables mutual enhancement between sensing and communication. Sensing derived channel information supports beam alignment, power allocation, and resource scheduling \cite{10988567}, while real-time sensing assists cell selection, mobility prediction, and handover management \cite{11007607}. Conversely, communication systems improve sensing through synchronization, carrier calibration, base station cooperation, and real-time feedback \cite{10124135}. With edge computing and network intelligence, these functions further support communication-assisted sensing \cite{11184332}. Architecturally, ISAC shares antenna arrays, radio frequency chains, spectrum resources, and baseband modules, improving spectrum and energy efficiency. It supports autonomous driving \cite{10806828}, smart factories, smart cities \cite{9945983}, and virtual or augmented reality through low latency environmental feedback.

\vspace{-1em}
\subsection{Motivations: From ISAC to MA-Enabled ISAC}
Traditional ISAC systems predominantly rely on fixed-position antennas (FPAs) with predetermined locations and array geometries. Although FPAs offer simple and mature implementations, their spatial degrees of freedom (DoFs) are mainly exploited through amplitude and phase adjustments, which limit the adaptability of the array geometry to dynamic propagation environments. Consequently, their performance can deteriorate in the presence of blockage, severe fading, and rapidly varying channels. Movable antenna (MA) provides a promising solution by introducing antenna position as an additional design DoF into ISAC systems \cite{Lyu2025MAISAC}. By dynamically adjusting antenna positions, MA arrays can reconfigure their spatial responses and adapt the effective channels to the surrounding propagation environment, thereby providing new opportunities for jointly optimizing communication and sensing performance.

The additional spatial DoF offered by MA has been shown to benefit various aspects of ISAC systems. In communication, antenna position optimization can reshape channel conditions, mitigate deep fading, and enhance beamforming and achievable rates. In sensing, MA-enabled arrays can adapt their spatial configurations to improve sensing accuracy and resolution \cite{Ma2024MASensing}, while reducing the Cram\'er-Rao bound (CRB) for parameter estimation in multiuser ISAC systems \cite{Qin2024MAISAC}. Related concepts have also been investigated in fluid antenna systems \cite{10505149,10694739,11080242} and pinching antenna systems \cite{11267231}, further demonstrating the potential of position-flexible antenna architectures for exploiting spatial diversity and improving wireless system performance. These advances indicate that MA-enabled ISAC can provide substantial gains in communication reliability, sensing capability, and measurement accuracy beyond conventional fixed-array designs.

Despite these promising results, MA-enabled ISAC also introduces a range of new research challenges that are fundamentally different from those of conventional ISAC systems. The additional position DoF couples antenna deployment with channel acquisition, parameter estimation, beamforming, and security, leading to new system models, optimization problems, and practical implementation issues. In particular, accurate channel and sensing parameter estimation becomes essential to optimize antenna positions and other design dimensions, while the reconfigurable spatial characteristics of MA also create new opportunities and challenges for physical-layer security. Existing studies have addressed these issues from different perspectives, but the relevant techniques and insights remain scattered across different research topics. A systematic tutorial that unifies the fundamentals, parameter estimation techniques, and security aspects of MA-enabled ISAC is therefore highly desirable.

\subsection{Related Works and Comparisons}
\begin{table*}[!t]
    \centering
    \renewcommand{\arraystretch}{1.2} % Adjust row spacing
    \resizebox{\textwidth}{!}{%
        \begin{threeparttable}
            \caption{{A Comparison of Contributions Between Our Tutorial and Relevant Works}}
            \label{contribution}
            \begin{tabular}{|c|c|>{\raggedright\arraybackslash}m{13.5cm}|c|c|c|c|c|}
                \hline
                \textbf{Year} & \textbf{Refs.} & \textbf{Contribution} & \textbf{A} & \textbf{B} & \textbf{C} & \textbf{D} & \textbf{E}  \\ \hline
                
                2025 & \cite{11050889} &
                Survey on full-duplex ISAC covering fundamentals, system architectures, hardware implementations, and future research directions. 
                & \emptycirc & \fullcirc & \emptycirc & \emptycirc & \emptycirc  \\ \hline
                
                2025 & \cite{11062661} &
                Survey on learning algorithms for ISAC covering transmitter and receiver design, resource allocation, wireless sensing, security, and future research directions.
                & \emptycirc & \fullcirc & \halfcirc & \halfcirc & \fullcirc \\ \hline
                
                2025 & \cite{10608156} &
                Survey on secure ISAC in 6G covering a unified Internet of Things (IoT) architecture, AI-enabled intelligent connectivity, security mechanisms and metrics, and future research directions.
                & \emptycirc & \fullcirc & \emptycirc & \emptycirc & \fullcirc \\ \hline
                
                2026 & \cite{11493774} &
                Survey on intelligent reflecting surfaces for ISAC covering system architectures, beamforming design, implementations, and future directions.
                & \emptycirc & \fullcirc & \fullcirc & \fullcirc & \halfcirc \\ \hline
                
                2023 & \cite{10012421} &
                Survey on ISAC signals toward 5G-A/6G covering signal design, signal processing, signal optimization, and future research directions.
                & \emptycirc & \fullcirc & \emptycirc & \emptycirc & \emptycirc \\ \hline
                
                2025 & \cite{11184827} &
                Review on localization in ISAC covering radar sensing topologies, localization techniques, performance metrics, energy-efficient antenna solutions, and open challenges.
                & \emptycirc & \fullcirc & \emptycirc & \fullcirc & \emptycirc \\ \hline
                
                2025 & \cite{11184506} &
                Survey on network-level ISAC covering cooperative architectures, stochastic-geometry-based performance analysis, distributed signaling and over-the-air synchronization, and future research challenges.
                & \emptycirc & \fullcirc & \emptycirc & \emptycirc & \emptycirc \\ \hline
                
                2024 & \cite{10536135} &
                Survey on communication-centric ISAC for 6G covering network vision, waveforms and channel modeling, sensing/localization techniques, and sensing-assisted communication.
                & \emptycirc & \fullcirc & \halfcirc & \fullcirc & \emptycirc \\ \hline
                
                2025 & \cite{11021487} &
                Survey on networked ISAC covering architectures, physical-to-network layer technologies, security and privacy issues, resilience, and emerging research directions.
                & \emptycirc & \fullcirc & \emptycirc & \emptycirc & \fullcirc \\ \hline
                
                2022 & \cite{9722893} &
                Survey on intelligent reflecting surface (IRS)-aided wireless communications focusing on channel estimation, practical passive beamforming under hardware constraints, and open design challenges.
                & \emptycirc & \emptycirc & \fullcirc & \emptycirc & \emptycirc \\ \hline
                
                2025 & \cite{10906511} &
                Tutorial on movable antennas for wireless networks covering fundamentals, field-response channel modeling, movement optimization, prototypes, and future research directions.
                & \fullcirc & \fullcirc & \emptycirc & \emptycirc & \fullcirc \\ \hline
                
                2025 & \cite{11142311} &
                Tutorial on six-dimensional movable antennas (6DMA) for 6G networks covering fundamentals, channel and hardware modeling, design and applications of rotatable/positionable 6DMA, and prototypes with future research directions.
                & \fullcirc & \fullcirc & \fullcirc & \fullcirc & \emptycirc  \\ \hline
                
                2025 & \cite{11175437} &
                Survey on fluid antenna systems covering application scenarios fundamentals including channel modeling and estimation networking technologies.
                & \fullcirc & \emptycirc & \fullcirc & \emptycirc & \emptycirc  \\ \hline
                
                2025 & \cite{11103480} &
                Proposal of ICENet, an adaptive implicit deep learning channel estimator for 6G that balances accuracy, complexity, and memory usage in high-mobility vehicle-to-everything (V2X) scenarios.
                & \emptycirc & \emptycirc & \fullcirc & \emptycirc & \emptycirc  \\ \hline
                
                2025 & \cite{10753482} &
                Tutorial on fluid antenna systems for 6G networks, focusing on channel modeling, channel estimation, core communication principles, novel multiple access techniques, hardware designs, and synergy with emerging technologies.
                & \fullcirc & \emptycirc & \fullcirc & \emptycirc & \emptycirc  \\ \hline
                
                2025 & \cite{11173626} &
                Overview of distributed movable antenna systems covering architectures, key design issues, optimization frameworks, and future deployment challenges.
                & \fullcirc & \fullcirc & \fullcirc & \emptycirc & \fullcirc \\ \hline
                
                2024 & \cite{10292797} &
                Survey on ISAC channels covering measurement methods and platforms, new channel features, modeling approaches, and future research directions.
                & \emptycirc & \fullcirc & \fullcirc & \emptycirc & \emptycirc \\ \hline
                
                2025 & \cite{11007274} &
                Overview of movable-antenna–enhanced wireless communications covering general MA architectures, practical mechanical/electronic implementation methods, and performance evaluation.
                & \fullcirc & \emptycirc & \emptycirc & \emptycirc & \emptycirc  \\ \hline
                
                2024 & \cite{10286328} &
                Overview of movable antennas for wireless communication covering applications, system architecture and channel modeling, performance gains, and key design challenges with potential solutions.
                & \fullcirc & \emptycirc & \fullcirc & \emptycirc & \emptycirc \\ \hline
                
                2022 & \cite{9755276} &
                Overview of sensing-aided secure ISAC integrating sensing, communication, and security, covering secure waveform design, robust and hardware-efficient architectures, and open challenges.
                & \emptycirc & \fullcirc & \emptycirc & \fullcirc & \fullcirc \\ \hline
                
                Now & \textbf{Ours} &
               Comprehensive tutorial on MA-enabled ISAC systems covering MA architectures, ISAC waveform design, joint channel and target parameter estimation, security mechanisms, and open challenges.
                & \fullcirc & \fullcirc & \fullcirc & \fullcirc & \fullcirc \\ \hline
            \end{tabular}
            \begin{tablenotes}
                \item \textbf{A}: MA \quad \textbf{B}: ISAC Waveform Design \quad \textbf{C}: Channel Parameter Estimation \quad \textbf{D}: Target Parameter Estimation \quad \textbf{E}: Security 
                \item \fullcirc\ Fully Included \quad \halfcirc\ Partially Included \quad \emptycirc\ Not Included
            \end{tablenotes}
        \end{threeparttable}}
        \vspace{-1em}
\end{table*}
Numerous existing works have explored ISAC systems from different perspectives, including system architectures, waveform design, channel modeling, localization, security, and networked sensing. We summarize and compare the studies relevant to this article in Table~\ref{contribution}.

For secure and networked ISAC, a comprehensive survey was provided in \cite{10608156}, where secure ISAC technologies were reviewed from the perspectives of system architecture, signal modeling, waveform optimization, and cross layer security design. In \cite{11021487}, a basic framework for networked ISAC was constructed, and security, privacy, and resilience issues were analyzed throughout the data lifecycle. The work in \cite{9755276} reviewed the evolution of multifunctional 6G wireless networks, discussed sensing centric, communication centric, and joint waveform design paradigms, and highlighted the security issue that sensing targets may also act as eavesdroppers. For localization, sensing, and channel modeling, \cite{11184827} reviewed localization techniques in ISAC systems, including radar topology types, adaptive positioning algorithms, and key performance parameters. The vision, techniques, and applications of ISAC were comprehensively discussed in \cite{10536135}, covering waveform design, channel estimation, resource allocation, and representative sensing modes. In \cite{10292797}, the characteristics, measurements, and modeling methods of ISAC channels were investigated, and the differences between ISAC channels and conventional communication channels were analyzed. For MA-related technologies, the tutorial in \cite{10906511} summarized the evolution, basic theory, channel acquisition, and antenna position optimization methods of MAs. Distributed MA systems were studied in \cite{11173626}, where the hierarchical architecture, performance benefits, and key design issues such as deployment, synchronization, channel estimation, and optimization were reviewed. In \cite{11007274}, three general MA architectures were proposed, including element-level, array level, and dual-scale architectures, and their implementation methods based on mechanical actuation and electronic control were discussed. In addition, fluid antenna systems and six-dimensional MAs (6DMAs) provide important references for position-reconfigurable antenna design. A tutorial on fluid antenna systems for 6G networks was presented in \cite{10753482}, covering communication theory, optimization methods, and hardware designs. The fundamentals and networking perspectives of fluid antenna systems were surveyed in \cite{11175437}, with emphasis on channel modeling, physical layer design, and networking issues. \cite{11142311} introduced 6DMAs, where both antenna positions and rotations can be adjusted to provide additional DoFs for future wireless communication and sensing systems.

Although prior studies have established important foundations for ISAC, MA, fluid antenna systems, and related technologies, a comprehensive tutorial on MA-enabled ISAC is still lacking. In particular, existing works have not provided a systematic treatment of its architectures, waveform design, antenna positioning, beamforming, parameter estimation, applications, and security. To address this gap, this article reviews the architectures, classifications, and advantages of MA, and surveys key techniques for MA-enabled ISAC, including system modeling, waveform design, antenna positioning, beamforming, and parameter estimation. We further examine security issues, with a focus on physical-layer security and MA-enabled secure ISAC. Finally, we highlight open problems and promising research directions for MA-enabled ISAC.

\subsection{Contributions}
% The integration of MAs into ISAC networks represents a transformative hardware paradigm for future 6G systems. While recent literature has extensively explored the fundamentals of ISAC and MA technologies independently, a systematic and comprehensive review bridging these two intersecting domains remains conspicuously absent. To fill this critical gap, this article presents an in-depth survey of the state-of-the-art in MA-enabled ISAC systems. By synthesizing existing research, we provide a holistic overview focusing on foundational architectural designs, parameter estimation, and security issues. The primary contributions of this comprehensive tutorial are summarized below.
The integration of MA into ISAC systems provides a promising hardware paradigm for enhancing the flexibility and performance of wireless networks. Although ISAC and MA technologies have been extensively investigated individually, a systematic treatment of their integration is still lacking. To address this gap, this article presents a comprehensive tutorial on MA-enabled ISAC systems, with a particular focus on their fundamentals, parameter estimation, and security issues. The main contributions are summarized as follows.

% To systematically review the research landscape of MA-enabled ISAC, Section II provides a comprehensive overview of the key theoretical foundations and technical frameworks of MA-enabled ISAC systems. Specifically, we first revisit the fundamental principles of ISAC, summarize representative ISAC waveform and beam design approaches, and review key performance metrics for both communication and sensing. We then introduce MA as a novel array paradigm, and systematically discuss their technical characteristics and the resulting spatial DoF from the perspectives of hardware architectures, mobility dimensionality (1D/2D/3D/6D MA), and comparisons with conventional fixed antenna arrays. Building upon this foundation, we focus on signal and simplified channel modeling methods, beamforming design, and antenna positioning and trajectory optimization algorithms for MA-enabled ISAC systems. We also highlight how antenna position reconfigurability provides new design freedoms for joint communication and sensing. Finally, we discuss representative application scenarios, which include digital agriculture, smart healthcare, smart homes, intelligent factories, vehicular networks, smart oceans, the low-altitude economy, and integrated air-space-ground networks, to illustrate the practical potential and future prospects of MA-enabled ISAC in diverse real-world environments.
First, we review the fundamentals and key techniques of MA-enabled ISAC systems. We first introduce MA as a new antenna array paradigm, covering their hardware architectures, mobility dimensions, spatial DoFs, and key differences from conventional FPAs. We then discuss signal and channel modeling, beamforming, antenna positioning, and trajectory optimization, with an emphasis on how antenna position reconfigurability can be exploited to enhance joint communication and sensing performance. Finally, we review representative application scenarios of MA-enabled ISAC, including digital agriculture, smart healthcare, smart homes, intelligent factories, vehicular networks, smart oceans, the low-altitude economy, and integrated air-space-ground networks.
% Furthermore, to provide essential prior information for optimizing the positions of MAs, we discussed in Section III the fundamentally important problem of parameter estimation in MA-ISAC systems. This unified framework not only facilitates the joint processing of communication and sensing signals but also enables the reuse of estimation results across different tasks. Specifically, we first noted the mathematical similarity between communication and sensing channels, then established a unified mathematical framework for parameter estimation applicable to both systems. Subsequently, we reviewed the parameter estimation algorithms available for MA systems, elaborating on their core principles and key steps. We provided a detailed introduction to tensor decomposition-based parameter estimation algorithms, which exploit the structural characteristics of high-dimensional received signals by decomposing them into factor matrices and performing channel estimation independently within these matrices, thereby enhancing algorithm performance. Unlike conventional approaches that suffer from parameter coupling, tensor decomposition decouples the estimation problem into multiple lower-dimensional subproblems, leading to improved accuracy and reduced complexity. Finally, we compared the computational complexity and estimation accuracy of various algorithms, offering a comprehensive summary of parameter estimation techniques in MA-enabled ISAC systems.

Second, we provide a systematic review of parameter estimation techniques for MA-enabled ISAC systems, which provide essential information for antenna position optimization. We identify the mathematical similarities between communication and sensing channels and establish a unified estimation framework for both tasks, enabling joint signal processing and the reuse of estimation results. We then review representative parameter estimation methods, highlighting their underlying principles and key procedures. In particular, tensor decomposition methods exploit the inherent structure of high-dimensional received signals by decomposing observations into factor matrices, thereby decoupling parameter estimation, improving estimation accuracy, and reducing computational complexity. Finally, we compare different estimation methods in terms of estimation accuracy and computational complexity.

% Moreover, to address the escalating security vulnerabilities in ISAC systems, we provide a holistic investigation into MA-enabled secure ISAC frameworks. We first establish the theoretical foundations of PLS  by reviewing information-theoretic metrics including secrecy capacity and outage probability alongside traditional enabling techniques. Building upon this basis, we characterize the MA-enhanced security framework to demonstrate how antenna position flexibility introduces unique spatial DoF. Furthermore, by jointly optimizing antenna positions and beamforming, MAs are shown to significantly outperform FPAs in enhancing legitimate links while enforcing precise null-steering and correlation control against eavesdroppers. 
% Finally, we evaluate MA-aided security across diverse mobility scenarios to highlight MAs as an adaptive and environment-aware defense mechanism capable of maintaining robust secrecy and covertness under high channel uncertainty and adversarial dynamics.

Third, we provide a systematic review of security issues in MA-enabled ISAC systems. We first review the fundamentals of physical-layer security (PLS), covering information-theoretic metrics such as secrecy capacity and secrecy outage probability, as well as representative security-enhancing techniques. We then examine MA-enabled security mechanisms and highlight how antenna position flexibility provides additional spatial DoFs for secure transmission. By jointly optimizing antenna positions and beamforming, MA can strengthen legitimate links and suppress eavesdroppers through adaptive null steering and channel correlation control. Finally, we evaluate the security performance of MA-enabled ISAC under different mobility scenarios, highlighting its potential to maintain high security in the presence of channel uncertainty and dynamic adversarial environments.

\begin{figure*}
    \centering
    \includegraphics[width=17cm]{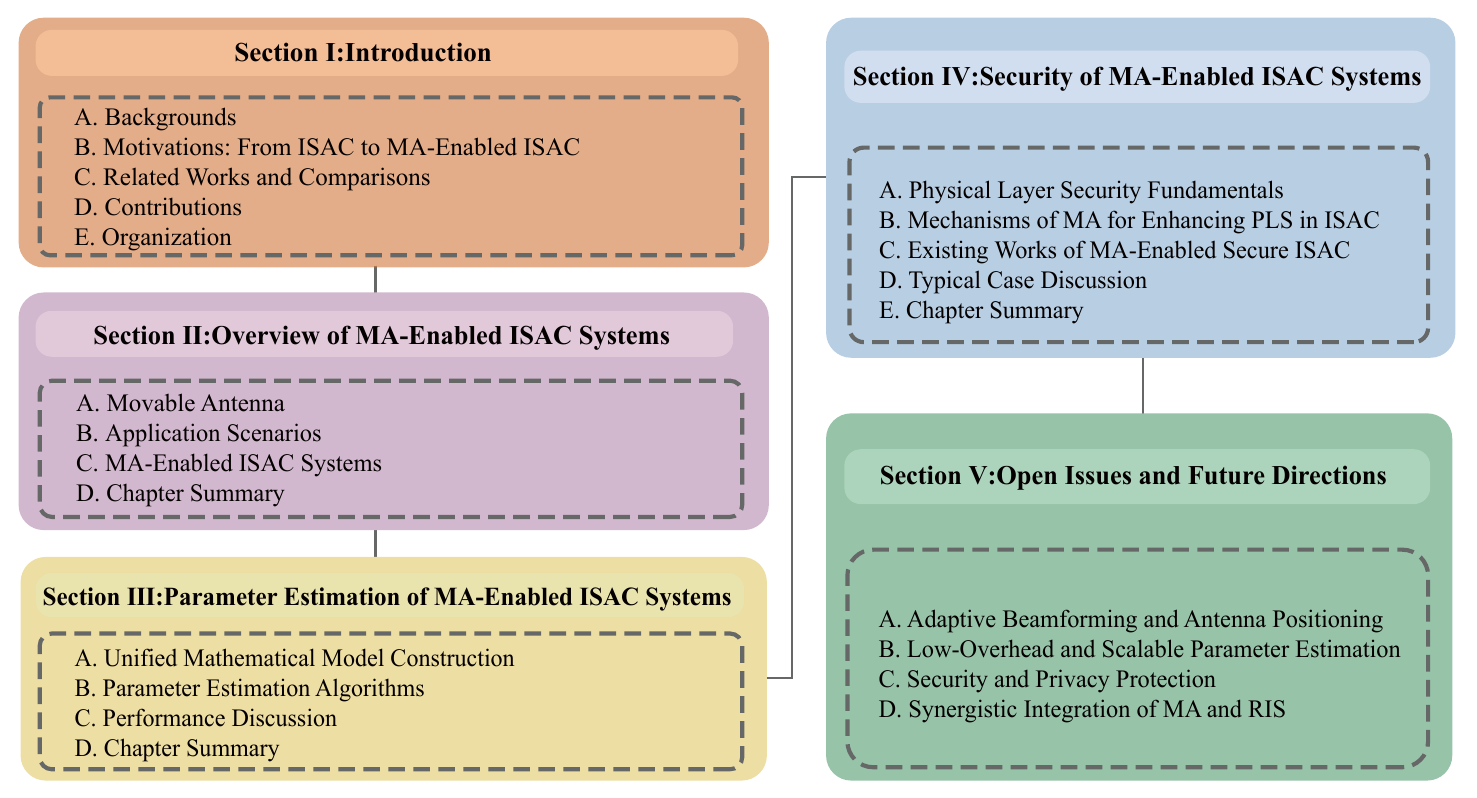}
    \caption{Organization of the tutorial.}
    \label{fig:organization}
    \vspace{-1em}
\end{figure*}

\subsection{Organization}
The organization of this tutorial is illustrated in Fig.~\ref{fig:organization}. Section I introduces the background, motivation, related work, and contributions of MA-enabled ISAC. Section II reviews the fundamentals and key techniques of MA-enabled ISAC, including MA architectures, signal and channel modeling, antenna positioning, beamforming, and representative applications. Section III focuses on parameter estimation, covering unified mathematical modeling, representative estimation algorithms, and performance comparisons. Section IV investigates security issues, including the fundamentals of physical-layer security, MA-enabled secure ISAC mechanisms, and representative case studies. Section V discusses open issues and future research directions, including adaptive beamforming and antenna positioning, low-overhead and scalable parameter estimation, security and privacy protection, and the synergistic integration of MAs and reconfigurable intelligent surfaces (RIS). Finally, Section VI concludes the tutorial.

\section{Overview of MA-Enabled ISAC Systems}
ISAC integrates wireless communication and radar sensing into a unified system by sharing spectrum, hardware, waveform, and signal-processing resources. Compared with separated communication and sensing infrastructures, ISAC can improve spectral efficiency, reduce hardware redundancy, and support simultaneous information transmission and environmental perception. However, conventional ISAC systems based on FPAs still suffer from limited spatial adaptability in dynamic or blockage-rich environments. To overcome this limitation, MA-enabled ISAC system introduces additional spatial DoF by dynamically adjusting antenna positions, thereby providing new opportunities for enhancing both communication and sensing performance.

\subsection{Movable Antenna}

By enabling individual antenna elements or subarrays to actively move and dynamically reconfigure their spatial arrangements in real time, MA systems facilitate intelligent, on-the-fly reconfiguration of the physical array geometry \cite{10716477,10746645,11418663}. This revolutionary capability is expected not only to significantly reduce the hardware scale and consequently the cost required to achieve comparable performance levels, but more importantly, to substantially expand the system DoF by leveraging the dynamic positioning of antenna elements/subarrays\cite{11226954,10709885}. Such enhanced flexibility empowers the system to dynamically reconfigure wireless channels, mitigate signal blockages, and precisely track targets in highly dynamic and complex scenarios, thereby offering a new perspective for the design and development of ISAC systems.

\subsubsection{Hardware Architecture}
\begin{figure}
    \centering
    \includegraphics[width=8.5cm]{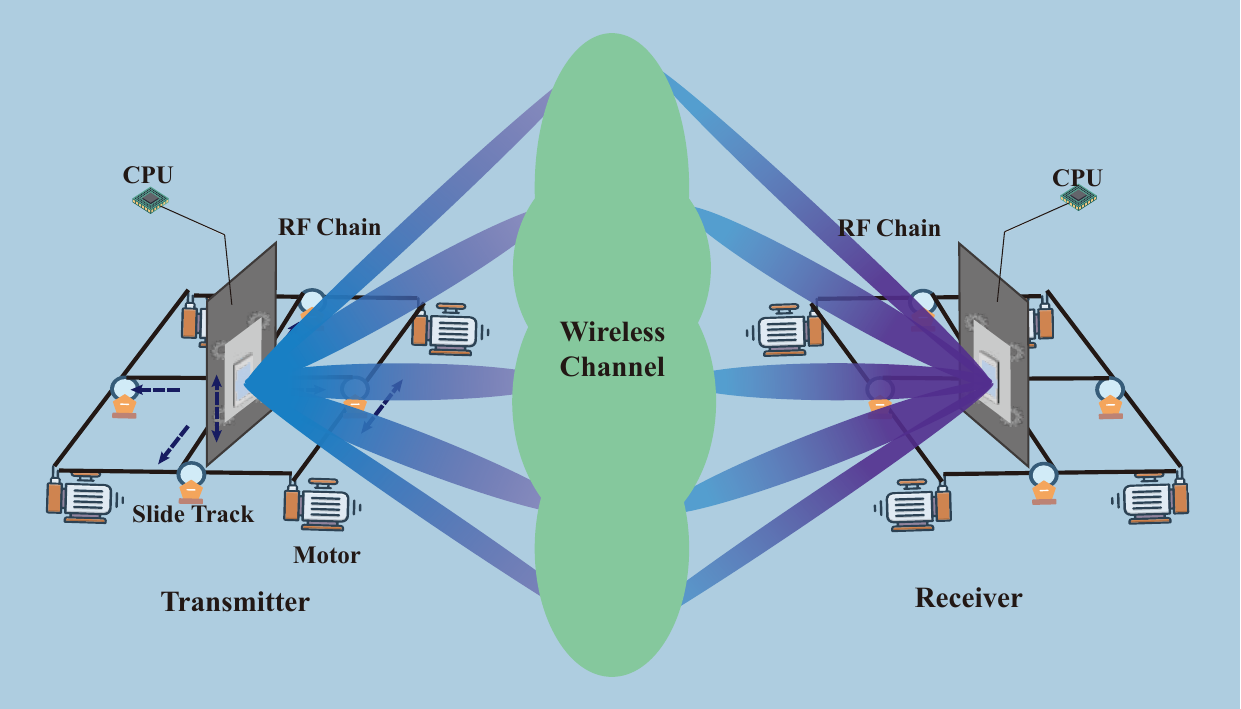}
    \caption{Hardware architectures of MA.}
    \label{fig:YUAN}
    \vspace{-1em}
\end{figure}

MA's core principle resides in optimizing the positions of antenna elements/subarrays in real time through mechanical or electronic means, enabling dynamic adaptation of the wireless channel to environmental changes and thus significantly enhancing system performance \cite{11345706}. There are diverse implementation methods for MA, which allow antenna elements to move dynamically within a specified aperture. Mechanically-driven MA systems bear structural similarities to conventional antennas and typically adopt flexible cables to connect MA elements to the radio frequency (RF) chain \cite{11095802}. A central processing unit (CPU) or field-programmable gate array (FPGA) is responsible for channel estimation and position optimization algorithms, transmitting pulse or analog signals to the driver, which in turn controls a stepper or servo motor. This motor drives a ball screw mechanism to move the antenna along linear or planar tracks, as shown in Fig. \ref{fig:YUAN}. To further enhance the spatial mobility of antenna elements, multi-DoF robotic arms can be employed. For example, a six-axis collaborative robotic arm can carry the antenna, enabling arbitrary pose adjustments within a spherical workspace. This approach extends the movement of the antenna element into 6D space, substantially improving deployment flexibility.

\subsubsection{Types of MA}

%\newline %新起一行
\indent Based on the above discussion, MA systems are next classified according to their spatial DoF.

\begin{figure*}
    \centering
    \includegraphics[width=17cm]{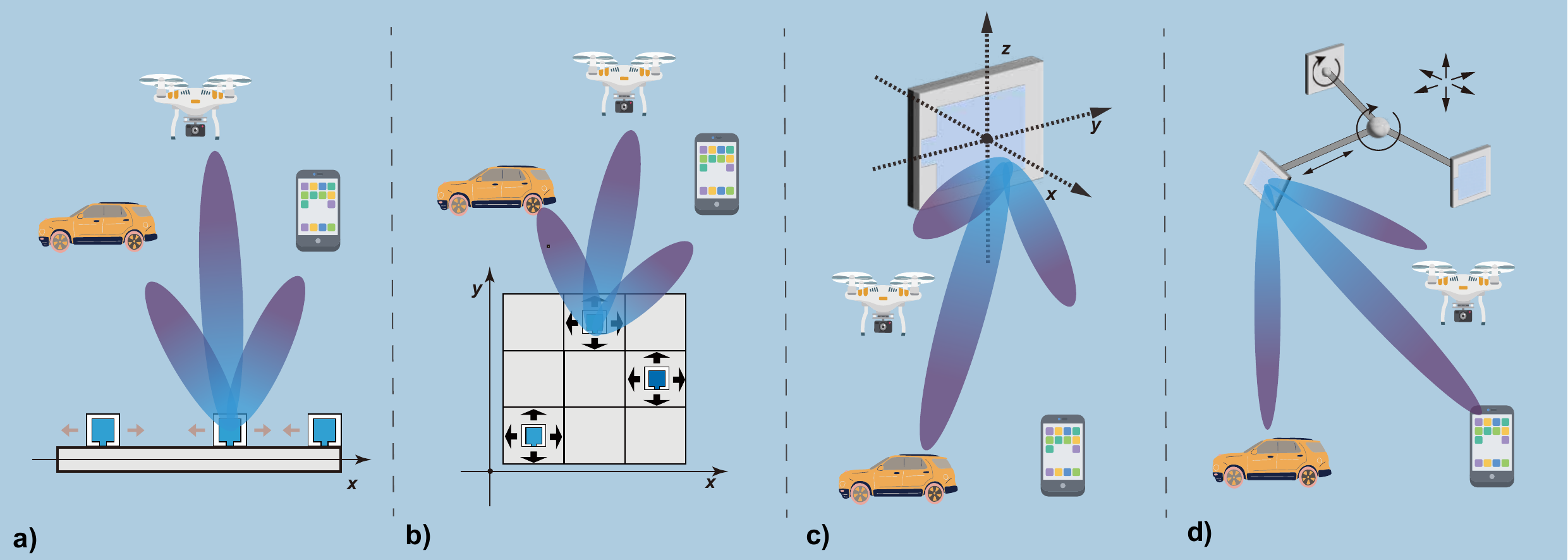}
    \caption{Types of MA: a) 1DMA; b) 2DMA; c) 3DMA; d) 6DMA.}
    \label{fig:FLfigure}
    \vspace{-1em}
\end{figure*}

\textbf{One-dimensional movable antennas (1DMA):} As shown in Fig. \ref{fig:FLfigure} a), it has $N$ MA elements, and the movable space for the MA elements is a 1D line segment of length $L$. Assuming the position of the $n$-th antenna element is ${x_n} \in \left[ {0,L} \right]$, and the physical AoA of the target-receiver line-of-sight (LoS) path is denoted as $\theta  \in \left[ {0,\pi } \right]$, the steering vector of the MA array can be expressed as 
\begin{equation}
\begin{aligned}
{\bf{a}}{_\text{1DMA}} \buildrel \over = {\left[ {{e^{j\frac{{2\pi }}{\lambda }{x_1}\cos \theta }},{e^{j\frac{{2\pi }}{\lambda }{x_2}\cos \theta }},...,{e^{j\frac{{2\pi }}{\lambda }{x_N}\cos \theta }}} \right]^{{T}}}.
\end{aligned}
\end{equation}
For 1DMA, \cite{10508218} transformed the antenna position selection problem in multiple-input single-output (MISO) systems into a shortest-path problem and developed efficient optimization algorithms. In \cite{Qin2024MAISAC}, 1DMA was applied to a multi-user ISAC system, where beamforming and antenna positions were jointly optimized to improve direction-of-arrival (DoA) estimation accuracy. From a practical perspective, 1DMA elements only need to slide along a straight line, offering a simple structure and easy deployment. However, they can only fine-tune the beam direction along that single axis, limiting flexibility, making them more suitable for application scenarios with singular directional control requirements.

\textbf{Two-dimensional movable antennas (2DMA):} As shown in Fig. \ref{fig:FLfigure} b), it consists of $N$ MA elements whose movable space is a rectangular region, allowing movement along both horizontal and vertical directions. Assuming the 2DMA movement region is continuous, the position of the $n$-th antenna element is denoted as ${{\bf{t}}_n}{\text{ = }}{\left[ {{x_n},{y_n}} \right]^{{T}}} \in {\cal C}$, and the physical elevation and azimuth angles of the receiver LoS path are denoted as $\theta  \in \left[ {0,\pi } \right]$ and $\phi  \in \left[ {0,\pi } \right]$ respectively. Then, the steering vector of the 2DMA array can be expressed as 
\begin{align}
\mathbf{a}_{\text{2DMA}} &=
\left[
e^{j\frac{2\pi}{\lambda}(x_1\sin\theta\cos\phi + y_1\cos\theta)},
e^{j\frac{2\pi}{\lambda}(x_2\sin\theta\cos\phi + y_2\cos\theta)},
\right. \nonumber\\
&\quad \left.
\cdots,
e^{j\frac{2\pi}{\lambda}(x_N\sin\theta\cos\phi + y_N\cos\theta)}
\right]^T .
\end{align}

For 2DMA, representative studies have mainly focused on exploiting planar movement for communication performance enhancement. In \cite{10504625}, the weighted sum-rate maximization problem in 2DMA-enhanced multi-user MIMO systems was studied through joint beamforming and antenna position optimization. \cite{10638767} further introduced 2DMA into a non-orthogonal multiple access (NOMA) downlink system and optimized power allocation together with antenna positions to improve channel capacity. Overall, compared to 1DMA, 2DMA offers greater mobility flexibility and stronger spatial gains, enabling the formation of more directive beams.

\textbf{Three-dimensional movable antennas (3DMA):} As shown in Fig. \ref{fig:FLfigure} c), it has $N$ MA elements. The movable space of the MA elements is a rectangular parallelepiped. The MA elements can move not only in the $x$ and $y$ axis directions but also along the $z$ axis. Assuming that the moving area of the 3DMA is continuous, the position of the $n$-th antenna element is ${{\bf{t}}_n}{\text{ = }}{\left[ {{x_n},{y_n},{z_n}} \right]^{T}} \in {\cal C}$. The physical elevation and azimuth angles of the target receiver LoS path are denoted as $\theta  \in \left[ {0,\pi } \right]$ and $\phi  \in \left[ {0,\pi } \right]$, respectively. The steering vector of the 3DMA array can be expressed as 
\begin{equation}
\begin{aligned}
\mathbf{a}_\text{3DMA} &= \left[ e^{j\frac{2\pi}{\lambda}\left( x_1\cos\theta\cos\phi + y_1\cos\theta\sin\phi + z_1\sin\theta \right)}, \dots, \right. \\
&\quad \left. e^{j\frac{2\pi}{\lambda}\left( x_N\cos\theta\cos\phi + y_N\cos\theta\sin\phi + z_N\sin\theta \right)} \right]^T.
\end{aligned}
\end{equation}
The application of 3DMA in multi-user communications was investigated in \cite{10354003}, where antenna positions were optimized to reduce the total transmit power. By introducing movement along the third spatial dimension, 3DMA provides more precise beam control and can achieve better communication and sensing performance than 1DMA and 2DMA due to the increased DoF.

\textbf{6DMA:} As shown in Fig. \ref{fig:FLfigure} d), compared with the aforementioned MAs, the antenna panel of the 6DMA is a uniform linear array or a uniform planar array instead of an MA element. The MA panel can not only achieve 3D movement but also realize 3D rotation through the contraction of the rod and the rotation motors at both ends of the rod. The movable space of the MA is extended to 3D. Assuming that the moving area of the 6DMA is continuous, the center coordinates of the $b$-th surface are ${{\bf{t}}_b}{\text{ = }}{\left[ {{x_b},{y_b},{z_b}} \right]^{{T}}} \in {\cal C}$, and the Euler angles are ${{\bf{u}}_b} = {\left[ {{\alpha _b},{\beta _b},{\gamma _b}} \right]^{{T}}}$. The physical elevation and azimuth angles of the target-receiver LoS path are denoted as $\theta  \in \left[ { -\pi,\pi } \right]$ and $\phi  \in \left[ { - {\pi  \mathord{\left/
 {\vphantom {\pi  2}} \right.
 \kern-\nulldelimiterspace} 2},{\pi  \mathord{\left/
 {\vphantom {\pi  2}} \right.
 \kern-\nulldelimiterspace} 2}} \right]$, respectively, and the corresponding unit direction vector is ${\bf{f}}{\text{ = }}\left[ {\cos \theta \cos \phi ,\cos \theta \sin \phi ,\sin \theta } \right].$
The position of the $b$-th antenna on the $n$-th surface in the local coordinate system is ${{\bf{r}}_n}$. It is transformed into the global coordinate system through the rotation matrix ${{\bf{r}}_{b,n}} = {{\bf{t}}_b} + {\bf{R}}\left( {{{\bf{u}}_b}} \right){{\bf{r}}_n}$. The steering vector of the 6DMA array can be expressed as 
\begin{equation}
\begin{aligned}
{\bf{a}}{_\text{6DMA}}\left( {{{\bf{t}}_b},{{\bf{u}}_b}} \right) \buildrel  \over = {\left[ {{e^{j\frac{{2\pi }}{\lambda }{{\bf{f}}^{{T}}}{{\bf{r}}_{b,1}}}},{e^{j\frac{{2\pi }}{\lambda }{{\bf{f}}^{{T}}}{{\bf{r}}_{b,2}}}},...,{e^{j\frac{{2\pi }}{\lambda }{{\bf{f}}^{{T}}}{{\bf{r}}_{b,N}}}}} \right]^{{T}}}.
\end{aligned}
\end{equation}
For 6DMA, \cite{10752873} proposed a base station (BS) architecture that jointly adjusts the 3D positions and rotations of distributed antenna surfaces to improve network capacity, especially in dense and uneven user distributions. In \cite{10737418}, 6DMA was further applied to wireless sensing, where both positions and orientations were optimized to reduce the CRB of DoA estimation. The coordination of rotatable antennas and RISs was studied in~\cite{zhang2026rotation}, where the BS antenna orientations, receive beamforming, and RIS phase shifts were jointly optimized for multi-user uplink communications. This design was extended in~\cite{zhang2026orientation} by further optimizing the RIS panel orientation under a mixed-field channel model. Furthermore, the authors in \cite{11306256} proposed a UAV-enabled passive 6DMA architecture and investigated the associated performance optimization problem. Due to the further increase of spatial DoF, 6DMA can obtain larger beam gain and more flexible beamforming. It is also worth noting that by combining the 3D movement and 3D rotation dimensions, MAs of any dimensionality from 1D to 6D can be customized.

\subsubsection{Comparison Between Conventional FPAs and MAs}
In this subsection, we systematically compare conventional FPAs with various types of MAs. Their key differences are summarized in Table \ref{table:ma111}. The following aspects will be elaborated: spatial DoF and dynamic channel adaptation capability, coverage and flexibility, performance gain, hardware resource efficiency, CSI acquisition complexity, and hardware actuation cost.
\begin{table*}[htbp]
\centering
\caption{Comparison of FPAs, 1D--3D MA, and 6D MA}
\renewcommand{\arraystretch}{1.4}
\label{table:ma111}
\begin{tabular}{|m{2.8cm}|m{4.2cm}|m{4.2cm}|m{4.2cm}|}
\hline
\textbf{Aspects} 
& \centering\textbf{FPAs} 
& \centering\textbf{1D--3D MA} 
& \centering\textbf{6D MA} \tabularnewline
\hline

\textbf{DoF and Adaptability} 
& Fixed positions with limited channel adaptability 
& Antennas move in 1D--3D space to adapt channels 
& Panel translation and rotation enable full-dimensional adaptation \\
\hline

\textbf{Coverage and Beam Control} 
& Coverage is predetermined and less flexible 
& Reconfigures coverage by antenna movement 
& Adjusts both position and orientation for flexible beam control \\
\hline

\textbf{Performance} 
& Limited by fixed array geometry 
& Improves SNR and capacity via position optimization 
& Further enhances performance through 6D alignment \\
\hline

\textbf{Hardware Efficiency} 
& Often requires more antenna elements 
& Reduces antenna demand using spatial DoF 
& Supports compact design through panel-level optimization \\
\hline

\textbf{Implementation Complexity} 
& Simple CSI estimation and low maintenance cost 
& Higher CSI and hardware complexity due to movement 
& Highest complexity due to 6D state tracking and actuation \\
\hline

\textbf{Applications} 
& Static or low-mobility scenarios 
& Dynamic coverage and user-aware communication 
& ISAC, positioning, and environment-aware communication \\
\hline

\end{tabular}
\vspace{-1em}
\end{table*}
\begin{itemize}
    \item \textbf{Spatial DoF and Dynamic Channel Adaptation:}
    Conventional FPAs have static array geometry, and their spatial DoF are fixed after deployment. In contrast, 1D--3D MAs can physically reposition antenna elements, enabling flexible array reconfiguration and adaptive channel exploitation. Meanwhile, 6D MAs move and rotate the entire antenna panel, thereby providing panel-level spatial control for dynamic channel optimization.

    \item \textbf{Coverage and Flexibility:}
    Conventional FPAs provide static coverage determined during installation, which limits their ability to serve mobile users or adapt to environmental changes. By adjusting their position and/or orientation within an authorized region, MAs can dynamically reshape coverage, focus signal energy, avoid obstacles, and track mobile users, leading to more flexible and resilient wireless networks.

    \item \textbf{Performance Gain:}
    MAs can exploit additional spatial optimization dimensions to improve communication performance. Through position or orientation adjustment, both 1D--3D and 6D MAs can achieve more effective beamforming toward target users or directions. This enhances signal strength, improves signal-to-noise ratio (SNR) and channel capacity, and provides notable performance gains over conventional FPAs.

    \item \textbf{Hardware Resource Efficiency:}
    In massive multiple-input multiple-output (MIMO) systems, conventional fixed antenna arrays often require many antenna elements, resulting in high hardware complexity and cost. By utilizing spatial mobility, MAs can achieve comparable capacity or coverage with fewer antenna elements. This improves hardware utilization efficiency and supports more compact and cost-effective system designs.

    \item \textbf{CSI Acquisition Complexity:}
    The mobility of MAs also introduces new channel acquisition challenges. To fully exploit their spatial adaptability, the system needs to estimate or predict channel information over the feasible movement region. As the movable range increases, the channel state space expands significantly, leading to higher estimation, modeling, and feedback overhead.

    \item \textbf{Hardware Actuation Cost:}
    MA systems require additional actuation and control mechanisms, such as motors, rails, rotation devices, and positioning controllers. These components enable precise movement of antenna elements or panels, but they also increase hardware cost, power consumption, and maintenance complexity compared with fully fixed antenna deployments. Recent delay-aware MA design further shows that jointly accounting for antenna movement delay and link quality can preserve throughput gains across different transmission-block durations and antenna speeds, while enabling the use of lower-speed actuators \cite{11082461}.

    \item \textbf{Future Potential:}
    Despite the challenges in CSI acquisition and hardware implementation, MAs show strong potential for future wireless systems. By enhancing spatial DoF, they can overcome key limitations of FPAs and support dynamic channel optimization, flexible coverage, and efficient resource utilization. Their integration with ISAC may further enable intelligent applications such as high-precision positioning, environment-aware communication, and network self-optimization.
\end{itemize}

\subsection{Application Scenarios}
\begin{figure*}
    \centering
    \includegraphics[width=15.5cm]{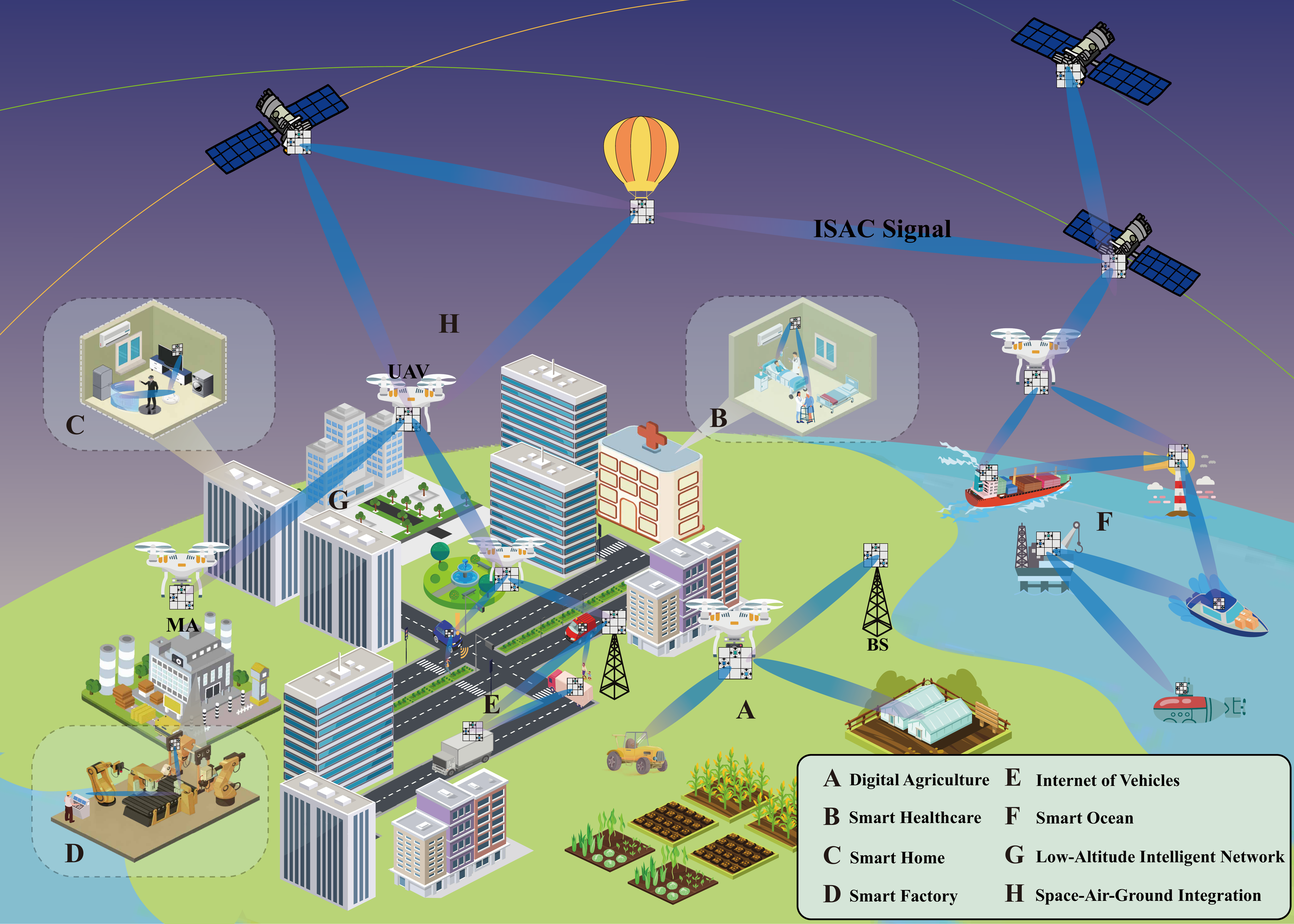}
    \caption{Application scenarios of MA-enabled ISAC systems.}
    \label{fig:scene3}
    \vspace{-1em}
\end{figure*}

Building on the aforementioned characteristics of MAs, this section presents representative application scenarios of MA-enabled ISAC. By exploiting flexible spatial reconfiguration, MA-enabled ISAC can support efficient communication and high-precision sensing in heterogeneous environments. Typical scenarios include digital agriculture, smart healthcare, smart homes, smart factories, the internet of vehicles (IoV), smart oceans, low-altitude intelligent networks, and integrated space-air-ground networks, as illustrated in Fig.~\ref{fig:scene3}.

\subsubsection{Digital Agriculture}

Digital agriculture integrates sensors, remote sensing, the Internet of Things (IoT), big data analytics, cloud computing, and AI to support intelligent agricultural management. It enables real-time monitoring, quantitative analysis, and dynamic regulation of crop environments and production factors, thereby improving efficiency and resource utilization. \cite{AVSAR2022102982} showed that wireless communication protocols differ in coverage, energy consumption, and data rate, which influences their suitability for smart agriculture. Low-power wide-area network (LPWAN) technologies, such as LoRaWAN and NB-IoT, are suitable for large-scale farmland monitoring because of their low power consumption and wide coverage. \cite{9903855} confirmed that sensor networks and IoT support intelligent agricultural processes, including tillage, sowing, irrigation, fertilization, harvesting, and livestock management. However, agricultural IoT still faces challenges in coverage, reliability, and sensing accuracy. MA-enabled ISAC can address these limitations by flexibly adjusting antenna positions and beam directions. For example, MA-integrated UAVs can collect crop growth, soil moisture, and pest information while maintaining stable data links, supporting precise and sustainable digital agriculture.

\subsubsection{Smart Healthcare}

Smart healthcare applies AI, IoT, big data, and cloud computing to provide digital, networked, and intelligent medical services. It supports disease prediction, assisted diagnosis, personalized treatment, remote healthcare, and health management through real-time data collection and analysis. Non-contact monitoring and secure data management are important research directions. \cite{s21217130} proposed a posture recognition method based on the micro-Doppler effect and ultra-wideband radar, which can identify daily activities of elderly people living alone without using cameras, improving both safety and privacy. \cite{https://doi.org/10.1155/2021/7608296} studied blockchain and IoT in healthcare, showing their potential in remote patient monitoring, drug traceability, and medical record management, while also pointing out challenges in interoperability, security, and standardization. These studies show that smart healthcare requires reliable communication, accurate sensing, and secure data collection. MA-enabled ISAC can adjust antenna positions to improve vital sign monitoring and data transmission in complex medical environments, providing timely and accurate support for safer healthcare services.

\subsubsection{Smart Home}

A smart home connects lighting, appliances, security systems, entertainment systems, and environmental control devices through network communication, automatic control, security protection, and audio-video processing technologies. It enables centralized management and intelligent control of home devices. Human-computer interaction and energy management are two major research directions. \cite{s23177523} proposed a gesture-recognition-based automation method that realizes accurate non-contact interaction through feature fusion and recurrent neural networks. \cite{KHAN2023100712} introduced an open connectivity foundation IoTivity-based indoor environmental control system that uses machine learning to predict energy consumption and thermal comfort, and applies an improved firefly algorithm for optimization, achieving up to 36.82\% energy savings. These studies show that smart homes are developing toward intelligent interaction and proactive energy management. MA-enabled ISAC can reduce multipath effects and occlusion by adjusting antenna positions, improving both communication stability and sensing accuracy. For example, MAs can adapt to occupant distribution, reducing energy use in unoccupied rooms while maintaining comfort in active areas.

\subsubsection{Smart Factory}

Smart factories apply IoT, AI, big data, digital twins, cloud computing, and 5G/6G communication to manufacturing. They enable interconnection and collaboration among equipment, production lines, workflows, and management systems. Sensors collect operational and environmental data, while AI and big data analytics support predictive maintenance and production optimization, helping reduce downtime and improve efficiency and quality. \cite{app12126258} proposed a virtual reality and online multi-user interaction framework based on metaverse and digital twin technologies, providing an immersive and scalable experimental platform. \cite{app13042156} designed a cyber-physical system and IoT-based smart factory architecture, and verified its effectiveness in improving production efficiency and data exchange through a drilling process case study. These studies show that smart factories are moving toward virtualization, intelligence, and efficiency. However, FPAs in automated workshops may suffer from metal blockage and multipath effects. MA-enabled ISAC can adjust antenna positions to sense robot trajectories and workpiece locations while maintaining reliable low-latency communication, improving production flexibility and operational efficiency.

\subsubsection{IoV}

The IoV connects vehicles with other vehicles, road infrastructure, pedestrians, and cloud platforms through onboard sensors, communication modules, and intelligent control systems. It enables real-time perception of vehicle status and road environments, supporting intelligent navigation, autonomous driving, traffic management, and information services. \cite{9650522} proposed low-complexity resource allocation algorithms based on channel gain priority and min-max channel gain matching for dense cellular V2X scenarios, reducing computational complexity while ensuring link stability and quality of service. \cite{10749940} analyzed V2X communication under 5G IoT integration, showing its value in road safety, congestion reduction, and energy saving through field pilots in smart cities and autonomous driving systems. MA-enabled ISAC provides new opportunities for the IoV by combining spatial flexibility and beam adaptability. It can support high-precision perception of dynamic vehicle behaviors, such as lane keeping and blind spot monitoring, while maintaining reliable low-latency communication. For example, roadside units can adjust MA positions to detect sudden obstacles and send warnings to nearby vehicles in real time.

\subsubsection{Smart Ocean}

Smart ocean relies on marine information collection, transmission, and cloud platforms to support intelligent ocean management and utilization. It integrates marine rights protection, ocean management, and resource development through industrial big data and internet big data, enabling information sharing and activity coordination. High-precision observation and intelligent management are essential for this scenario. With the surface water and ocean topography satellite mission, \cite{https://doi.org/10.1029/2023GL107652} showed that Ka-band radar interferometers can provide high-resolution water-surface elevation observations, revealing the role of small-scale ocean eddies and fronts in climate regulation and material transport. \cite{9750063} proposed a collaborative detection method based on graph neural networks and multi-agent reinforcement learning for underwater autonomous vehicles, improving marine pollution monitoring and path optimization. MA-enabled ISAC can enhance communication reliability and sensing accuracy in complex marine environments through flexible antenna adjustment. For example, in illegal fishing monitoring, it can reduce sea-surface multipath interference, sense ship locations and trajectories, and transmit real-time data to command centers for coordinated response.

\subsubsection{Low-Altitude Intelligent Network}

A low-altitude intelligent network uses communication, navigation, and remote sensing technologies in low-altitude airspace, generally below 1000 meters and extendable to 3000 meters when necessary. It enables intelligent perception, networking, collaborative control, and airspace applications. Its development depends on reliable UAV communication and collaboration. \cite{10040631} proposed an ISAC channel estimation method considering jitter effects to address attitude and motion instability in UAV communications, improving estimation accuracy and reducing complexity under a joint perception, communication, and control framework. \cite{10007855} proposed an age-of-information-based optimization strategy for multi-UAV-assisted wireless sensor networks, reducing average and peak information age through clustering, matching, and trajectory planning. MA-enabled ISAC can further enhance low-altitude intelligent networks. With flexible MA spatial configuration, it can mitigate channel fluctuations in dynamic environments and integrate efficient communication with high-precision sensing in multi-UAV collaboration. For example, in logistics obstacle avoidance, MA-enabled ISAC helps UAVs perceive obstacle positions and trajectories, share avoidance information through stable links, and support safe collaborative path planning.

\subsubsection{Integrated Space-Air-Ground Network}
An integrated space-air-ground network combines satellites, aerial platforms, and terrestrial networks into a three-dimensional communication and sensing architecture. It uses wide-area satellite coverage, aerial platform flexibility, and terrestrial network processing capability to provide continuous and reliable services for transportation, urban management, the low-altitude economy, and emergency response. \cite{9849675} discussed beamforming for improving communication capacity and reliability in 6G from the perspective of antenna array technology, and analyzed challenges in antenna design, multiple access, and network-layer optimization. \cite{Xiao2024SAGIN6G} reviewed the evolution of integrated space-air-ground networks, covering millimeter-wave and terahertz communication, edge computing, spectrum sharing, and mobility management, and demonstrated their potential in water resource monitoring and search and rescue. MA-enabled ISAC can improve communication and sensing collaboration in satellite-UAV-ground networks by flexibly adjusting MA spatial positions. For example, in emergency rescue, it can help LEO satellites and aerial drones perceive disaster areas, establish temporary communication links, and share key data through ground base stations, improving rescue reliability and real-time response.

\subsection{MA-Enabled ISAC Systems}
To comprehensively explore the design mechanisms of MA-enabled ISAC systems, this section delves into three perspectives: system modeling, antenna positioning, and beamforming design.
\subsubsection{Signal and Channel Models}
\begin{figure}[t]
    \centering
    \includegraphics[width=8.5cm]{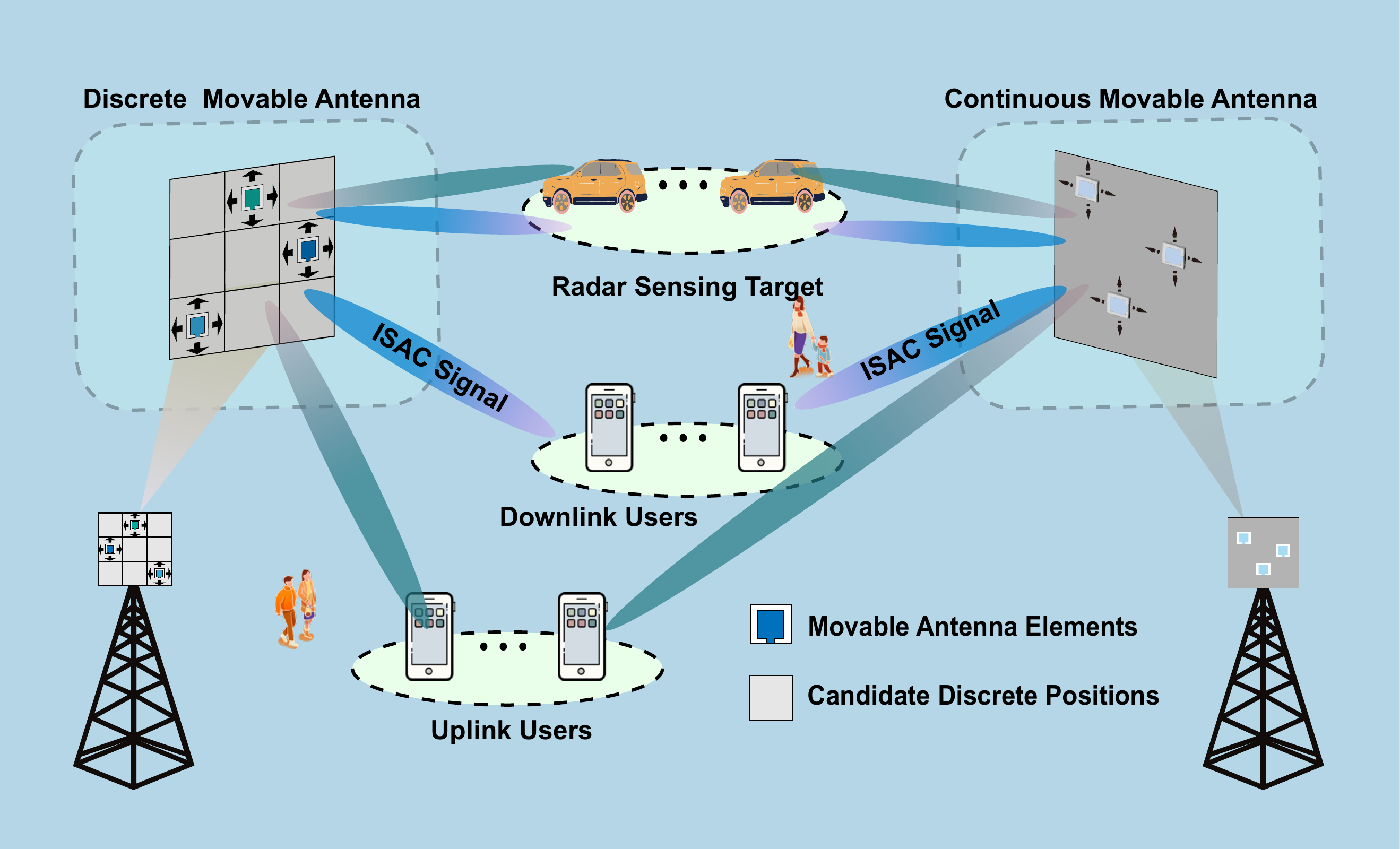}
    \caption{Discrete MA-enabled ISAC systems and continuous MA-enabled ISAC systems.}
    \label{fig:MA2}
    \vspace{-1em}
\end{figure}

In MA-enabled ISAC systems, the construction of the system model serves as the foundation for subsequent resource allocation and performance evaluation. According to the mobility paradigm of MA elements, existing system models can generally be categorized into two types: discrete MA-enabled ISAC systems and continuous MA-enabled ISAC systems, as shown in Fig.~\ref{fig:MA2}. Discrete modeling restricts the positions of MA elements to a finite set of candidate positions, which is more compatible with practical hardware implementation, whereas continuous modeling allows MA elements to move within a prescribed continuous region and therefore provides higher spatial DoF. In the following, a 2D MA-enabled ISAC system is considered as an illustrative example.

For ease of presentation, assume that the BS is equipped with $N_{\mathrm t}$ transmit MA elements. The BS transmits downlink ISAC signals to communicate with $D$ single-antenna downlink users while sensing one radar target. The narrowband ISAC signal transmitted by the BS is denoted by $\mathbf{x}\in\mathbb{C}^{N_{\mathrm t}\times 1}$ and can be expressed as
\begin{equation}
\mathbf{x}
=
{\sum}_{d=1}^{D}\mathbf{w}_{d}s_{d}
+
\mathbf{s}_{0},
\end{equation}
where $\mathbf{w}_{d}\in\mathbb{C}^{N_{\mathrm t}\times 1}$ denotes the transmit beamforming vector associated with the $d$-th downlink user. The symbol $s_{d}$ denotes the information transmitted to the $d$-th user and satisfies $\mathbb{E}\{|s_{d}|^{2}\}=1$. Moreover, $\mathbf{s}_{0}\in\mathbb{C}^{N_{\mathrm t}\times 1}$ denotes the dedicated sensing signal transmitted toward the radar target, whose covariance matrix is defined as $\mathbf{W}_{0}\triangleq\mathbb{E}\{\mathbf{s}_{0}\mathbf{s}_{0}^{H}\}\succeq\mathbf{0}$. The sensing signal $\mathbf{s}_{0}$ is assumed to be independent of $\{s_{d}\}_{d=1}^{D}$.

For the discrete MA-enabled ISAC system, let $\mathbf{P}_{\mathrm t}=[\mathbf{p}_{\mathrm t,1},\ldots,\mathbf{p}_{\mathrm t,N}]\in\mathbb{R}^{2\times N}$ denote the candidate-position matrix, where $\mathbf{p}_{\mathrm t,n}=[x_{\mathrm t,n},y_{\mathrm t,n}]^{T}$ is the $n$-th candidate position. The actual position of the $n_{\mathrm t}$-th transmit MA element is denoted by $\mathbf{t}_{\mathrm t,n_{\mathrm t}}=[t_{\mathrm x,n_{\mathrm t}},t_{\mathrm y,n_{\mathrm t}}]^{T}$ and is determined by
$\mathbf{t}_{\mathrm t,n_{\mathrm t}}=\mathbf{P}_{\mathrm t}\mathbf{b}_{\mathrm t,n_{\mathrm t}}$, where
$\mathbf{b}_{\mathrm t,n_{\mathrm t}}
=[b_{\mathrm t,n_{\mathrm t}}[1],\ldots,b_{\mathrm t,n_{\mathrm t}}[N]]^{T}\in\{0,1\}^{N}$
is the binary position-selection vector satisfying
$\sum_{n=1}^{N}b_{\mathrm t,n_{\mathrm t}}[n]=1$
for all $n_{\mathrm t}\in\{1,\ldots,N_{\mathrm t}\}$.
Let $\theta,\phi\in[-\pi/2,\pi/2]$ denote the elevation and azimuth angles of the radar target, respectively. Taking the first transmit MA element as the phase reference, the normalized transmit steering vector is expressed as
\begin{equation}
\mathbf{a}_{\mathrm t}(\theta,\phi)
=
\frac{1}{\sqrt{N_{\mathrm t}}}
\left[
1,
e^{j\frac{2\pi}{\lambda}\Delta_{\mathrm t,2}(\theta,\phi)},
\ldots,
e^{j\frac{2\pi}{\lambda}\Delta_{\mathrm t,N_{\mathrm t}}(\theta,\phi)}
\right]^{T},
\end{equation}
where
$\Delta_{\mathrm t,n_{\mathrm t}}(\theta,\phi)
=
(t_{\mathrm x,n_{\mathrm t}}-t_{\mathrm x,1})\cos\theta\sin\phi
+
(t_{\mathrm y,n_{\mathrm t}}-t_{\mathrm y,1})\sin\theta$
denotes the propagation-distance difference between the $n_{\mathrm t}$-th transmit MA element and the reference element.

The downlink channel matrix between the $D$ users and the $N_{\mathrm t}$ transmit MA elements is
$\mathbf{H}
=
[\mathbf{h}_{1}(\mathbf{t}_{\mathrm t,1}),\ldots,
\mathbf{h}_{N_{\mathrm t}}(\mathbf{t}_{\mathrm t,N_{\mathrm t}})]
\in\mathbb{C}^{D\times N_{\mathrm t}}$,
where
$\mathbf{h}_{n_{\mathrm t}}(\mathbf{t}_{\mathrm t,n_{\mathrm t}})
=
[h_{n_{\mathrm t},1}(\mathbf{t}_{\mathrm t,n_{\mathrm t}}),\ldots,
h_{n_{\mathrm t},D}(\mathbf{t}_{\mathrm t,n_{\mathrm t}})]^{T}
\in\mathbb{C}^{D\times1}$
denotes the channel vector associated with the $n_{\mathrm t}$-th transmit MA element. To characterize the channels over all candidate positions, define the extended channel matrix as
$\widehat{\mathbf{H}}
=
[\widehat{\mathbf{H}}_{1},\ldots,\widehat{\mathbf{H}}_{N_{\mathrm t}}]
\in\mathbb{C}^{D\times NN_{\mathrm t}}$,
where
$\widehat{\mathbf{H}}_{n_{\mathrm t}}
=
[\mathbf{h}_{n_{\mathrm t}}(\mathbf{p}_{\mathrm t,1}),\ldots,
\mathbf{h}_{n_{\mathrm t}}(\mathbf{p}_{\mathrm t,N})]
\in\mathbb{C}^{D\times N}$.
For $L_{p}$ propagation paths between the $n_{\mathrm t}$-th transmit MA element and the $d$-th user, the channel coefficient corresponding to the $n$-th candidate position is modeled as
\begin{equation}
h_{n_{\mathrm t},d}(\mathbf{p}_{\mathrm t,n})
=
{\sum}_{l_{p}=1}^{L_{p}}
\sqrt{\alpha_{d,l_{p}}}
e^{j\frac{2\pi}{\lambda}\Delta_{\mathrm t,n,l_{p}}},
\label{bxindao}
\end{equation}
where
$\Delta_{\mathrm t,n,l_{p}}
=
(x_{\mathrm t,n}-x_{\mathrm t,1})
\cos\theta_{d,l_{p}}^{D}\sin\phi_{d,l_{p}}^{D}
+
(y_{\mathrm t,n}-y_{\mathrm t,1})
\sin\theta_{d,l_{p}}^{D}$,
and $\alpha_{d,l_{p}}$ denotes the path power gain of the $l_{p}$-th propagation path. The position-selection matrix is
\begin{equation}
\mathbf{B}_{\mathrm t}
=
\begin{bmatrix}
\mathbf{b}_{\mathrm t,1} & \mathbf{0}_{N} & \cdots & \mathbf{0}_{N}\\
\mathbf{0}_{N} & \mathbf{b}_{\mathrm t,2} & \cdots & \mathbf{0}_{N}\\
\vdots & \vdots & \ddots & \vdots\\
\mathbf{0}_{N} & \mathbf{0}_{N} & \cdots & \mathbf{b}_{\mathrm t,N_{\mathrm t}}
\end{bmatrix}
\in\{0,1\}^{NN_{\mathrm t}\times N_{\mathrm t}},
\end{equation}
where $\mathbf{0}_{N}\in\mathbb{R}^{N\times1}$. Accordingly, the channel matrix associated with the selected MA positions is given by
$\mathbf{H}=\widehat{\mathbf{H}}\mathbf{B}_{\mathrm t}$.
Let
$\widehat{\mathbf{h}}_{d}
\triangleq[\widehat{\mathbf{H}}]_{d,:}
\in\mathbb{C}^{1\times NN_{\mathrm t}}$
denote the $d$-th row of the extended channel matrix. The received signal at the $d$-th downlink user is then expressed as
\begin{equation}
\begin{aligned}
y_{d,\mathrm D}^{\mathrm{DL}}
={}&
\widehat{\mathbf{h}}_{d}
\mathbf{B}_{\mathrm t}
\mathbf{w}_{d}s_{d}
+
{\sum}_{\substack{d'=1\\d'\neq d}}^{D}
\widehat{\mathbf{h}}_{d}
\mathbf{B}_{\mathrm t}
\mathbf{w}_{d'}s_{d'}
\\
&+
\widehat{\mathbf{h}}_{d}
\mathbf{B}_{\mathrm t}
\mathbf{s}_{0}
+
n_{d},
\end{aligned}
\end{equation}
where
$n_{d}\sim\mathcal{CN}(0,\sigma_{d}^{2})$
denotes additive white Gaussian noise (AWGN) at the receiver.

Unlike the discrete position configuration, the continuous MA-enabled ISAC system allows the antenna elements to be deployed at arbitrary positions within a prescribed feasible region, thereby enabling fine-grained spatial configuration and continuous beamforming. Let
$\mathbf{T}_{\mathrm t}
=
[\mathbf{t}_{\mathrm t,1},\ldots,\mathbf{t}_{\mathrm t,N_{\mathrm t}}]
\in\mathbb{R}^{2\times N_{\mathrm t}}$
denote the transmit MA position matrix, where
$\mathbf{t}_{\mathrm t,n_{\mathrm t}}
=
[t_{\mathrm x,n_{\mathrm t}},t_{\mathrm y,n_{\mathrm t}}]^{T}
\in\mathcal{C}_{\mathrm t}$
and $\mathcal{C}_{\mathrm t}$ denotes the feasible movement region. To avoid physical overlap and excessive mutual coupling, the distance between any two transmit MA elements is constrained by
$\|\mathbf{t}_{\mathrm t,i}-\mathbf{t}_{\mathrm t,j}\|_{2}\geq d_{\min}$
for all $i\neq j$, where $d_{\min}$ denotes the minimum allowable inter-element distance. Taking the first transmit MA element as the phase reference, the normalized transmit steering vector is expressed as
\begin{equation}
\mathbf{a}_{\mathrm t}(\theta,\phi)
=
\frac{1}{\sqrt{N_{\mathrm t}}}
\left[
1,
e^{j\frac{2\pi}{\lambda}\Delta_{\mathrm t,2}(\theta,\phi)},
\ldots,
e^{j\frac{2\pi}{\lambda}\Delta_{\mathrm t,N_{\mathrm t}}(\theta,\phi)}
\right]^{T},
\end{equation}
where
$\Delta_{\mathrm t,n_{\mathrm t}}(\theta,\phi)
=
(t_{\mathrm x,n_{\mathrm t}}-t_{\mathrm x,1})
\cos\theta\sin\phi
+
(t_{\mathrm y,n_{\mathrm t}}-t_{\mathrm y,1})
\sin\theta$.

The downlink channel matrix associated with the continuous MA configuration is
$\mathbf{H}(\mathbf{T}_{\mathrm t})
=
[\mathbf{h}_{1}(\mathbf{t}_{\mathrm t,1}),\ldots,
\mathbf{h}_{N_{\mathrm t}}(\mathbf{t}_{\mathrm t,N_{\mathrm t}})]
\in\mathbb{C}^{D\times N_{\mathrm t}}$,
where
$\mathbf{h}_{n_{\mathrm t}}(\mathbf{t}_{\mathrm t,n_{\mathrm t}})
=
[h_{n_{\mathrm t},1}(\mathbf{t}_{\mathrm t,n_{\mathrm t}}),\ldots,
h_{n_{\mathrm t},D}(\mathbf{t}_{\mathrm t,n_{\mathrm t}})]^{T}
\in\mathbb{C}^{D\times1}$
denotes the channel vector between the $n_{\mathrm t}$-th transmit MA element and the $D$ downlink users. For the $d$-th downlink user, the channel coefficient associated with the $n_{\mathrm t}$-th transmit MA element is modeled as
\begin{equation}
h_{n_{\mathrm{t}},d}
\left(\mathbf{t}_{\mathrm{t},n_{\mathrm{t}}}\right)
=
\sum_{l_p=1}^{L_p}
\sqrt{\alpha_{d,l_p}}
\exp\left(
j\frac{2\pi}{\lambda}
\Delta_{\mathrm{t},n_{\mathrm{t}},l_p}
\right).
\end{equation}
where
$\Delta_{\mathrm t,n_{\mathrm t},l_{p}}
=
(t_{\mathrm x,n_{\mathrm t}}-t_{\mathrm x,1})
\cos\theta_{d,l_{p}}^{D}\sin\phi_{d,l_{p}}^{D}
+
(t_{\mathrm y,n_{\mathrm t}}-t_{\mathrm y,1})
\sin\theta_{d,l_{p}}^{D}$.
Let
$\mathbf{h}_{d}(\mathbf{T}_{\mathrm t})
\triangleq[\mathbf{H}(\mathbf{T}_{\mathrm t})]_{d,:}
\in\mathbb{C}^{1\times N_{\mathrm t}}$
denote the channel row vector between the transmit MA array and the $d$-th downlink user. The received signal at the $d$-th user is then expressed as
\begin{equation}
\begin{aligned}
y_{d,\mathrm C}^{\mathrm{DL}}
={}&
\mathbf{h}_{d}(\mathbf{T}_{\mathrm t})
\mathbf{w}_{d}s_{d}
+
{\sum}_{\substack{d'=1\\d'\neq d}}^{D}
\mathbf{h}_{d}(\mathbf{T}_{\mathrm t})
\mathbf{w}_{d'}s_{d'}
\\
&+
\mathbf{h}_{d}(\mathbf{T}_{\mathrm t})
\mathbf{s}_{0}
+
n_{d},
\end{aligned}
\end{equation}
where
$n_{d}\sim\mathcal{CN}(0,\sigma_{d}^{2})$
denotes the AWGN at the $d$-th downlink user.

In summary, the system models of MA-enabled ISAC systems provide the theoretical foundation and operational framework for subsequent algorithm design and performance optimization. The differences between the discrete and continuous models are mainly reflected in their optimization variables and solution strategies. In the discrete model, the spatial positions of the MA elements are represented by binary position-selection vectors or matrices, which better align with practical deployment requirements but generally result in combinatorial optimization problems. In contrast, the continuous model treats the physical coordinates of the MA elements as continuous optimization variables, enabling fine-grained position or trajectory control and providing higher spatial DoF. Therefore, the appropriate modeling approach should be selected according to the performance requirements, hardware constraints, available computational resources, and system complexity of the considered application.
\subsubsection{Antenna Positioning}

In MA-enabled ISAC systems, antenna positioning is a key technique for enabling dynamic spatial control. Existing MA positioning methods can be mainly classified into mathematical optimization-based algorithms, heuristic methods, and machine learning-based algorithms.

Mathematical optimization-based algorithms formulate MA positioning as a structured optimization problem and solve it using tools such as convex optimization, fractional programming, gradient projection, and alternating optimization (AO). For example, \cite{11086422} employed boundary traversal search and Rosen gradient projection for MA position optimization in ISAC systems. In \cite{11038923}, fractional programming and constrained grid search were used for joint beamforming and MA position optimization. Similarly, \cite{Lyu2025MAISAC} adopted fractional programming, AO, and projected gradient ascent to improve system performance. These methods provide theoretical formulations and reliable performance gains, but their complexity may become high in large-scale or non-convex scenarios.

Heuristic methods are suitable for high-dimensional, discrete, or non-differentiable MA positioning problems. They obtain high-quality suboptimal solutions through rule-based or intelligent search strategies. For instance, \cite{11075964} proposed a two-layer stochastic positioning algorithm combined with AO and greedy matching for near-field ISAC. A multi-stage grid refinement strategy was developed in \cite{10978811}, while \cite{10508218} converted MA positioning into a graph-based shortest path problem. These approaches offer good adaptability to complex optimization landscapes with moderate computational burden.

Machine learning-based algorithms learn the relationship between MA configurations and system performance from data, making them attractive for dynamic and imperfect CSI scenarios. Representative methods include meta learning, deep reinforcement learning, and actor-critic frameworks. For example, \cite{amhaz2025enhancingcomprsmaperformancemovable} applied gradient-based meta learning for large-scale MA optimization, while \cite{xiu2025movableantennaaidedcooperativeisac} and \cite{11162069} adopted deep reinforcement learning methods for cooperative ISAC and UAV-assisted data collection. Although these methods offer adaptability and generalization potential, their performance depends on data quality, model design, and training complexity.

\subsubsection{Beamforming Design}

Beamforming is another fundamental technique in MA-enabled ISAC systems, as it controls the spatial distribution of transmitted or received signals and directly affects communication quality and sensing performance. Existing beamforming designs can be broadly divided into mathematical optimization-based algorithms and machine learning-based algorithms.

Mathematical optimization-based algorithms typically formulate beamforming as a constrained optimization problem and solve it using AO, semidefinite relaxation, successive convex approximation, fractional programming, or penalty dual decomposition. For example, \cite{Qin2024MAISAC} proposed an AO-based joint beamforming and MA positioning method for multi-target ISAC systems. In \cite{guo2025movableantennaenhancednetworked}, transmit beamforming, power allocation, receive filters, and MA configurations were jointly optimized for full-duplex networked ISAC. Moreover, \cite{11174346} developed a double-loop algorithm based on penalty dual decomposition, while \cite{10496515} used an AO framework combining semidefinite relaxation and majorization minimization. These methods achieve reliable performance improvements, but may suffer from high computational complexity in large-scale and highly coupled ISAC systems. Beyond single-cell settings, joint MA-position and beamforming optimization has also been shown to effectively suppress inter-cell interference and simplify transmitter design in interference networks \cite{10595399}.

Machine learning-based algorithms aim to learn efficient beamforming policies from data and are suitable for complex dynamic scenarios. Deep reinforcement learning, model-driven learning, data-driven learning, and hybrid learning have been widely explored to reduce computational cost while maintaining competitive performance. For instance, \cite{bai2025ageinformationminimizationuavenabled} used the soft actor-critic algorithm for joint UAV trajectory and beamforming optimization. In \cite{10857301}, a model-driven learning method was developed by unrolling the alternating direction method of multipliers for joint waveform and reconfigurable intelligent surface (RIS) beamforming design. In addition, \cite{10902060} proposed a deep learning framework for dual-functional symbol-level precoding. Overall, learning-based methods improve real-time adaptability, but their effectiveness depends on training data, network architecture, and convergence stability. Representative studies on antenna positioning and beamforming design for MA-enabled ISAC systems, together with their application scenarios and main methods, are summarized in Table~\ref{tab:positioning_beamforming}.

% Make the X column vertically centered
\renewcommand{\tabularxcolumn}[1]{m{#1}}

\begin{table*}[t]
\centering
\caption{Representative Studies on Antenna Positioning and Beamforming Design for MA-Enabled ISAC}
\label{tab:positioning_beamforming}
\renewcommand{\arraystretch}{1.4}
\setlength{\tabcolsep}{5pt}

\begin{tabularx}{\textwidth}{
|>{\raggedright\arraybackslash}m{1.0cm}
|>{\raggedright\arraybackslash}m{2.6cm}
|>{\raggedright\arraybackslash}m{5.0cm}
|>{\raggedright\arraybackslash}X|
}
\hline

\textbf{Ref.} &
\textbf{Design Aspect} &
\textbf{Scenario} &
\textbf{Main Method} \\
\hline

\cite{11086422} &
Antenna Positioning &
MA-enabled ISAC systems &
Boundary traversal search and Rosen gradient projection \\
\hline

\cite{11038923} &
Antenna Positioning &
MA-enabled ISAC systems &
Fractional programming and constrained grid search \\
\hline

\cite{11075964} &
Antenna Positioning &
Near-field MA-enabled ISAC systems &
Two-layer stochastic positioning, alternating optimization, and greedy matching \\
\hline

\cite{xiu2025movableantennaaidedcooperativeisac} &
Antenna Positioning &
Cooperative MA-enabled ISAC systems &
Deep reinforcement learning \\
\hline

\cite{Qin2024MAISAC} &
Beamforming Design &
Multi-target MA-enabled ISAC systems &
Alternating optimization-based joint beamforming and MA positioning \\
\hline

\cite{guo2025movableantennaenhancednetworked} &
Beamforming Design &
Full-duplex networked MA-enabled ISAC systems &
Joint optimization of transmit beamforming, power allocation, receive filters, and MA configurations \\
\hline

\cite{11174346} &
Beamforming Design &
MA-enabled ISAC systems &
Double-loop penalty dual decomposition \\
\hline

\cite{10496515} &
Beamforming Design &
MA-enabled ISAC systems &
Alternating optimization, semidefinite relaxation, and majorization-minimization \\
\hline

\end{tabularx}
\end{table*}

\subsection{Chapter Summary}
This section first reviewed the hardware architectures and classifications of MA and highlighted their key differences from conventional FPAs. By dynamically adjusting antenna positions, MA can provide additional spatial DoF, enabling greater spatial flexibility, coverage, and adaptability in dynamic wireless environments. Representative application scenarios of MA-enabled ISAC were then discussed, including digital agriculture, smart healthcare, smart homes, smart factories, IoV, smart oceans, low-altitude intelligent networks, and integrated space-air-ground networks. Finally, recent studies on signal and channel modeling, antenna positioning, and beamforming for MA-enabled ISAC systems were reviewed, highlighting their potential to enhance communication performance, sensing accuracy, and interference management.

\section{Parameter Estimation of MA-Enabled ISAC Systems}
This section focuses on parameter estimation in MA-enabled ISAC systems, covering both communication and sensing parameters. To fully exploit the spatial DoFs of MA, accurate channel and sensing information is essential for antenna position optimization and joint communication-sensing design. While exhaustive antenna movement can acquire channel and sensing information over all possible positions, it incurs prohibitive training and measurement overhead as the moving regions grow. To address this challenge, MA-enabled ISAC systems can recover the required information from a limited number of sampled positions by exploiting the inherent spatial correlation and structural sparsity of wireless channels and sensing signals \cite{10753482}. This motivates the efficient estimation of key communication and sensing parameters from received signals rather than measuring every possible antenna position. Existing parameter estimation methods can be broadly categorized into least-squares (LS), subspace-based, compressed sensing (CS), and tensor decomposition methods, each offering different tradeoffs between estimation accuracy and computational complexity.

\subsection{Unified Mathematical Model Construction}
Consider a MA-enabled ISAC system, where a BS equipped with \( N_\text{BS}^\text{t}+N_\text{BS}^\text{r} \) MAs serves an uplink user equipped with $N^\text{t}_\text{UE}$ for communication and simultaneously detects $R$ targets for radar sensing. The transmitter sends signals over the channel with the dual purpose for communicating an information message to the users, and enabling sensing of targets or the environment by generating back-scattered signals for the sensors. The  received signal at the BS can be expressed by
\begin{equation}
\begin{aligned}
        \mathbf{Y}_{k,m}=\mathbf{H}_{k,m}\mathbf{X} +\mathbf{G}_{k,m}\mathbf{S}+\mathbf{N},
\end{aligned}
\end{equation}
where \( \mathbf{H}_{k,m}\) is the spatially correlated MIMO communication channel between BS and the user, and \( \mathbf{G}\) denotes the radar sensing channel between BS and the targets. $\mathbf{X}=\left[\mathbf{X}_\text{p},\mathbf{X}_\text{d}\right]$ is the communication transmitting signal which contains shared training pilots and data symbols. Similarly, $\mathbf{S}$ denotes the sensing transmitting signal, which is identical to the training pilot $\mathbf{X}_\text{p}$ in the communication system. Expanding the exponential factors in (\ref{bxindao}) yields the explicit structure of the channel matrix, which contains the information of various key parameters.  Specifically, the time domain communication channel at the $m$-th time slot in the $n$-th time block on $k$-th subcarrier can be represented as 
\begin{equation}
    \mathbf{H}_{k,m}={\sum}_{p = 1}^{P}\mathbf{f}_\text{c}(\mathbf{r}_n;\theta_p) \boldsymbol{\Sigma}(\alpha_p, \tau_p, \nu_p)\mathbf{g}_\text{c}(\mathbf{t}_n;\phi_p)^T,
    \label{tongxin}
\end{equation}
where $\mathbf{g}_\text{c} \in \mathbb{C}^{N^\text{t}_\text{UE}\times 1}$ and $\mathbf{f}_\text{c}\in \mathbb{C}^{N^\text{r}_\text{BS}\times 1}$ denote the steering vectors in MA-enabled ISAC system, and $\boldsymbol{\Sigma}$ denotes the path response as
\begin{equation}
    \boldsymbol{\Sigma}(\alpha_p, \tau_p, \nu_p) = \alpha_{p}e^{j2\pi \nu_{p}(\tau_p+(m-1)N_\text{s}T_\text{s})-j2\pi \frac{k}{K_\text{t}}f_s\tau_p},
\end{equation}
with $\alpha$, $\theta$, $\phi$, $\nu$, $\tau$ being the channel gain, AoA, AoD, Doppler shift, and time delay, respectively. $N_s$ denotes the number of symbols in a time slot and $T_s$ denotes symbol duration.
Similarly, the sensing channel is modeled as
\begin{equation}
    \begin{aligned}
    \mathbf{G}_{k,m}={\sum}_{r = 1}^{R}\mathbf{f}_\text{s}(\mathbf{r}_n;\theta_r) \boldsymbol{\Sigma}(\beta_r, V_r, D_r)\mathbf{g}_\text{s}(\mathbf{t}_n;\phi_r)^T,
    \label{ganzhi}
    \end{aligned}
\end{equation}
where $\mathbf{f}_\text{s}\in \mathbb{C}^{N^\text{r}_\text{BS}\times 1}$ and $\mathbf{g}_\text{s}\in \mathbb{C}^{N^\text{t}_\text{BS}\times 1}$ are defined analogously to those in the communication channel . Echo signals reflected from targets are captured by spatially distributed \(N^\text{r}_\text{BS}\) receive antennas at the BS, which mitigate self-interference resulting from transmitter leakage\cite{8827589}.  $\beta_r$, $V_r$ and $D_r$ denote the reflection coefficient, velocity and distance of the target. From a physical standpoint, the time delay and frequency shift of the received echo correspond directly to the target’s range and radial velocity, respectively, which provides the foundational relationship for parameter estimation in sensing systems. 

From a communication theory perspective, rather than estimating the complete channel matrix \(\mathbf{H}_{k,m}\), one may pursue the estimation of the underlying physical parameters, such as angles, time delays, and Doppler shifts, which fully characterize the channel\cite{9296833}. These parameters are fundamentally aligned with those required for target parameter estimation and establish an intrinsic link to the range, velocity, and angular parameters of the sensing targets\cite{10403776}. From (\ref{tongxin}) and (\ref{ganzhi}), it is notably identified that the communication and sensing channels share a similar mathematical structure. This commonality establishes a basis for developing a unified algorithmic framework to estimate the unknown channel and target parameters. Therefore, the same training pilots can be transmitted and processed using unified parameter estimation approaches applied to the received signals, simultaneously achieving both channel estimation and target sensing. Next, we take the communication system as an example to introduce various parameter estimation methods.

\subsection{Parameter Estimation Algorithms}
\subsubsection{LS-based Algorithm} The LS method is one of the most fundamental and widely applied deterministic approaches in the field of parameter estimation\cite{ 1033876,10233765,11012849}. In parameter estimation, the classical LS method directly computes the channel response matrix, from which important parameter information such as delay and Doppler shift can be further recovered. It does not depend on assumptions regarding the probability distributions of the signal or noise. Instead, it solely aims to achieve an optimal fit to the observed data under the given model structure. In \cite{10233765}, the authors introduced an LS-based technique to estimate channel parameters in multi-paths mmWave MA systems. Specifically, it initially acquires channel gain estimates at several ports, which are then employed both for full channel reconstruction and for selecting the optimal antenna port, thereby massively reducing the required channel estimation count. The authors of \cite{11012849} introduced a maximum likelihood-based angle estimation scheme for MA systems. The likelihood function is first derived for arbitrary MA placements. Subsequently, the AoD-AoA pair that maximizes this function is identified as the angles associated with the underlying channel.
 
A direct and straightforward approach to estimate \( \mathbf{H}_{k,m} \) requires the user to transmit orthogonal pilot sequences $\mathbf{X}_{k,m}$ while switching the radiating elements synchronously across all \( N^\text{t}_\text{UE} \) ports of the transmitting side. The key idea of the LS-based algorithm is to solve the following optimization problem
\begin{equation}
   \hat{ \mathbf{H}}_{k,m} = \arg\min \|\mathbf{Y}_{k,m} - \mathbf{H}_{k,m}\mathbf{X}_{k,m}\|_F^2.
\end{equation}
 Upon receiving the pilot sequences at the BS, the LS estimate of \( \mathbf{H}_{k,m} \) can be obtained by post-multiplying the normalized received signal matrix by the pseudo-inverse of the corresponding pilot sequence. 
 
 However, to simultaneously achieve channel parameter estimation in the communication system and target parameter estimation in the sensing system, we need to further extract the corresponding parameter information from the channel matrix. Due to the matched filtering property of the Fourier transform, a complex exponential signal, after undergoing Fourier transformation, produces an impulsive peak at its corresponding frequency. Since delay and Doppler shift correspond to complex exponential frequencies in the frequency domain and time domain, respectively, utilizing the inverse fast Fourier transform (IFFT) and fast Fourier transform (FFT) can yield peaks at the coordinates ($\tau_p$, $\nu_p$), thereby enabling parameter extraction. Other angle-domain parameters can also be obtained using correlation-based methods. It is evident that such a parameter estimation strategy incurs substantial computational overhead. Consequently, this approach is only suitable for channel estimation and impractical for parameter estimation. Therefore, adopting an LS-based parameter estimation algorithm in an ISAC system is far from cost-effective.

\subsubsection{Subspace-Based Algorithm}
The subspace-based algorithm is essentially a direct mathematical realization of the subspace concept in parameter estimation\cite{1143830, 7549}. Its core lies in explicitly decomposing the statistical space of the received signal into two orthogonal complementary subspaces and leveraging this geometric structure for parameter estimation. The multiple signal classification (MUSIC) algorithm remains the preferred choice for some certain practical systems due to its solid theoretical foundation. 
\paragraph{MUSIC Algorithm}
The MUSIC algorithm operates by decomposing the covariance matrix of the received signal into a signal subspace \(\mathbf{U}_s\) and a noise subspace \(\mathbf{U}_n\). Based on the orthogonality between these two subspaces, the target parameters can be estimated from the MUSIC spectrum \(P(\mathbf{z})\), where \(\mathbf{z}\) denotes the vector of parameters of interest. As an example, the authors in \cite{Ma2024MASensing} employed the  MUSIC algorithm to extract spatial parameter information from the received signal, and subsequently optimized the antenna positions with the objective of minimizing the CRB of the estimated parameters. Besides \cite{Ma2024MASensing}, the authors in \cite{11162470} also employed MUSIC for user localization under the near-field propagation condition with MAs. In summary, MUSIC algorithm generally involves the following steps:
(i) Construct the joint steering vector $\mathbf{a}(\overline{\mathbf{z}})$, where $\overline{\mathbf{z}}$ is the search vector that sweeps over all possible values of the parameters to be estimated.
(ii) Compute the covariance matrix of the received signal.
(iii) Perform eigenvalue decomposition (EVD) on the covariance matrix and arrange the eigenvalues in descending order: $\mathbf{R} = \mathbf{U}\boldsymbol{\Lambda}\mathbf{V}^H = \mathbf{U}_s\boldsymbol{\Lambda}_s\mathbf{V}_s^H + \mathbf{U}_n\boldsymbol{\Lambda}_n\mathbf{V}_n^H$. (iv) Extract the noise subspace $\mathbf{U}_n = \mathbf{U}_{[:, P+1:\text{end}]}$ and construct the MUSIC spectrum, where $\mathbf{U}_{[:, P+1:\text{end}]}$ denotes the submatrix of $\mathbf{U}$ consisting of columns $P+1$ through the last column. To be specific, the detailed procedure of the MUSIC algorithm is elaborated as follows. In step (i), the joint steering vector $\mathbf{a}(\overline{\mathbf{z}})$ is constructed, where $\overline{\mathbf{z}}$ denotes the search vector that traverses all possible values of the parameters of interest. To enable multi-parameter joint estimation, the steering vector can be constructed via Kronecker product, i.e., $\mathbf{a}({\boldsymbol{\Theta}})\otimes\mathbf{a}({\boldsymbol{\Phi}})$, which captures the coupled structure among multiple parameters, allowing the algorithm to resolve them simultaneously. However, this operation inevitably leads to an expansion of the codebook size constituting $\overline{\mathbf{z}}$, thereby increasing computational complexity and storage requirements, especially when the parameter space is high-dimensional or finely discretized. In step (ii), the sample covariance matrix is computed as $\mathbf{R} = \frac{1}{KM}\sum_{k,m}\mathbf{Y}_{k,m} \mathbf{Y}^H_{k,m}$. It serves as the foundation for subspace decomposition, as the signal and noise subspaces are extracted from the eigen-structure of $\mathbf{R}$. Next, in step (iii), EVD is performed on the covariance matrix. Based on the principle that the signal subspace is spanned by eigenvectors corresponding to the largest eigenvalues, the decomposition is expressed as
\begin{equation}
    \mathbf{R} = 
\begin{bmatrix} \mathbf{U}_s, \mathbf{U}_n \end{bmatrix}
\begin{bmatrix}
\boldsymbol{\Gamma}_s & \mathbf{0} \\
\mathbf{0} & \boldsymbol{\Gamma}_n
\end{bmatrix}
\begin{bmatrix}
\mathbf{U}_s^H \\
\mathbf{U}_n^H
\end{bmatrix},
\end{equation}
where $\boldsymbol{\Gamma}_s$ is a diagonal matrix containing the eigenvalues associated with the signal subspace, and $\boldsymbol{\Gamma}_n$ is the diagonal matrix of eigenvalues corresponding to the noise subspace. In practice, the number of signal sources $\mathbf{R}$ determines the partitioning: the eigenvectors corresponding to the $\mathbf{R}$ largest eigenvalues form the signal subspace $\mathbf{U}_s$, while the remaining eigenvectors constitute the noise subspace $\mathbf{U}_n$. In step (iv), the noise subspace $\mathbf{U}_n$ is extracted and used to construct the MUSIC spatial spectrum
\begin{equation}
    P(\mathbf{z}) = \frac{1}{\mathbf{a}^H({\mathbf{z})\mathbf{U}_n\mathbf{U}^H_n\mathbf{a}(\mathbf{z})}}.
\end{equation}
For the true parameter vector $\mathbf{z}$, the steering vector $\mathbf{a}(\mathbf{z})$ lies in the signal subspace and thus satisfies $\mathbf{a}(\mathbf{z})^H \mathbf{U}_n \mathbf{U}_n^H \mathbf{a}(\mathbf{z}) = 0$. Conversely, for any other vector $\tilde{\mathbf{z}}$ in the search space, $\mathbf{a}(\tilde{\mathbf{z}})$ is not orthogonal to $\mathbf{U}_n$, yielding a non-zero denominator and a finite spectrum value. This mechanism transforms the parameter estimation problem into a geometric orthogonality test, where the true signal parameters correspond to the peaks of the spatial spectrum. By sweeping through the parameter space and identifying these peaks, the parameter estimates are obtained as 
\begin{equation}
    \left(\hat{\boldsymbol{\Theta}}, \hat{\boldsymbol{\Phi}}, \hat{\boldsymbol{\nu}}, \hat{\boldsymbol{\tau}}\right) = \text{arg} \max _{\hat{\boldsymbol{\Theta}}, \hat{\boldsymbol{\Phi}}, \hat{\boldsymbol{\nu}}, \hat{\boldsymbol{\tau}}} P(\hat{\boldsymbol{\Theta}}, \hat{\boldsymbol{\Phi}}, \hat{\boldsymbol{\nu}}, \hat{\boldsymbol{\tau}}).
\end{equation}
After the angle, Doppler shift, and delay parameters are estimated, the complex gain $\alpha$ can be obtained via the LS method.
\paragraph{Estimation of Signal Parameters via Rotational Invariance Techniques (ESPRIT) Algorithm}
The ESPRIT algorithm is a high-resolution parameter estimation method that can be viewed as a specialized variant within the broader MUSIC framework \cite{4393459, 506629}. While MUSIC exploits the orthogonality between the noise subspace and the steering vectors to construct a spatial spectrum, ESPRIT leverages the rotational invariance property inherent to translationally invariant arrays, thereby transforming parameter estimation into a generalized eigenvalue problem that eliminates the need for spectral peak search and avoids grid mismatch issues. The principle of ESPRIT relies on an array structure comprising two identical subarrays with a fixed interval. By adjusting the positions of the MA array, the shift invariance requirement of the ESPRIT algorithm can be satisfied. However, antenna movement introduces corresponding time overhead, which may render it unsuitable for estimation in time-varying channels. Under this configuration, the steering vectors of the two subarrays satisfy $\mathbf{a}_2(\mathbf{z}) = \mathbf{a}_1(\mathbf{z}) \boldsymbol{\Phi}$, where $\boldsymbol{\Phi}$ is a diagonal matrix containing the phase shifts induced by the displacement, directly encoding the parameters of interest. The algorithm proceeds as follows: (i) Compute the covariance matrix $\mathbf{R}$ from the received signals and perform eigenvalue decomposition to obtain the signal subspace $\mathbf{U}_s$, which spans the same column space as the steering matrix. (ii) Partition $\mathbf{U}_s$ into two submatrices $\mathbf{U}_{s_1}$ and $\mathbf{U}_{s_2}$ according to the subarray structure. (iii) Exploiting the rotational invariance property, solve the invariance equation $\mathbf{U}_{s_1} = \mathbf{U}_{s_2} \boldsymbol{\Psi}$ for the rotation operator $\boldsymbol{\Psi}$. (iv) Compute the eigenvalues of $\boldsymbol{\Psi}$, from which the desired parameters are extracted.

Both ESPRIT and MUSIC are founded on subspace decomposition, rely on eigenvalue decomposition of the covariance matrix, and require knowledge of the number of signal sources. Their core distinctions lie in the following aspects. First, MUSIC exploits orthogonality between the noise subspace and steering vectors to construct a spatial spectrum requiring peak search, whereas ESPRIT directly utilizes the rotational invariance of the signal subspace, converting parameter estimation into an algebraic eigenvalue problem without spectral search. Next, MUSIC imposes no constraints on array geometry and is applicable to arbitrary configurations, while ESPRIT requires translational invariance.

\subsubsection{CS-Based Algorithm}
{Owing to the sparse distribution of finite channel paths in the angular domain, compressed sensing techniques are suited for the acquisition of MA channels, particularly for AoA and AoD estimation\cite{10236898, 10497534, 11148216}. The parameter estimation problem is transformed into an optimization problem of recovering a sparse vector from an underdetermined linear system. By employing greedy pursuit algorithms such as orthogonal matching pursuit (OMP), it is possible to estimate the underlying sparse parameters from the support set of the solution. In \cite{11148216}, a joint multi-domain compressed sensing strategy was employed for ISAC processing in UAV networks to reduce pilot overhead for joint channel estimation and radar sensing in MA systems. The compressed sensing-based algorithm consists of the following key steps: (i) Construct an overcomplete dictionary matrix with $G$ search grids. (ii) Perform the atom selection step in the constructed search dictionary until the channel parameter vector $\mathbf{z}$ associated with each path is screened out. (iii) Select the estimating results from the support set. 

To be specific, in step (i), the overcomplete dictionary matrix which contains all the possible combinations of the parameters can be given as $\mathbf{A}(\tilde{\mathbf{z}})\in\mathbb{C}^{Q_\text{MS}N_sKM\times G}$.  We have \( \mathbf{A}(\mathbf{z}) \mathbf{x} \approx \mathbf{A}(\tilde{\mathbf{z}}) \bar{\mathbf{x}}\), where \( \bar{\mathbf{x}}^t \in \mathbb{C}^{G} \) is a sparse vector with \( Q \) nonzero elements corresponding to \( \mathbf{x}\)\cite{9103231, 1580791}. In step (ii), atom selection is performed via greedy iteration: initialize $\mathbf{r}^{(0)} = \mathbf{y}$, $\Lambda^{(0)} = \emptyset$, $\mathbf{x}^{(0)} = \mathbf{0}$. At each iteration, select the atom with the largest inner product with the current residual, add its index to the support set $\Lambda^{(k)} = \Lambda^{(k-1)} \cup \{i_k\}$. The sparse coefficients are updated by solving a LS problem on the subspace spanned by the selected atoms, ensuring the optimal fit to the observed signal under the current support set
\begin{equation}
    \mathbf{x}_{\Lambda^{(k)}}^{(k)} = \arg \min_{\mathbf{z}} \| \mathbf{y} - \mathbf{D}_{\Lambda^{(k)}} \mathbf{z} \|_2^2 = \left( \mathbf{D}_{\Lambda^{(k)}}^H \mathbf{D}_{\Lambda^{(k)}} \right)^{-1} \mathbf{D}_{\Lambda^{(k)}}^H \mathbf{y}.
\end{equation}
Finally, the currently estimated signal is subtracted from the original observation to generate a new residual, which will then be used for the next round of correlation calculations with dictionary atoms to capture the remaining signal components $ \mathbf{r}^{(k)} = \mathbf{y} - \mathbf{D}_{\Lambda^{(k)}} \mathbf{x}_{\Lambda^{(k)}}^{(k)}$. The algorithm stops after $P$ iterations or when the residual energy falls below a predefined threshold. Once terminated, the non-zero elements of $\mathbf{x}$ correspond to estimated parameter values, while their indices in the dictionary map to specific parameter locations. Finally, in step (iii), the positions of the nonzero entries in the sparse vector $\mathbf{x}$ serve as indices to determine the estimated parameter values from the sparse dictionary. 

Note that the OMP algorithm is sensitive to the mutual correlation of the measurement matrix. High correlation may lead to incorrect atom selection. To ensure that the essential structure of the sparse signal remains undistorted after measurement and that the measurement vectors corresponding to different atoms maintain sufficient distinguishability, the OMP algorithm requires the measurement matrix to satisfy the restricted isometry property (RIP). Moreover, its computational complexity grows rapidly with signal dimensionality and dictionary size, limiting its efficiency in large‑scale problems. To overcome these drawbacks, several improved algorithms, such as regularized OMP (ROMP)\cite{5419092} and compressive sampling matching pursuit (CoSaMP) \cite{NEEDELL2009301} have been developed to enhance robustness and efficiency.

\subsubsection{Tensor Decomposition-Based Algorithm}
Noting that the received signal inherently contains multi‑dimensional information in frequency, time and spatial domains, it possesses a natural high‑dimensional structure. However, the above-mentioned algorithms collapse these dimensions through vectorization or matricization, which disrupts the inherent multidimensional structure and results in unsatisfactory estimation results and high computational complexity. To address these issues, tensor decomposition-based channel estimation algorithms, which fully exploit the high-dimensional property in MA-enabled ISAC systems, have attracted considerable research attention\cite{10659325,10839618, 10643882}. Tensor decomposition-based parameter estimation provides an advanced framework for processing multidimensional signals and data. It treats the received multidimensional data directly as a tensor and employs tensor decomposition models to extract physically interpretable parameter factors. Through tensor decomposition, we can extract the factor matrices that embed parameter information from each dimension, thereby alleviating the computational burden associated with joint multi-parameter estimation \cite{7914672, 6573422}.

Aggregating the received signals from the $K$ pilot subcarriers and $M$ time slots yields a fourth-order canonical polyadic (CP) decomposition tensor representation, which can be written as $\mathcal{Y}$, i.e.,
\begin{equation}
\mathcal{Y}={\sum}_{p=1}^P\left(\mathbf{a}_p\circ\mathbf{b}_p\circ\mathbf{c}_p\circ\mathbf{d}_p\right)+\mathcal{N},
\end{equation}
where $\circ$ denotes the outer product, and $\mathcal{N}$ represents the noise tensor. The factor matrices $ \mathbf{A}, \mathbf{B}, \mathbf{C}$, and $\mathbf{D}$ can be obtained by collecting the factor vectors. The  parameter information  is contained by these matrices, which are defined as
\begin{equation}
        \mathbf{a}_p = \mathbf{f}(\mathbf{r}_n;\theta_p),
\end{equation}
\begin{equation}
        \mathbf{b}_p = \Big[1,e^{j2\pi \nu_pN_\text{s}T_\text{s}},\ \cdots\ ,\ e^{j2\pi \nu_p(M-1)N_\text{s}T_\text{s}}\Big]^T,
\end{equation}
\begin{equation}
         \mathbf{c}_p = \alpha_pe^{j2\pi\tau_p\nu_p}\!\Big[e^{-j2\pi(f_\text{s}\tau_p)/K_\text{t}},\cdots,e^{-j2\pi K(f_\text{s}\tau_p)/K_\text{t}}\Big]^T,
\end{equation}
\begin{equation}
         \mathbf{d}_p = \mathbf{X}_\text{p}^T\mathbf{g}(\mathbf{t}_n;\phi_p).
\end{equation}

The factor vectors associated with each rank-one component are directly related to the parameters of the corresponding source along different dimensions, thereby naturally decoupling the parameters and enabling their direct estimation. We then introduce two representative tensor-based CP decomposition methods, namely the alternating least squares (ALS) and the structured canonical polyadic decomposition (SCPD).

\paragraph{ALS}
This method formulates the tensor decomposition problem as an optimization problem by iteratively solving for the factor matrices until the convergence criteria are met, at which point the resulting solution is regarded as the estimate of the factor matrices. This optimization problem can be efficiently solved through alternating minimization, which iteratively reduces the data fitting error by updating one factor matrix while keeping the remaining fixed \cite{7914672}. The iterative update procedure is outlined as follows
\begin {align}
    \hat{\mathbf{A}}_{(t+1)}&=\text{arg}\min_{\hat {\mathbf{A}}}\Big\|\mathbf{Y}_{(1)}\!-\!\hat{\mathbf{A}}(\mathbf{D}_{(t)}\odot \mathbf{C}_{(t)} \odot\mathbf{B}_{(t)})^T\Big\|_F^2,    \label{alstendec}\\
    \hat{\mathbf{B}}_{(t+1)}&=\text{arg}\min_{\hat {\mathbf{B}}}\Big\|\mathbf{Y}_{(2)}-\hat{\mathbf{B}}(\mathbf{D}_{(t)}\odot \mathbf{C}_{(t)} \odot\mathbf{A}_{(t+1)})^T\Big\|_F^2,\nonumber\\
    \hat{\mathbf{C}}_{(t+1)}&=\text{arg}\min_{\hat {\mathbf{C}}}\Big\|\mathbf{Y}_{(3)}\!-\hat{\mathbf{C}}(\mathbf{D}_{(t)}\odot \mathbf{B}_{(t+1)} \odot\mathbf{A}_{(t+1)})^T\!\Big\|_F^2,\nonumber\\
    \hat{\mathbf{D}}_{(t+1)}&=\!\text{arg}\min_{\hat {\mathbf{D}}}\!\Big\|\mathbf{Y}_{(4)}\!-\hat{\mathbf{D}}(\mathbf{C}_{(t+1)}\odot \mathbf{B}_{(t+1)} \odot\mathbf{A}_{(t+1)})^T\!\Big\|_F^2,\nonumber
\end {align}
where $\mathbf{Y}_{(n)}$ denotes the mode-$n$ unfold of tensor and  $\odot$ denotes the Khatri-Rao product. The ALS algorithm exhibits strong generality, as it is applicable to any CP decomposition problem without imposing specific structural requirements on the factor matrices. Moreover, it readily accommodates various constraints by allowing each LS step to be easily formulated as a constrained LS problem, thereby rendering it suitable for diverse and flexible application scenarios \cite{doi:10.1137/07070111X}. Although the ALS is a standard solver for CP decomposition, it is iterative and computationally intensive, and is often prone to converging to local optima. Moreover, while the ALS-based approach is theoretically straightforward and simple in operation, it fails to incorporate the inherent structural constraints of the factor matrices and imposes limitations on the maximum number of resolvable targets \cite{10403776}. Thus, we consider another algorithm that leverages the inherent Vandermonde structure of the factor matrices arising from the uniform linear array geometry and orthogonal frequency-division multiplexing (OFDM) subcarrier spacing.  Consequently, the parameters can be retrieved via closed-form eigenvalue decomposition, which is significantly more stable than iterative searches.
\paragraph{Structured Canonical Polyadic Decomposition (SCPD)}
 To overcome the limitations of ALS algorithm, we introduce an SCPD algorithm that directly computes the factor matrices via matrix algebra, thereby avoiding the iterative optimization process. This approach avoids potential convergence problems and enhances the stability of tensor decomposition. The SCPD algorithm consists of three key steps: (i) Apply spatial smoothing to address rank deficiency in the factor matrices. (ii) Extract the factor matrices that exhibit Vandermonde structure using an ESPRIT-inspired method based on rotational invariance. (iii) Recover the remaining factor matrices by exploiting Kronecker product relationships.

Specifically, some factor matrices of a tensor may be rank-deficient (e.g., $\mathbf{A}$ and $\mathbf{D}$), leading to rank deficiency in the mode-unfolded matrices of the tensor. This is undesirable for tensor decomposition \cite{6573422}.  Therefore, prior to performing tensor decomposition, step (i) must be applied to resolve this issue. Spatial smoothing is first performed by partitioning a uniform linear array into overlapping subarrays, which increases the tensor order and resolves the rank deficiency issue \cite{890366, 6573422}.  By exploiting the Vandermonde structure inherent in $\mathbf{B}$ and $\mathbf{C}$, spatial smoothing maps the factor matrices to $\mathbf{B}^{(k_1)}(\mathbf{B}^{(l_1)})^T \in \mathbb{C}^{k_1\times l_1}$ and $\mathbf{C}^{(k_2)}(\mathbf{C}^{(l_2)})^T$, where $k_1,l_1$ are the selected pairs of the spatial smoothing technique. This technique effectively alleviates the rank deficiency problem by increasing the order of the tensor. Consequently, we obtain $\mathbf{X}^{[3]} = \left(\mathbf{A} \odot \mathbf{B}^{(k_1)} \odot \mathbf{C}^{(k_2)}\right) \left(\mathbf{B}^{(l_1)} \odot \mathbf{C}^{(l_2)} \odot \mathbf{D}\right)^T$. The spatial smoothing operation can be implemented by constructing a cyclic selection matrix $\mathbf{J}$\cite{10403776}, which is formed by assembling multiple block matrices according to a predefined structure, which yields
\begin{equation}
    \begin{aligned}
         \mathbf{J}\mathbf{Y}^{[3]}\!=\!\left(\!\mathbf{A} \!\odot \!\mathbf{B}^{(k_1)} \!\odot\! \mathbf{C}^{(k_2)}\!\right) \!\boldsymbol{\Lambda}\!\left(\!\mathbf{B}^{(l_1)} \!\odot \!\mathbf{C}^{(l_2)} \!\odot\! \mathbf{D}\!\right)^T,
    \end{aligned}
\end{equation}
where $\boldsymbol{\Lambda}$ denotes a diagonal scaling matrix. After smoothing, in step (ii), we adopt an ESPRIT-inspired approach based on rotational invariance, which involves setting up and solving a generalized eigenvalue problem, to estimate the generators of the Vandermonde factor matrices $\mathbf{B}$ and $\mathbf{C}$. To be specific, let $\mathbf{X}^{[3]} = \mathbf{U} \boldsymbol{\Sigma} \mathbf{V}^H$ denote the singular value decomposition (SVD) of $\mathbf{X}^{[3]}$, where the first $P$ principal singular vectors of $\mathbf{U}$ and $\mathbf{V}$ form the basis of the signal subspaces. Through SVD, the row space and column space of the received signal are separated. Moreover, because the spatially smoothed received signal no longer suffers from rank deficiency, there exists a nonsingular matrix $\mathbf{M} \in \mathbb{C}^{R \times R}$ such that $\mathbf{UM}=\mathbf{A} \odot \mathbf{B}^{(k_1)} \odot \mathbf{C}^{(k_2)}$. Noting the structure of the Vandermonde matrix and the properties of the Khatri-Rao product, we obtain
\begin{equation}
\begin{aligned}
    \underbrace{\mathbf{A}\odot\underline{\mathbf{B}}^{(k_1)}\odot\mathbf{C}^{(k_2)}}_{ \mathbf{U}_1}\mathbf{Z}_b = \underbrace{\mathbf{A}\odot\overline{\mathbf{B}}^{(k_1)}\odot\mathbf{C}^{(k_2)}}_{\mathbf{U}_2}.\\
\end{aligned}
\end{equation}
From the relationship between $\mathbf{U}_1$ and $\mathbf{U}_2$, performing an eigenvalue decomposition yields the diagonal matrix $\mathbf{Z}_b$ containing the Vandermonde generators. After obtaining $\mathbf{B}$, the matrices $\mathbf{A}$ and $\mathbf{C}$ can be further separated by exploiting the properties of the Kronecker product and the Khatri-Rao product. Notice that $\mathbf{C}$ also exhibits a Vandermonde structure. Therefore, a similar extraction procedure can be employed to separate $\mathbf{C}$ from $\mathbf{A} \odot \mathbf{C}$. Owing to the inherent scaling ambiguity of tensor decomposition, it is reasonable to assume, without loss of generality, that the columns of $\mathbf{B}$ and $\mathbf{C}$ are normalized to unit norm. Using the estimated $\mathbf{B}$ and $\mathbf{C}$, the remaining factor matrices can be estimated via a projection operation. Taking the acquisition of the factor matrix $\mathbf{D}$ as an example, the remaining factor matrices can be obtained by
\begin{equation}
   \hat{\mathbf{d}}_p \!=\! \left(\frac{(\hat{\mathbf{b}}^{(l_1)}_p)^H}{(\hat{\mathbf{b}}^{(l_1)}_p)^H \hat{\mathbf{b}}^{(l_1)}_p} \!\otimes\! \frac{(\hat{\mathbf{c}}^{(l_2)}_p)^H}{(\hat{\mathbf{c}}^{(l_2)}_p)^H \hat{\mathbf{c}}^{(l_2)}_p} \otimes\mathbf{I}_{N_s}\right)\mathbf{V}^* \boldsymbol{\Sigma} \mathbf{n}_p.
\end{equation}

Table~\ref{table:ma} summarizes the characteristics of various parameter estimation algorithms and lists related research papers that provide specific algorithmic details.

\begin{table*}[htbp]
\centering
\caption{Representative Studies on Parameter Estimation for MA-Enabled ISAC}
\renewcommand{\arraystretch}{1.4}
\label{table:ma}
\begin{tabular}{|m{0.8cm}|m{4cm}|m{3.2cm}|m{7.5cm}|}
\hline
\textbf{Refs.} & \textbf{Research Background}  & \textbf{Estimation Method} & \textbf{Key Features} \\ 
\hline

\cite{11012849} 
& Angular parameter estimation for MA MIMO systems.
& Maximum likelihood-based algorithm 
& High computational complexity. Unsuitable for real-time dynamic scenarios.
\\ \hline

 \cite{Ma2024MASensing} 
 & Sensing parameter estimation for MA position optimization.
 & MUSIC-based algorithm 
 & Super-resolution and pilot-free properties, yet with a heavy reliance on a large number of snapshots.
 \\ \hline

 \cite{10497534} 
 & Channel parameter estimation for MA MIMO systems and measurement position setups.
 & CS-based algorithm
 & Sparse-channel-suitable properties, but sensitive to dictionary grid mismatch and orthogonality conditions.
 \\ \hline

\cite{11148216} 
& UAV parameter estimation for MA-Enabled ISAC systems.
& Joint multi-domain CS-based algorithm 
& Utilizes joint-sparsity features, but suffers from iterative complexity and slow convergence.
\\ \hline

\cite{10659325} 
&  Channel parameter estimation for MA MIMO systems.
& Tensor decomposition-based algorithm 
& Fully exploits the multidimensional structure, yielding high accuracy.
\\ \hline

\end{tabular}
\vspace{-1em}
\end{table*}

\vspace{-1em}
\subsection{Performance Analysis}
In this subsection, we take a communication system as an example to concretely demonstrate the performance of different parameter estimation algorithms. Since the channel models of communication systems and sensing systems exhibit similar mathematical structures, we will take the performance of various estimation algorithms in communication systems as examples. 

\subsubsection{Estimation Accuracy Analysis}
 We consider a BS equipped with $N_\text{BS} = {\text{12}}$ transmit MAs and a MS equipped with $N_\text{MS} = {\text{12}}$ receive MAs. The number of RF chains at the BS and MS are $Q_\text{BS} = {\text{10}}$ and $Q_\text{MS} = {\text{10}}$. The carrier frequency is set to $f_\text{c} = {\text{28}}$ GHz with the system bandwidth of $f_s = {\text{100}}$ MHz, which is divided into $K_t = {\text{2048}}$ subcarriers. Thus, the spacing between adjacent subcarriers is $\Delta f = {\text{48.8}}$ kHz. Let $M = \text{14}$ time slots in each time block and $N_s = \text{10}$ OFDM symbols in each time slot. We focus on the estimation performance of the reconstructed MA channel for these parameter estimation algorithms. In MA-enabled ISAC systems, the corresponding channels between all candidate positions at the transmitting and receiving sides can be reconstructed using the results obtained by these parameter estimation algorithms. The algorithm performance of the reconstructed channel is subsequently evaluated using the normalized mean square error (NMSE)\cite{11069254}, defined as
\begin{equation}
        \text{NMSE}_{\mathbf{H}} = \frac{\sum^M_{m=1}\sum^K_{k=1}\Big \lVert\mathbf{H}_{k,m} - \widehat{\mathbf{H}}_{k,m}\Big \rVert_F^2}{MK\sum^M_{m=1}\sum^K_{k=1}\Big \lVert \mathbf{H}_{k,m}\Big \rVert_F^2}.
\end{equation}
The superiority of the tensor decomposition-based algorithm is evident from the NMSE results shown in Fig. \ref{mseh}, where it demonstrates a significant advantage over the existing baseline algorithms. Additionally, both the MUSIC and OMP algorithms exhibit sensitivity to noise. This limitation, together with the disadvantages discussed in the preceding analysis, results in inferior performance. In contrast, the tensor decomposition-based estimation algorithm imposes more relaxed structural constraints on the model and exhibits notable robustness against noise. The SCPD algorithm achieves the best estimation accuracy in low SNR scenarios, as shown in Fig. \ref{mseh}. 

\begin{figure}
    \centering
    \includegraphics[width=0.9\linewidth]{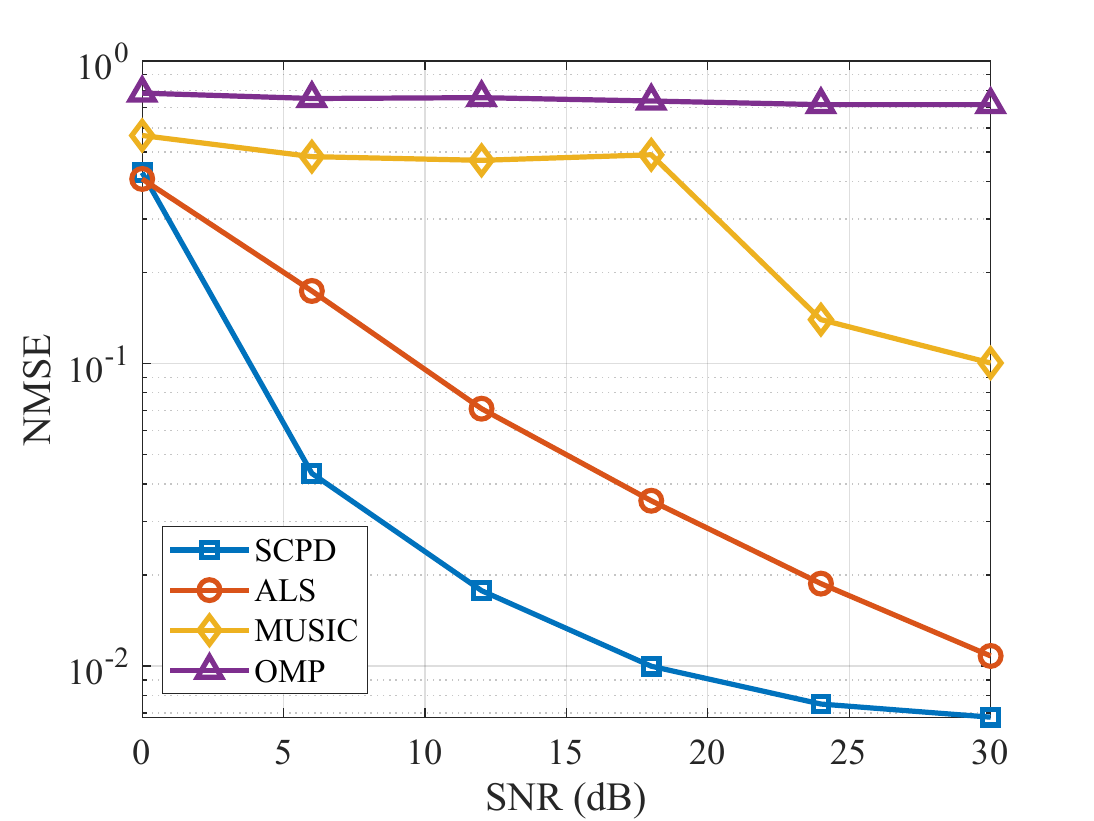}
    \caption{NMSE of channel reconstruction achieved by SCPD, ALS, MUSIC and OMP algorithms.}
    \label{mseh}
    \vspace{-1em}
\end{figure}
\begin{figure}
    \centering
    \includegraphics[width=0.9\linewidth]{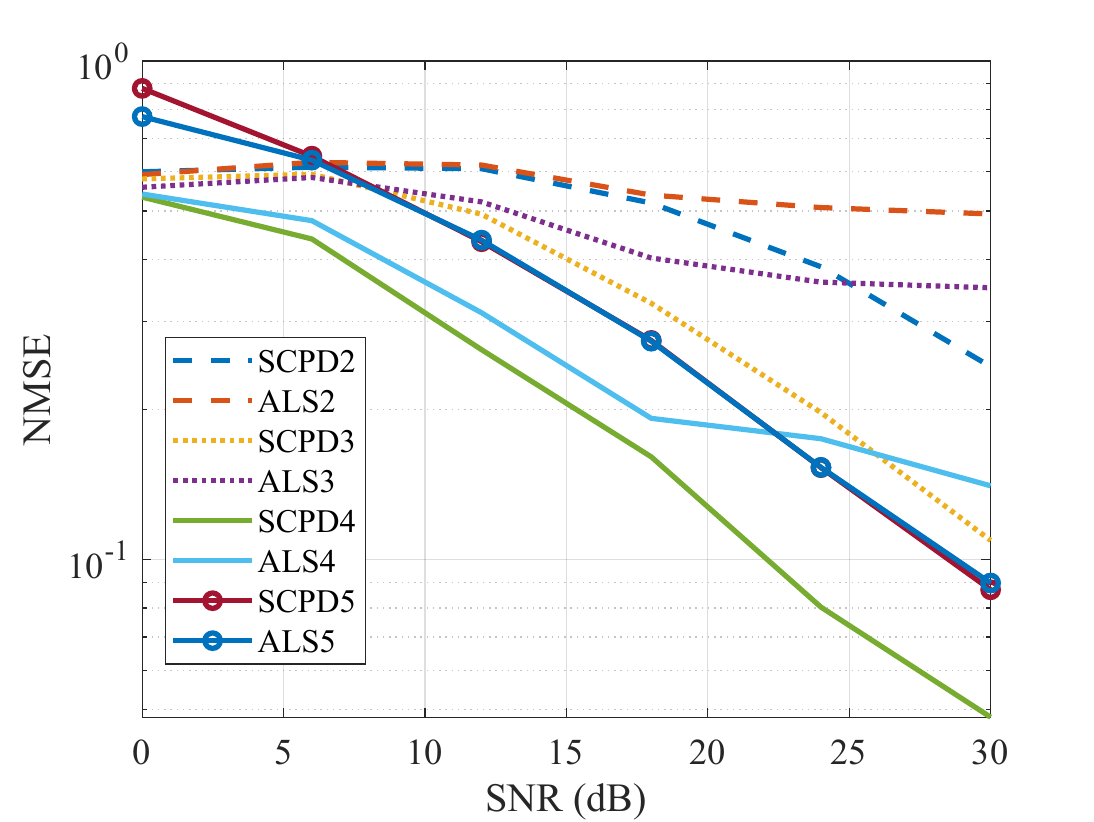}
    \caption{NMSE of channel reconstruction achieved by the tensor decomposition algorithm with different numbers of paths.} 
    \label{2345}
    \vspace{-1em}
\end{figure}

The performance of the tensor-based parameter estimation algorithm under different numbers of paths is further evaluated as shown in Fig. \ref{2345}. Both excessively large and small numbers of paths affect estimation accuracy, which is consistent with intuitive understanding. When the number of paths is too low, the rank of the precoding matrix decreases, which impairs the accuracy of the LS solution and consequently degrades the estimation accuracy. When the number of paths is too large, the number of columns of the factor matrix increases, thereby affecting the orthogonality among column vectors and also leading to inaccurate results. It is also observable that when the number of paths is \text{5}, the accuracy of the two estimation algorithms does not differ significantly. It can be anticipated that as the path number increases further, the estimation accuracy will change only marginally. However, it should be noted that as the number of paths increases, the uniqueness condition for tensor decomposition becomes difficult to guarantee, which may cause tensor decomposition-based estimation algorithms to fail. It can be proven that once the dimension such as the number of RF chains, subcarriers, and time slot are fixed, the maximum number of paths that can be determined by the tensor decomposition-based method is consequently established.

\subsubsection{Computational Complexity Analysis}
For ALS algorithm, a single iteration requires $\mathcal{O}\{Q_\text{MS}N_sKMR\}$ operations, dominated by the tensor dimensions. With $N_\text{iter}$ iterations, the total complexity is $\mathcal{O}\{N_\text{iter}Q_\text{MS}N_sKMR\}$. Parameter extraction from the factor matrices incurs an additional complexity of  $\mathcal{O}\{RGK\}$, where $G$ denotes the search grid precision. For the SCPD algorithm, the dominant computational cost lies in the spatial smoothing step, which involves constructing a cyclic selection matrix through $l_1l_2$ iterations. This step has complexity $\mathcal{O}\{k_1 k_2 KMQ_\text{BS}^2 N_s l_1 l_2 R\}$. Notably, performing spatial smoothing only on the factor matrix $\mathbf{B}$ reduces complexity to $\mathcal{O}\{l_1 k_1 KM Q_\text{MS} N_\text{s}\}$, albeit at the expense of the estimation accuracy. The subsequent singular value decomposition and eigenvalue decomposition steps incur lower complexity and are therefore omitted here. For the MUSIC algorithm, the total complexity is $\mathcal{O}\{K^2 Q_\text{MS} N_\text{s} M + G K (K - R) + K^3\}$, where the spectral peak search term $\mathcal{O}\{G K (K - R)\}$ typically dominates. For the OMP algorithm, the dominant complexity arises from the iterative selection of atoms. With $R$ iterations, each requiring a correlation operation between the residual and the measurement matrix, the complexity is approximately $\mathcal{O}\{RN_\text{grid}L\}$, where $N_\text{grid}$ is the number of grid points in the dictionary and $L$ is the measurement vector length. In the context of MA-enabled ISAC parameter estimation, this simplifies to  $\mathcal{O}\{RN_\text{grid}Q_\text{MS}N_sKM\}$.

In summary, the SCPD algorithm exhibits lower complexity than ALS when the factor matrix dimensions are small, as ALS complexity scales with the number of iterations $N_\text{iter}$. However, as dimensions increase, the SCPD complexity grows rapidly due to the multiplicative terms $l_1 l_2$ and $k_1 k_2$ in the spatial smoothing step. MUSIC complexity is largely governed by the spectral search grid size $G$ and the number of subcarriers $K$, while OMP complexity scales linearly with the number of targets $R$ and grid size $G$, making it particularly efficient when $R$ is small.
\subsubsection{CRB Analysis}
As is well known, the CRB provides a lower bound on the variance of any unbiased estimator\cite{Chi2006}. The CRB results illustrate the behavior of the resulting bounds and therefore it is appropriate to evaluate our proposed tensor decomposition-based approach against the CRB results.
The log-likelihood function
of $\mathbf{p}$ can be expressed as
\begin{equation}
    \begin{aligned}
        L(\mathbf{p}) &= -Q_\text{MS}N_\text{s}KM\ln{(\pi\sigma_n^2)} \!-\! \frac{1}{\sigma_n^2}\big\|\mathbf{Y}_{(n)}\!-\!\mathbf{Z}_{(n)}\Big\|_F^2,
\end{aligned}
\end{equation}
where $\mathbf{Z}_{(n)}$ denotes the mode-$n$ unfolding of noiseless signal tensor $\mathcal{Z}$. The calculation of the complex Fisher information matrix (FIM) for $\mathbf{p}$ is discussed in \cite{942635} and the complex FIM for $\mathbf{p}$ can be obtained by taking the expectation of the second-order partial derivatives of the log-likelihood function, which can be given as
\begin{equation}
    \mathbf{\Omega}(\mathbf{p}) = \mathbb{E}\Big[\left(\frac{\partial L(\mathbf{p})}{\partial \mathbf{p}}\right)\left(\frac{\partial L(\mathbf{p})}{\partial \mathbf{p}}\right)^H\Big].
\end{equation}
Hence, the CRB is given by the inverse of the matrix $\mathbf{\Omega}(p)$, such that $\text{CRB}(\mathbf{p}) = \mathbf{\Omega}(\mathbf{p})^{-1}$.

\subsection{Chapter Summary}
This section reviewed representative parameter estimation methods for MA-enabled ISAC systems, including LS, subspace-based, CS-based, and tensor decomposition-based approaches. By exploiting different characteristics of MA-enabled ISAC signals, such as signal-noise subspace properties, parameter sparsity, and multidimensional data structures, these methods provide different trade-offs among estimation accuracy, computational complexity, and prior knowledge requirements. In particular, tensor decomposition-based methods can effectively exploit the multidimensional structure of ISAC signals to estimate communication and sensing parameters, thereby providing useful information for exploiting the spatial DoFs of MA. Nevertheless, computational complexity, initialization sensitivity, and robustness under practical channel conditions remain important challenges. Overall, the selection of a parameter estimation method should be tailored to the signal characteristics, estimation requirements, and spatial flexibility of MA-enabled ISAC systems.

\section{Security of MA-Enabled ISAC Systems}
As ISAC systems integrate communication and sensing functions over shared wireless resources, ensuring the confidentiality of communication information has become an important concern. PLS provides an effective approach to protecting wireless transmissions against potential eavesdropping attacks. Owing to the inherent broadcast nature of wireless channels, ISAC systems are susceptible to eavesdropping, since communication signals and sensing waveforms may both be observed by unintended receivers. Conventional FPA architectures, however, have limited spatial flexibility due to their static antenna configurations, which may restrict the ability to adapt the communication and sensing links to different security requirements. MA can provide an additional spatial DoF by allowing antenna positions to be adjusted within a predefined moving region. By properly selecting antenna positions, the resulting channel characteristics can be modified to improve the legitimate link and reduce the quality of eavesdropping links, while maintaining the desired sensing performance. Therefore, MA technology provides an additional design dimension for advancing secure ISAC systems.

\begin{figure}[t]
	\centering
	\includegraphics[width=0.95\linewidth]{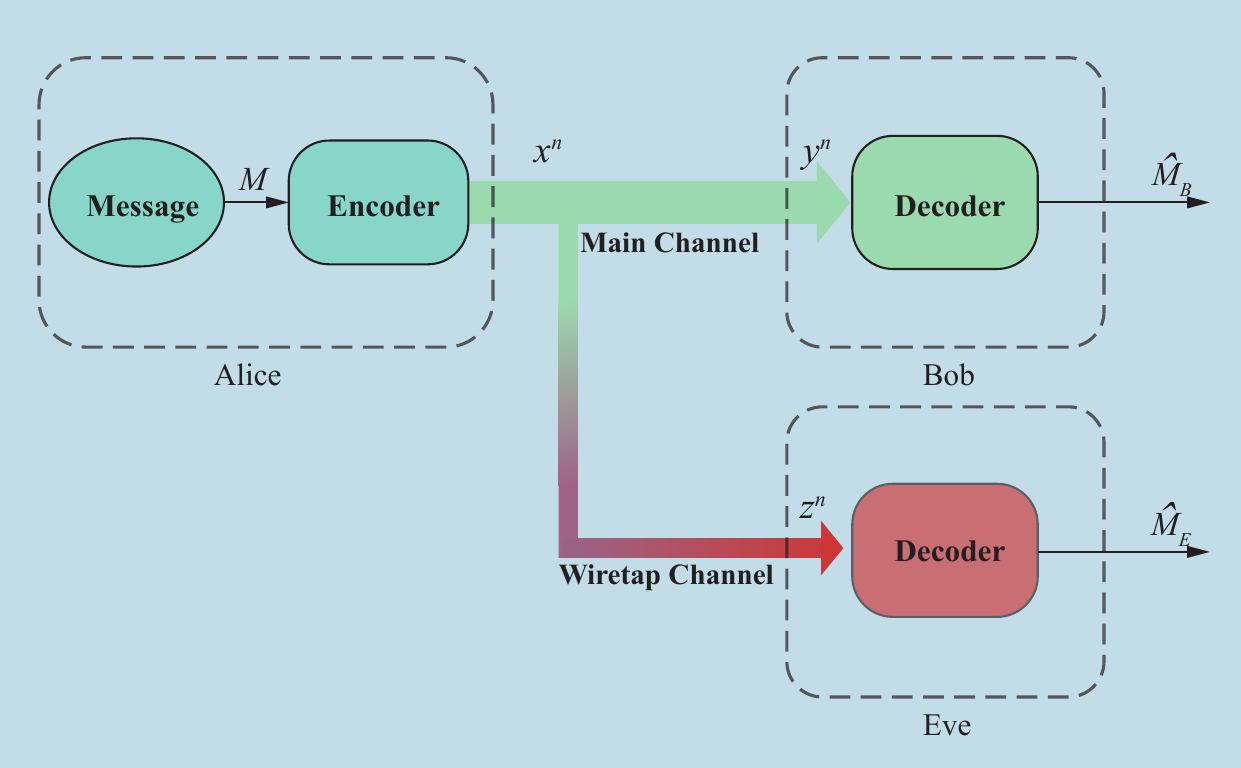}
	\caption{Wireless wiretap system model.}
	\label{fig:PLS}
    \vspace{-1em}
\end{figure}

\subsection{Physical Layer Security  Fundamentals}
\subsubsection{Introduction to PLS}
In ISAC, a single transmitted waveform serves the dual purpose of delivering information to communication users and illuminating targets for sensing. Because of the broadcast nature of the wireless medium, any leakage not only compromises data confidentiality but also exposes the system's sensing intentions and parameters. To address these critical security challenges, PLS has emerged as a promising solution that leverages the intrinsic physical characteristics of wireless channels, including fading, noise, and diversity, to enable information-theoretic secure transmission without relying on higher-layer encryption \cite{6739367}.

The foundational concept of PLS traces back to Wyner's pioneering work in 1975 \cite{6772207}, which introduced the wiretap channel model and advanced the field of information-theoretic security. In this model, a legitimate transmitter (Alice) communicates with a legitimate receiver (Bob) in the presence of an eavesdropper (Eve). Both the main and Eve channels are typically modeled as discrete memoryless channels.
To provide a clear conceptual framework, the fundamental architecture of such a system is illustrated in Fig.~\ref{fig:PLS}, which delineates the information flow from the transmitter to the intended and unauthorized receivers.
Wyner defined the secrecy capacity as the maximum achievable transmission rate at which a message can be transmitted reliably to the intended receiver while remaining perfectly confidential from the Eve. This theoretical framework remains central to modern PLS research and its application in wireless networks.

\subsubsection{Wiretap Channel Models and Secrecy Capacity}

\paragraph{Wyner's Wiretap Channel Model}
Wyner's wiretap channel model is a foundational achievement in modern PLS theory \cite{6772207}. This model describes a classic communication scenario where a sender, Alice, aims to transmit a confidential message to an intended receiver, Bob, while her transmission is simultaneously intercepted by Eve, through another channel. Wyner's pivotal contribution was demonstrating the existence of a fundamental performance limit known as the secrecy capacity. This limit represents the maximum information rate at which Alice can reliably transmit to Bob while ensuring that Eve cannot glean any useful information \cite{1055892}. For ISAC, the presence of a sensing target further complicates the channel model, as the same waveform must simultaneously convey information and extract environmental parameters, potentially creating additional leakage paths to Eve.

\paragraph{Gaussian Wiretap Channel}
The Gaussian wiretap channel is a significant extension of Wyner's model to environments characterized by AWGN \cite{1055917}. In the Gaussian wiretap channel, the signals received by Bob and Eve can be expressed as
\begin{equation}
    y_b = \sqrt{P}h_bx + n_b,
\end{equation}
\begin{equation}
    y_e = \sqrt{P}h_ex + n_e,
\end{equation}
where $P$ is the transmit power, $h_b$ and $h_e$ are the main and wiretap channels respectively, while $n_b$ and $n_e$ are mutually independent AWGN with noise powers $\sigma_{b}^{2}$ and $\sigma_{e}^{2}$.
The secrecy capacity for the Gaussian wiretap channel is given by
\begin{equation}
    C_{s} = \max\left\{ \log_{2}\left(1 + \gamma_{b}\right) - \log_{2}\left(1 + \gamma_{e}\right), 0 \right\},
\end{equation}
where $\gamma_{b} = P|h_{b}|^{2}/\sigma_{b}^{2}$ and $\gamma_{e} = P|h_{e}|^{2}/\sigma_{e}^{2}$ denote the received signal-to-noise ratios (SNRs) at Bob and Eve, respectively. This expression reveals the core principle of PLS: when the quality of the main channel is superior to that of the wiretap channel, the system can achieve a strictly positive secure transmission rate. Furthermore, a greater channel quality disparity yields a higher secrecy capacity. The importance of the Gaussian channel lies not only in its analytical tractability but also in providing a robust approximation for performance analysis in practical wireless systems.
In ISAC, this channel disparity must be engineered while simultaneously maintaining sufficient sensing requirements toward the target, creating a trade‑off among communication, sensing, and security.

\paragraph{Secrecy Capacity in Fading Channels}
In practical wireless communication systems, channels typically exhibit time-varying fading characteristics unlike static Gaussian channels. The randomness inherent in fading channels introduces both unique opportunities and complex challenges for PLS \cite{4035982}. On the one hand, fading can momentarily degrade the main channel quality below that of the wiretap channel, leading to a security outage. On the other hand, the variability of fading can be exploited opportunistically: when the main channel experiences a deep fade but Eve's channel suffers even more severe degradation, the system can still achieve a high secure transmission rate.

In fading channels, defining the secrecy capacity necessitates accounting for the statistical properties of the channel. If the transmitter possesses instantaneous CSI, it can adapt the transmission rate based on current channel conditions to achieve secure transmission perfectly matched to the instantaneous channel quality. Conversely, if the transmitter only acquires the channel statistics without instantaneous CSI, a fixed transmission rate must be employed. In this case, the system's security performance is fundamentally characterized by statistical metrics such as ergodic secrecy capacity and secrecy outage probability \cite{4529264}.
For ISAC, the instantaneous CSI required for rate adaptation must also support adaptive sensing waveform design, so security and sensing adaptivity are tightly coupled.

\subsubsection{Performance Metrics of PLS}

\paragraph{Ergodic Secrecy Capacity}
Ergodic secrecy capacity is a paramount performance metric for PLS in fading channels \cite{4626059}. It represents the time average of the maximum achievable secure transmission rate over all possible fading states. Assuming the transmitter can track instantaneous channel variations and adapt the transmission rate accordingly, the ergodic secrecy capacity is defined as the statistical expectation of the secrecy capacity over the fading distribution \cite{4529282}
\begin{equation}
    \bar{C}_{s} = \mathbb{E}_{h_{b},h_{e}}\left\{ \left[\log_{2}(1+\gamma_{b}) - \log_{2}(1+\gamma_{e})\right]^{+} \right\},
\end{equation}
where $[x]^{+} = \max\{x, 0\}$. Ergodic secrecy capacity is particularly suitable for delay-tolerant services and is utilized to determine the long-term average secure transmission rate.

\paragraph{Secrecy Outage Probability}
When the transmitter cannot obtain instantaneous CSI, especially regarding Eve's channel, it must transmit confidential information at a predetermined fixed rate. Due to random channel fading fluctuations, the instantaneous secrecy capacity might occasionally fall below the target rate $R_{s}$, thereby causing a security outage. The secrecy outage probability (SOP) is defined as the probability that the instantaneous secrecy capacity drops below this target secrecy rate \cite{4529264}
\begin{equation}
    P_{\text{SOP}}(R_{s}) = \Pr\{C_{s} < R_{s}\}.
\end{equation}
SOP is a critical indicator for measuring the security reliability of a system. In practical system design, the SOP is strictly required to remain below a specified threshold for a given target secrecy rate. Research indicates that the spatial correlation between the main and wiretap channels significantly impacts SOP performance, depending heavily on specific system parameters and operating regimes.
In ISAC design, a sensing outage probability should be simultaneously constrained, reflecting the dual reliability requirements of the integrated system.

\paragraph{Secrecy Energy Efficiency}
With the proliferation of green communication paradigms, secrecy energy efficiency (SEE) has become an essential optimization objective in PLS design \cite{6199997}. SEE is defined as the ratio of the secrecy rate to the total power consumption, measured in bits per Joule
\begin{equation}
    \eta_{\text{SEE}} = \frac{R_{s}}{P_{\text{total}}},
\end{equation}
where $P_{\text{total}}$ encompasses both the transmit power and the hardware circuit power consumption. The SEE metric gracefully balances security performance with energy expenditure and is particularly relevant for energy-sensitive scenarios such as IoT devices and UAV networks. In MA systems, the physical actuation of antennas naturally consumes energy, demanding a rigorous trade-off between the enhanced secrecy rates, the sensing performance gains, and the increased mobility energy consumption.

\subsubsection{Techniques for Achieving PLS}
 A diverse range of techniques has been proposed to enhance PLS, including beamforming, artificial noise (AN), cooperative interference, and directional modulation \cite{7467419}.  In the context of ISAC, PLS must be jointly designed with sensing functionality to protect both the communicated data and the sensing process itself.

\paragraph{Beamforming Technology}
Beamforming serves as a core technique for enhancing PLS in multi-antenna systems. Its fundamental principle is to exploit the spatial selectivity of antenna arrays to precisely focus transmitted signal energy toward the legitimate user while simultaneously creating deep nulls or low-gain regions in the direction of the Eve. Given knowledge of the main channel CSI, the transmitter can design a beamforming vector $\mathbf{w}$ to maximize the received power at the legitimate user while strictly minimizing information leakage. 
In ISAC, the same beamforming vector must also illuminate the sensing target with sufficient gain. This design task is often formulated as a constrained optimization that balances communication signal-to-interference-plus-noise ratio (SINR), sensing beampattern matching, and Eve nulling.

\paragraph{Artificial Noise Injection}
AN technology actively jams Eve's reception by injecting specific synthesized noise components into the transmitted signal, leveraging prior knowledge of the main channel to guarantee that the noise does not degrade the legitimate user's reception \cite{Lyu2025MAISAC}. The core concept is to intelligently allocate transmit power between the information signal and the AN while confining the noise strictly within the null space of the main channel. Specifically, the transmitter designs a beamforming vector $\mathbf{w}$ for the information signal and generates an AN vector $\mathbf{z}$ whose covariance matrix $\mathbf{C} = \mathbb{E}\{\mathbf{z}\mathbf{z}^{H}\}$ satisfies $\mathbf{h}_{b}^{H}\mathbf{C}\mathbf{h}_{b} = 0$. Consequently, the received SINR at the legitimate user remains $\gamma_{b} = |\mathbf{h}_{b}^{H}\mathbf{w}|^{2}/\sigma_{b}^{2}$, while the SINR at the Eve is severely degraded to $\gamma_{e} = |\mathbf{h}_{e}^{H}\mathbf{w}|^{2}/(\sigma_{e}^{2} + \mathbf{h}_{e}^{H}\mathbf{C}\mathbf{h}_{e})$.
 In ISAC, the AN must be carefully engineered to avoid corrupting the sensing echo, e.g., by projecting it onto the orthogonal complement of the sensing channel subspace.

\paragraph{Cooperative Relaying and Jamming}
Cooperative relaying technology introduces multiple distributed nodes to transmit collaboratively, fundamentally expanding the spatial dimensions available for achieving PLS \cite{5352243}. In networks plagued by Eves, relay nodes can assume versatile roles: they can participate in cooperative beamforming to bolster the legitimate signal, act as cooperative jammers transmitting AN to thoroughly confuse the Eve, or function as trusted relays that securely decode and forward confidential information. Cooperative relaying is particularly advantageous in scenarios where the direct communication link is entirely absent or of unacceptably poor quality.
When integrated with sensing, relays can also assist in target localization, forming a distributed ISAC network that simultaneously enhances sensing coverage and secure communication.

\vspace{-1em}
\subsection{Mechanisms of MA for Enhancing PLS in ISAC}
MA technology introduces unprecedented position DoF at the transmitter, enabling the real-time physical reconfiguration of wireless channels and providing a transformative dimension for enhancing PLS in ISAC \cite{10318061}. This section systematically elaborates on the mechanisms by which MAs elevate PLS from three distinct perspectives: spatial channel customization, joint position and beamforming optimization, and collaborative design with AN.

\subsubsection{Spatial Channel Customization}~{}

\paragraph{Legitimate Channel Enhancement and Sensing Beampattern Shaping} The fundamental advantage of MAs lies in their ability to physically alter the channel response by dynamically adjusting antenna positions, thereby improving the received SINR for the legitimate user and simultaneously shaping the sensing beampattern. Based on the field-response channel model \cite{10318061}, the channel vector from a transmitter equipped with $N$ MAs to a single-antenna legitimate user, Bob, can be expressed as
\begin{equation}
    \mathbf{h}_{b}(\mathbf{t}) = \left[ h_{b}(\mathbf{t}_{1}), h_{b}(\mathbf{t}_{2}), \ldots, h_{b}(\mathbf{t}_{N}) \right]^{T},
\end{equation}
where $\mathbf{t}_{n} = [x_{n}, y_{n}]^{T} \in \mathcal{C}$ denotes the 2D coordinates of the $n$-th MA within the continuous movable region $\mathcal{C}$. The channel response at each spatial position is determined by multipath superposition
\begin{equation}
    h_{b}(\mathbf{t}) = \sum_{\ell=1}^{L_{b}} \sqrt{\frac{\mu_{b}}{L_{b}}} \sigma_{b,\ell} e^{-j\frac{2\pi}{\lambda}\mathbf{t}^{T}\mathbf{\rho}_{b,\ell}},
\end{equation}
where $L_{b}$ is the number of multipath components, $\mu_{b}$ represents the distance-dependent path loss, $\sigma_{b,\ell}$ is the complex small-scale fading coefficient, and $\mathbf{\rho}_{b,\ell}$ is the angle-dependent steering vector. 

By optimizing the MA positions, the signal components originating from different paths can be coherently combined in the direction of the legitimate user, while the radiated power in the sensing direction of interest is maximized to ensure high target illumination. 
The dynamic capability of MAs to navigate the spatial continuum for gain optimization and interference suppression is visually demonstrated through the channel power gain maps in Fig. \ref{channel_map}.
Compared to traditional FPAs, MAs actively bypass deep fading spatial pockets and autonomously select coordinates with optimal channel conditions \cite{10354003}. As illustrated in Fig. \ref{channel_map}(a) and \ref{channel_map}(b), the optimized MA positions enhance the total channel gain for Bob while simultaneously suppressing the gain for Eve. 

\paragraph{Dynamic Null Steering for the Eve's Channel} While simultaneously enhancing the legitimate channel and maintaining sensing illumination, MAs can synthesize deep spatial nulls in the precise direction of the Eve through position optimization. For Eve, the channel response vector is
\begin{equation}
    \mathbf{h}_{e}(\mathbf{t}) = \left[ h_{e}(\mathbf{t}_{1}), h_{e}(\mathbf{t}_{2}), \ldots, h_{e}(\mathbf{t}_{N}) \right]^{T}.
\end{equation}
By jointly optimizing the physical coordinates of each MA, the signals radiated from different antennas can be forced to destructively interfere exactly at Eve's location, satisfying the critical null-steering condition \cite{10416363}
\begin{equation}
    {\sum}_{n=1}^{N} w_{n} h_{e}(\mathbf{t}_{n}) \approx 0,
\end{equation}
where $w_{n}$ represents the transmit beamforming coefficient. This active spatial degradation dramatically widens the performance gap between the legitimate and wiretap channels without compromising the sensing performance.

\begin{figure}[t!]
	\centering
	\subfloat[]{
		\includegraphics[width=0.45\textwidth]{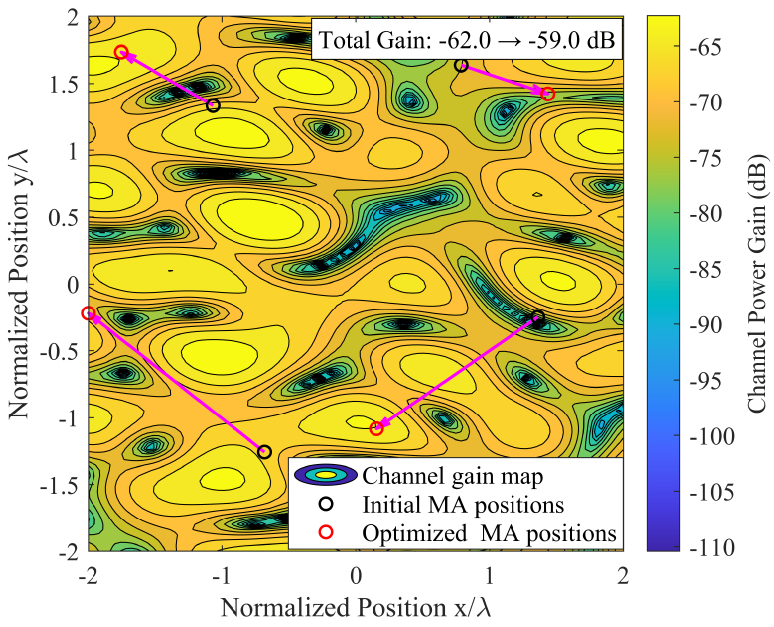}
		\label{DL_channel_map}
	}
	\hfill
	\subfloat[]{
		\includegraphics[width=0.45\textwidth]{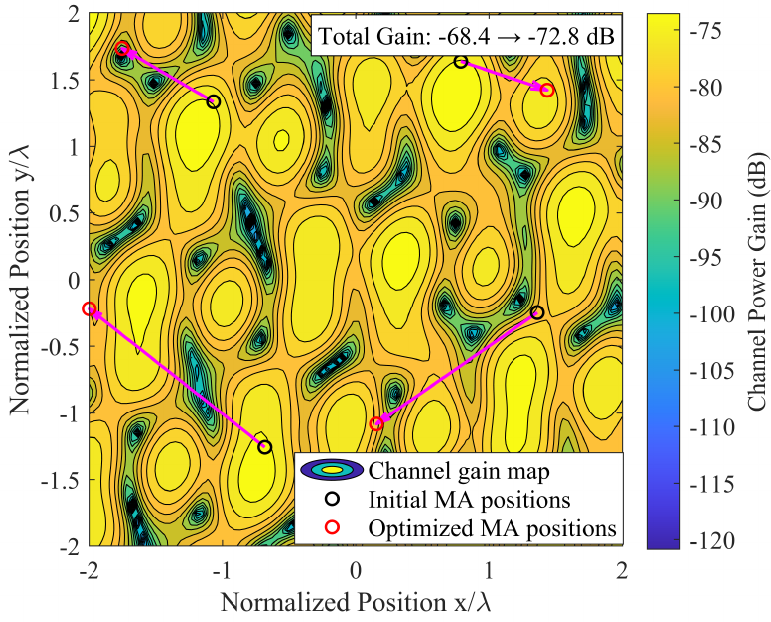}
		\label{EVE_channel_map}
	}
	\caption{Channel power gains: (a) between transmit MA and single FPA of the Bob, (b) between transmit MA and single FPA of the Eve.}
	\label{channel_map}
    \vspace{-1em}
\end{figure}

\paragraph{Channel Correlation Control} The spatial correlation between the main channel and the wiretap channel heavily dictates PLS performance. When these channels are highly correlated, beamforming designed to service the main channel inevitably enhances the wiretap channel as well. MAs provide a revolutionary dimension for directly manipulating this channel correlation. By continuously adjusting antenna positions, the spatial characteristics of the two channels can be physically orthogonalized. This correlation control is particularly indispensable when the Eve is angularly close to the legitimate user, a challenging scenario where traditional FPA systems struggle to distinguish between the two entities \cite{10839251}.
Moreover, MA-driven decorrelation can simultaneously decouple the sensing channel from the wiretap channel, ensuring that the probing waveform focused on the target does not leak useful sensing information to an adversary.

\subsubsection{Joint Antenna Position and Beamforming Optimization}
Jointly optimizing both the antenna positions and the beamforming weights forms the core technical approach to fully realizing the security gains offered by MAs in ISAC. Consider a typical MA-assisted secure ISAC system where the transmitter, Alice, is equipped with $N$ MAs and aims to serve a communication user Bob, detect a sensing target, and combat an Eve \cite{11086422}. The transmitted signal encompasses both an information signal and AN
\begin{equation}
    \mathbf{x} = \mathbf{w}s + \mathbf{z},
\end{equation}
where $\mathbf{w} \in \mathbb{C}^{N \times 1}$ is the beamforming vector, $s$ is the information symbol satisfying $\mathbb{E}\{|s|^{2}\} = 1$, and $\mathbf{z}$ is the AN vector with covariance matrix $\mathbf{C} = \mathbb{E}\{\mathbf{z}\mathbf{z}^{H}\}$. The corresponding received signals at Bob and Eve are
\begin{equation}
    y_{b} = \mathbf{h}_{b}(\mathbf{t})^{H}(\mathbf{w}s + \mathbf{z}) + n_{b},
\end{equation}
\begin{equation}
    y_{e} = \mathbf{h}_{e}(\mathbf{t})^{H}(\mathbf{w}s + \mathbf{z}) + n_{e}.
\end{equation}

A typical optimization problem aiming to maximize the achievable secrecy rate $\mathcal{R}_{s}$ while guaranteeing a minimum sensing accuracy can be formulated as follows
\begin{align}
    \max_{\mathbf{t}, \mathbf{w}, \mathbf{C} \succcurlyeq \mathbf{0}} \quad & \mathcal{R}_{s}(\mathbf{t}, \mathbf{w}, \mathbf{C}) \\ 
    \text{s.t.} \quad & \|\mathbf{w}\|^{2} + \text{tr}(\mathbf{C}) \le P, \\ 
    & \text{CRB}(\mathbf{t},\mathbf{w}) \le \gamma_{\text{sens}}, \\
    & \mathbf{t}_{n} \in \mathcal{C}, \quad \forall n, \\ 
    & \|\mathbf{t}_{n} - \mathbf{t}_{m}\| \ge D_{\min}, \quad \forall n \ne m,
\end{align}
where $P$ is the total available transmit power, $\gamma_{\text{sens}}$ is the maximum tolerable CRB for sensing, and $D_{\min}$ denotes a minimum physical distance between any two antennas to avoid severe mutual coupling. If the AN is strictly projected onto the null space of the legitimate channel, the constraint $\mathbf{h}_{b}(\mathbf{t})^{H}\mathbf{C}\mathbf{h}_{b}(\mathbf{t}) = 0$ must also be explicitly satisfied. Since the position variables are highly coupled with the beamforming coefficients across a continuous non-convex space, AO frameworks are widely adopted to decouple and iteratively solve this intricate problem \cite{10978811}.

Recent literature has extensively demonstrated that MA arrays provide substantial secrecy rate gains over conventional FPA systems without sacrificing sensing performance. For instance, in scenarios encompassing multiple Eves, the joint optimization of MA positions and AN projection has been shown to substantially suppress eavesdropping capabilities while maintaining a desired sensing beampattern \cite{10416363}. Furthermore, in near-field MIMO ISAC, MAs unlock the ability to perform precise range-domain focusing, successfully isolating a legitimate user from an Eve located at the exact same angle but at a different distance, all while achieving high-resolution target localization.

\subsubsection{Collaborative Design of AN and Antenna Positions}
While AN is a proven technique for securing communications, its effectiveness in traditional FPA systems is strictly bounded by the fixed spatial layout of the array. The introduction of MAs provides a crucial new dimension for AN covariance design, especially within ISAC where the AN must not degrade sensing performance \cite{10684758}. By physically shifting antenna positions, the spatial radiation pattern of the AN can be aggressively sculpted to maximize interference precisely at the Eve's location while placing minimal energy in the sensing target's direction. The transmitter can jointly optimize the positions $\mathbf{t}$ and the covariance matrix $\mathbf{C}$ to solve the following maximization problem
\begin{align}
    \max_{\mathbf{t}, \mathbf{C} \succcurlyeq \mathbf{0}} \quad & \mathbf{h}_{e}(\mathbf{t})^{H}\mathbf{C}\mathbf{h}_{e}(\mathbf{t}) \\ 
    \text{s.t.} \quad & \mathbf{h}_{b}(\mathbf{t})^{H}\mathbf{C}\mathbf{h}_{b}(\mathbf{t}) = 0, \\ 
    & \mathbf{a}(\mathbf{t},\theta_{\text{target}})^{H}\mathbf{C}\mathbf{a}(\mathbf{t},\theta_{\text{target}}) \le \epsilon, \\
    & \text{tr}(\mathbf{C}) \le P_{\text{AN}}, \\ 
    & \mathbf{t} \in \mathcal{C},
\end{align}
where $\mathbf{a}(\mathbf{t},\theta_{\text{target}})$ is the steering vector toward the sensing target, and $\epsilon$ limits the AN power leaked in that direction to preserve sensing accuracy.
This formulation elegantly illustrates how MAs amplify AN effectiveness while respecting sensing constraints: even when the noise covariance is mathematically constrained by the null space of the legitimate channel, continuous antenna position adjustments can physically manipulate the orientation of this null space, allowing the noise energy to be overwhelmingly projected toward the Eve without contaminating the sensing echo.

\subsection{Existing Works of MA-Enabled Secure ISAC}
The spatial agility offered by MAs has been extensively leveraged to enhance PLS. It has been conclusively shown that adjusting MA positions can intentionally degrade the channel quality of unauthorized receivers. A pioneering methodology combining projected gradient ascent with alternating optimization was proposed in \cite{10416363} to obtain high-quality suboptimal solutions for secure communication under regional antenna position constraints. Furthermore, through the joint optimization of MA placement and AN covariance, transmitters can actively and precisely suppress Eves \cite{10684758}. Recent findings on MA-aided directional modulation \cite{11260943} and spatial correlation decoupling \cite{11175171} verify that MAs can break the inherent spatial correlation between legitimate users and Eves, ensuring positive secrecy capacities even under severe angular alignment conditions.

Concurrently, the paradigm of ISAC has advanced significantly, yet it faces unique PLS vulnerabilities due to its dual-functional transmission nature \cite{9737357, 9606831}. The deep integration of sensing and communication exposes the network to severe eavesdropping threats, necessitating comprehensive multi-functional architectures \cite{9755276, 10614082}. Conventional defense mechanisms, which rely heavily on static-antenna beamforming to thwart hybrid-colluding Eves \cite{10872934, 10971420}, exhibit severe performance bottlenecks. Specifically, achieving an optimal tradeoff between target detection probability and secure communication rates becomes highly restrictive when constrained by imperfect CSI and angular alignment \cite{10693719}. 

To overcome these fundamental limitations, MAs have been introduced as a powerful enabler for secure ISAC. Preliminary frameworks have demonstrated that by optimally deploying MAs, spatial flexibility can simultaneously guarantee covertness and sensing precision \cite{10839251, 11223111}. More recently, MAs have been exploited in diverse ISAC architectures to further enhance secrecy performance. In \cite{11033708}, MAs at the user side were jointly optimized with transceiver beamforming of the BS side and the reflection coefficients of the IRS, maximizing the sum rate under sensing SNR and secrecy leakage constraints. In \cite{Ma2024MASensing}, MAs were deployed at the BS of an IRS-assisted ISAC system, where the secrecy rate is maximized by jointly designing the transmit beamforming, sensing covariance matrix, IRS phase shifts, and MA positions under beampattern and practical constraints. These results confirm that harnessing MA-enabled spatial DoF can significantly boost the secrecy performance of ISAC systems. Recent extensions have also investigated MA-secured ISAC in near-field environments \cite{11075964} and UAV-assisted networks \cite{11351324}.

Despite these promising developments, realizing secure transmission mechanisms in highly complex MA-aided ISAC environments remains a critical open challenge. Most existing works focus on static scenarios or basic performance tradeoffs, leaving the collaborative optimization of sensing and secure communication in the presence of dynamic Eves an unsolved problem, which motivates the joint beamforming and antenna positioning design proposed in this paper.

\vspace{-1em}
\subsection{Typical Case Discussion}
The effectiveness of MA technologies in securing ISAC systems is intrinsically linked to the mobility patterns of both the legitimate transceivers and potential Eves  \cite{10318061, 10416363}. Depending on whether the BS and the Eve are static or mobile, four representative scenarios emerge. In each, the dual‑function ISAC transmitter must simultaneously deliver secure communication, perform target sensing, and neutralize eavesdropping, all while exploiting the spatial flexibility of MAs. In the following, we elaborate on these four scenarios by integrating conceptual discussions with concrete case studies, encompassing system models, optimization frameworks, and numerical insights.

\subsubsection{Static BS and Static Eve}

In scenarios where both the BS and the Eve are stationary, large-scale channel parameters such as distance, angle, and Rician factor remain quasi-static over extended periods \cite{10318061}. This stability is typical in indoor environments like offices or residential homes, where user devices and sensing targets can also be considered static on a macro scale. In such contexts, the primary challenge for FPA systems is the limited spatial DoF available to distinguish between a legitimate user, a sensing target, and an Eve that may be angularly proximate.

\begin{figure}[t!]
	\centering
	\includegraphics[height=55mm]{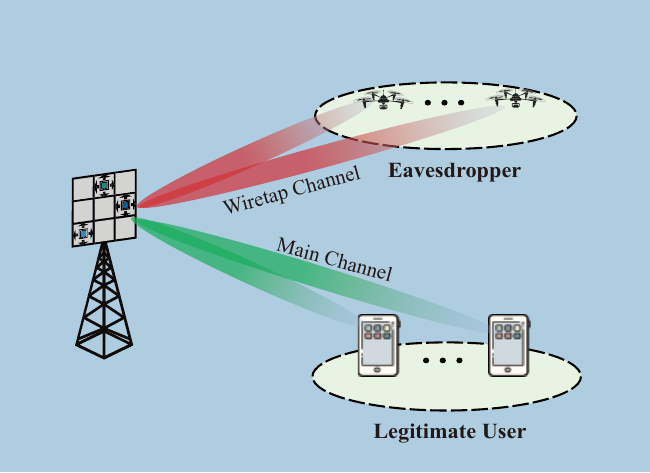}
	\caption{MA-enabled secure ISAC under static BS and static Eve conditions.}
	\label{fig:static_PLS}
    \vspace{-1em}
\end{figure}

\begin{figure}[t!]
	\centering
	\includegraphics[height=57mm]{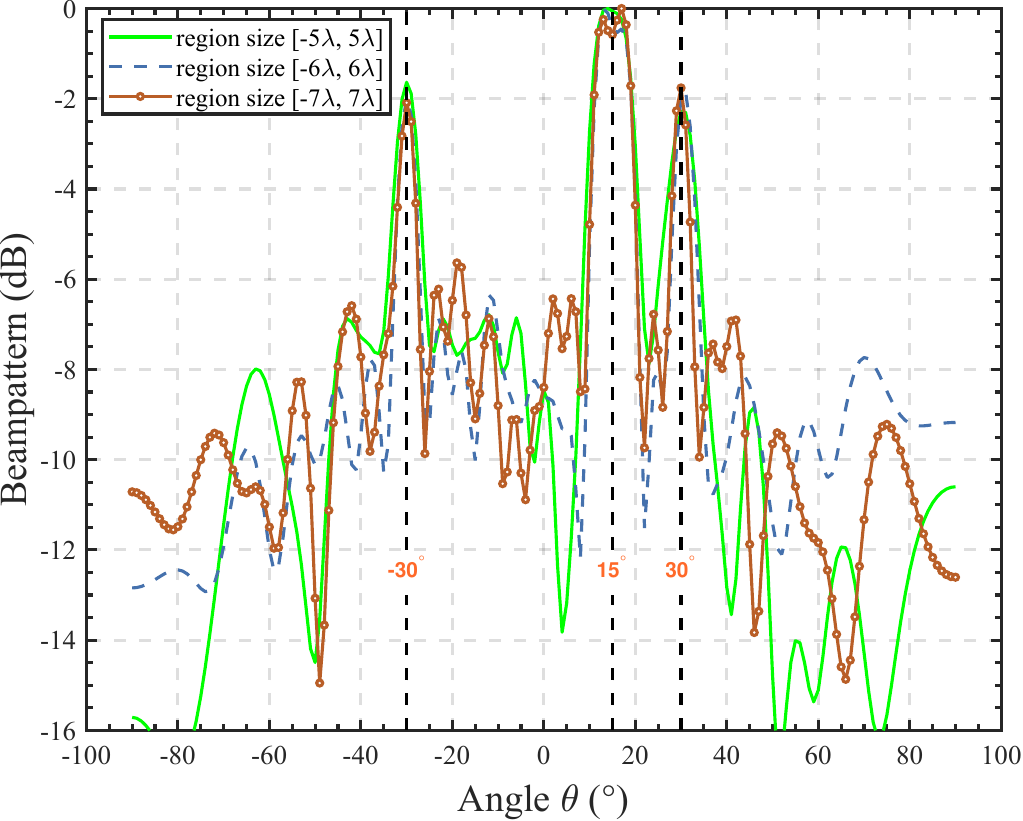}
	\caption{Beampattern versus the moving region size of the MA.}
	\label{fig:Beampattern_size}
    \vspace{-1em}
\end{figure}
\begin{figure}[t!]
	\centering
	\includegraphics[height=57mm]{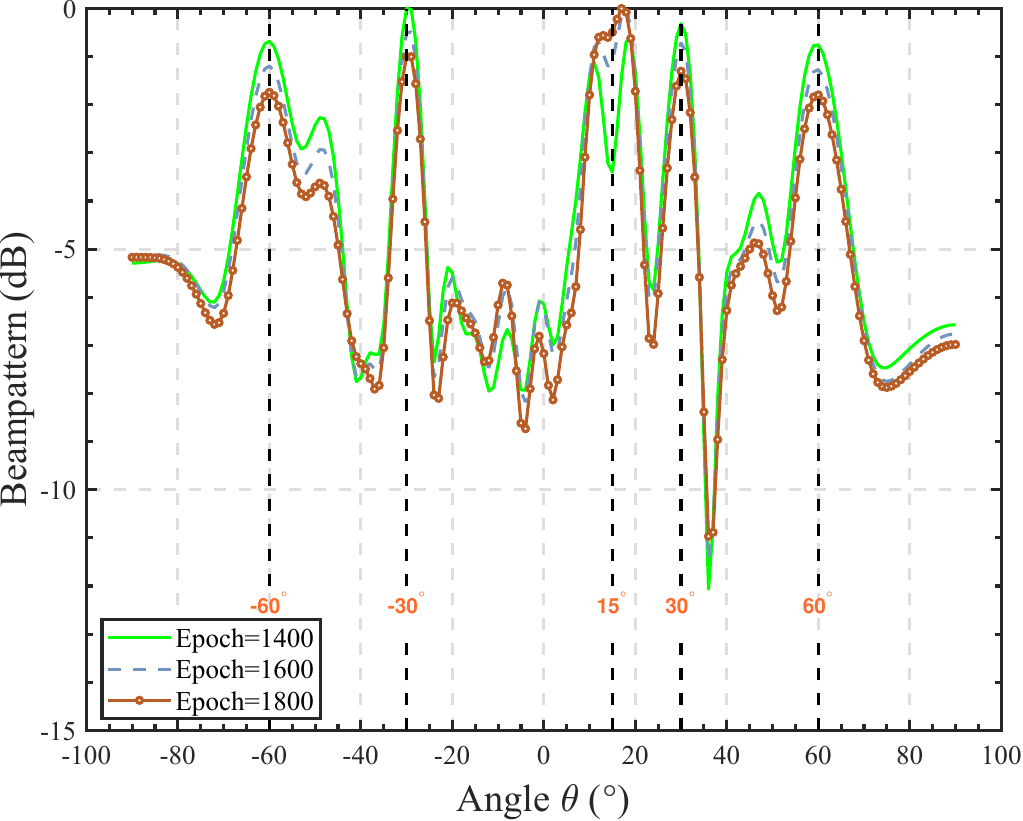}
	\caption{Beampattern versus epoch (Legitimate users at $\pm \text{30}^{\circ}$, $\pm \text{60}^{\circ}$ and Eve at $-\text{15}^{\circ}$).}
	\label{fig:Beampattern_epoch}
    \vspace{-1.5em}
\end{figure}

MA offers a paradigm-shifting solution by introducing an additional DoF through precise antenna position optimization \cite{10416363}. As conceptually illustrated in Fig.~\ref{fig:static_PLS}, an MA-empowered secure ISAC system under static conditions can perform a one-time deployment optimization of the MA positions based on the steady-state channel components. By finely adjusting these positions, the MA array can drastically reduce channel correlation with the Eve, effectively suppressing information leakage, while maintaining a high‑gain beam toward the sensing target and a strong communication link to Bob, even in the absence of instantaneous Eve CSI \cite{10318061}.

To quantify this joint security and sensing performance, let $\text{SINR}_{d}^\text{D}$ denote the received SINR at the $d$-th legitimate downlink user. The corresponding communication rate is 
\begin{equation}
    R_{d}^\text{D} = \log_{2}(1 + \text{SINR}_{d}^\text{D}),  \forall d \in \{1, 2, \dots, D\},
\end{equation}
where $D$ represents the total number of downlink users in the system.
Simultaneously, the corresponding eavesdropping rate for the downlink user $d$ can be expressed as
\begin{equation}
    R_{d}^\text{ED} = \log_{2}(1 + \gamma_{d}^\text{ED}),
\end{equation}
where $\gamma_{d}^\text{ED}$ represents the SNR at the Eve when attempting to decode the $d$-th user's message.

By maximizing the difference between these two rates, the system's sum secrecy rate is formulated as 
\begin{equation}
    R_\text{SSR} = {\sum}_{d=1}^{D} \left[ R_{d}^\text{D} - R_{d}^\text{ED} \right]^{+}.
\end{equation}

For sensing, we discretize the angular domain $\left[-{\pi}/{2}, {\pi}/{2}\right]$ into $Q$ directions and define the ideal beam pattern $P_{\mathrm{d}}(\theta_l)$ as
\begin{equation}
    P_{\text{d}}(\theta_l) =
    \begin{cases}
    1, & \theta_{\text{t}} - \Delta \le \theta_l \le \theta_{\text{t}} + \Delta, \\
    0, & \text{otherwise},
    \end{cases}
\end{equation}
where $\theta_{\text{t}}$ is the angle of the sensing target and $2\Delta$ is the desired beamwidth. The actual beam pattern generated by the MA array is given by $\boldsymbol{\alpha}^{\mathrm{H}}(\mathbf{t}, \theta) \mathbf{R}_{\mathrm{X}} \boldsymbol{\alpha}(\mathbf{t}, \theta)$, where $\boldsymbol{\alpha}(\mathbf{t}, \theta)$ is the steering vector toward angle $\theta$ evaluated at the antenna positions $\mathbf{t}$, and $\mathbf{R}_{\mathrm{X}} = \mathbb{E}\{\mathbf{x}\mathbf{x}^{\mathrm{H}}\}$ is the transmit signal covariance matrix. The beam pattern matching quality is then measured by the mean squared error (MSE) between the ideal and actual beam patterns
\begin{equation}
   \text{MSE}_{\text{bp}} \!=\! \frac{1}{Q} {\sum}_{q=1}^Q \left| \eta P_{\text{d}}(\theta_q) \!-\! \boldsymbol{\alpha}^{\mathrm{H}}(\mathbf{t}, \theta_q) \mathbf{R}_{\mathrm{X}} \boldsymbol{\alpha}(\mathbf{t}, \theta_q) \right|^2,
\end{equation}
where $\eta \geq 0$ is a scaling factor and $\theta_q$ refers to the $q$-th sample point in the angular domain. The sensing performance is therefore guaranteed by constraining $\text{MSE}_\text{bp} \le \gamma_{\text{bp}}$, where $\gamma_{\text{bp}}$ is a predefined threshold that ensures adequate target illumination and reliable echo processing.

The spatial advantages of the MA array are further corroborated by the array's radiation characteristics. As demonstrated in Fig.~\ref{fig:Beampattern_size}, expanding the continuous moving region size of the MA directly influences the precision of the generated beampatterns, which simultaneously serve communication and sensing. Furthermore, Fig.~\ref{fig:Beampattern_epoch} highlights the temporal evolution of the beampattern across different training epochs, showcasing the system's ability to maintain high gain toward legitimate users (e.g., at $\pm \text{30}^{\circ}$ and $\pm \text{60}^{\circ}$) and the Eve (e.g., at $-\text{15}^{\circ}$).

\subsubsection{Static BS and Dynamic Eve}

The assumption of a static Eve is frequently violated in practice by mobile adversaries, such as Eves mounted on vehicles or malicious UAVs. In this more challenging scenario, visually represented in Fig.~\ref{fig:dynamic_PLS}, the BS remains fixed but must contend with a dynamically changing eavesdropping channel. The key enabling factor for the BS is the integration of sensing capabilities, effectively transforming the infrastructure into an ISAC platform capable of actively tracking mobile targets \cite{9171304}. The mobile Eve itself becomes the sensing target, allowing the BS to reuse the same probing waveform for both target tracking and secure communication.

\begin{figure}[t!]
	\centering
	\includegraphics[height=53mm]{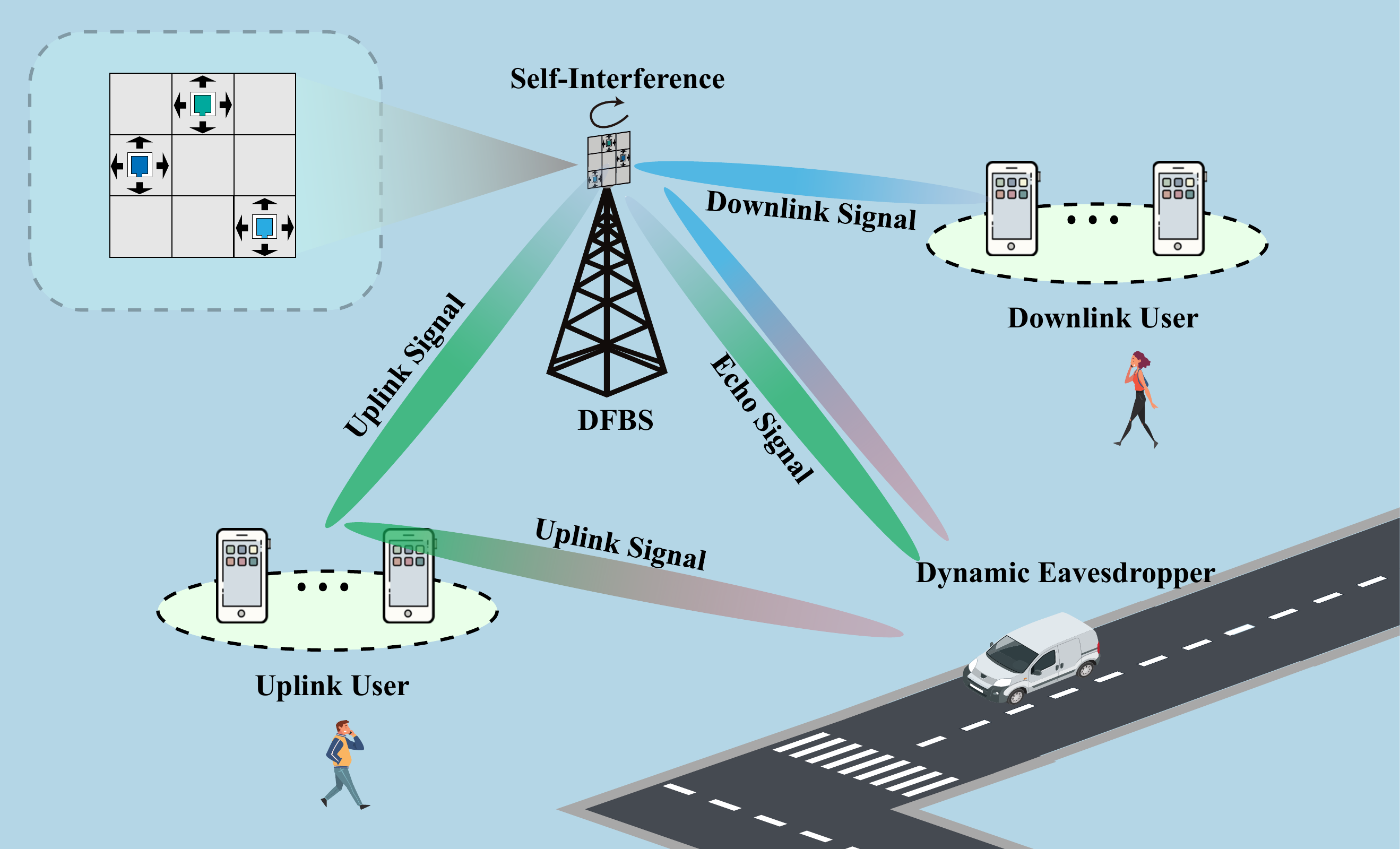}
	\caption{MA-enabled secure ISAC under static BS and dynamic Eve conditions.}
	\label{fig:dynamic_PLS}
    \vspace{-1.5em}
\end{figure}

The BS estimates the state of the mobile Eve by processing reflected echoes through a matched filter. By reconstructing this state information at the $i$-th time slot, the BS accurately obtains Eve's kinematic parameters. Specifically, the Eve's true state $\xi[i]$ comprises the round-trip delay $\hat{\tau}[i]$, Doppler frequency shift $\hat{\mu}[i]$, elevation angle $\hat{\theta}[i]$, and azimuth angle $\hat{\phi}[i]$. The relationship between these observation parameters and the true state can be established as \cite{11419928}
\begin{equation}
    \hat{\tau}[i] = \frac{2d_\text{ME}[i]}{c} + q_{\tau[i]},
\end{equation}
\begin{equation}
    \hat{\mu}[i] = \frac{2f_{c}\dot{\mathbf{l}}_\text{Eve}^{T}[i](\mathbf{l}_\text{Eve}[i] - \mathbf{l}_\text{MA})}{c d_\text{ME}[i]} + q_{\mu[i]},
\end{equation}
\begin{equation}
    \sin\hat{\theta}[i] = \frac{z_\text{Eve}[i]}{d_\text{ME}[i]} + q_{\sin\theta[i]},
\end{equation}
\begin{equation}
    \cos\hat{\theta}[i] = \frac{\sqrt{(x_\text{Eve}[i] - x_\text{MA})^{2} + (y_\text{Eve}[i] - y_\text{MA})^{2}}}{d_\text{ME}[i]} + q_{\cos\theta[i]},
\end{equation}
\begin{equation}
    \sin\hat{\phi}[i] = \frac{y_\text{Eve}[i]}{\sqrt{(x_\text{Eve}[i] - x_\text{MA})^{2} + (y_\text{Eve}[i] - y_\text{MA})^{2}}} + q_{\sin\phi[i]},
\end{equation}
where $c$ is the speed of light and $f_{c}$ is the carrier frequency. The term $d_\text{ME}[i]$ represents the instantaneous Euclidean distance between the MA base station and Eve. The vectors $\mathbf{l}_\text{MA} = [x_\text{MA}, y_\text{MA}, z_\text{MA}]^{T}$ and $\mathbf{l}_\text{Eve}[i] = [x_\text{Eve}[i], y_\text{Eve}[i], z_\text{Eve}[i]]^{T}$ denote the 3D Cartesian coordinates of the base station and the Eve, respectively, while $\dot{\mathbf{l}}_\text{Eve}[i]$ denotes Eve's velocity vector. Finally, the $q$ terms account for the corresponding measurement noise components.

Given the inherent nonlinearity of this measurement model, techniques like the extended Kalman filter (EKF) are widely adopted to linearize the system and predict Eve's motion state alongside its future CSI \cite{9171304}. The updated MSE matrix, which serves as a reliable metric for evaluating tracking accuracy, evolves according to
\begin{equation}
    \mathbf{M}[i] = (\mathbf{I} - \mathbf{K}[i]\mathbf{F}_{i})\mathbf{M}[i|i-1],
\end{equation}
where $\mathbf{M}[i|i-1]$ is the prior predicted MSE matrix, $\mathbf{I}$ is the identity matrix, $\mathbf{K}[i]$ denotes the optimal Kalman gain, and $\mathbf{F}_{i}$ represents the state transition Jacobian matrix evaluated at the current state estimate.

Armed with this predicted CSI, the BS can proactively optimize its transmission strategy via a dual-timescale approach: periodically adjusting large-scale MA positions to account for Eve's macro-motion and to maintain sensing focus, while executing real-time beamforming to dynamically synthesize deep nulls at each time slot. This tight interplay between sensing and secure communication epitomizes the ISAC advantage. 

\begin{figure}[t!]
	\centering
	\includegraphics[height=53mm]{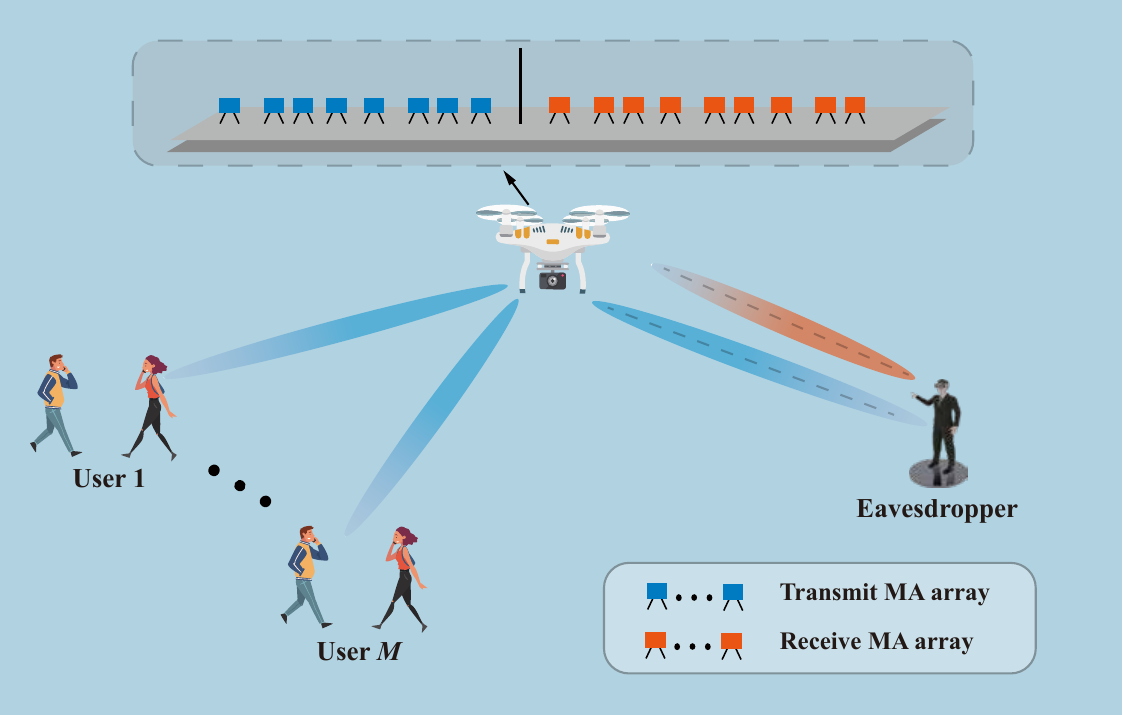}
	\caption{MA-enabled secure ISAC under dynamic BS and static Eve conditions.}
	\label{fig:UAV_static}
    \vspace{-1.5em}
\end{figure}
\subsubsection{Dynamic BS and Static Eve}

Deploying BSs on dynamic platforms, such as UAVs, introduces macroscopic mobility. As depicted in Fig.~\ref{fig:UAV_static}, a dual-function ISAC UAV is deployed to serve $M$ downlink users and detect a single-point target while navigating a static Eve. While the inherent high maneuverability of UAVs helps establish strong LoS links for legitimate users and provides flexible sensing geometries, it concurrently exposes the transmitted signal to the static Eve if not rigorously managed \cite{10040631}. 
To address this vulnerability, a dual-spatial-scale transmission protocol is essential \cite{11351324}
\begin{itemize}
    \item{Macro-deployment optimization:} The UAV's 3D trajectory and hovering locations are optimized over a large scale to remain in a highly favorable region for the legitimate user and the sensing target, while maximizing the distance to the Eve.
    \item{Micro-configuration optimization:} With the macroscopic position temporarily fixed, the onboard MA array is continuously fine-tuned on a wavelength scale to steer a sharp spatial null toward the adversary and simultaneously form a sensing beam toward the target.
    \item{Real-time beamforming:} Instantaneous CSI is leveraged to perform dynamic beamforming within each time slot, balancing secure communication and sensing requirements.
\end{itemize}

To formalize this in a 3D Cartesian coordinate system, let the UAV fly at a fixed altitude $H$. The vertical AoD $\theta_{m}[n]$ from the UAV to user $m$ located at the 2D horizontal coordinates $\mathbf{q}_{m}$ during time slot $n$ is geometrically determined by
\begin{equation}
    \theta_{m}[n] = \arcsin\left(\frac{H}{\sqrt{\|\mathbf{q}[n] - \mathbf{q}_{m}\|^{2} + H^{2}}}\right),
\end{equation}
where $\mathbf{q}[n]$ denotes the instantaneous 2D horizontal projection of the UAV's location.

Consequently, accounting for the signal propagation phase difference, the dynamic transmit array response vector $\overline{\mathbf{h}}_{m}[n]$ from the UAV to user $m$ is formulated as \cite{10643473}
\begin{equation}
    \overline{\mathbf{h}}_{m}[n] \!=\! \left[ e^{j\frac{2\pi}{\lambda}x_{1}[n]\sin(\theta_{m}[n])}, \ldots, e^{j\frac{2\pi}{\lambda}x_{N_{t}}[n]\sin(\theta_{m}[n])} \right]^{T},
\end{equation}
where $\lambda$ is the carrier wavelength, $N_{t}$ is the total number of transmit MAs onboard the UAV, and $x_{k}[n]$ specifies the 1D local continuous position of the $k$-th MA relative to the array's geometric center.

Similarly, adopting a single-path propagation model for the LoS-dominant aerial sensing channel, the steering vectors for the transmit array $\mathbf{a}[n]$ and the receive array $\mathbf{b}[n]$ toward a sensing target with an AoD of $\theta_{T}[n]$ are given respectively by
\begin{equation}
    \mathbf{a}[n] = \left[ e^{j\frac{2\pi}{\lambda}x_{1}[n]\sin(\theta_{T}[n])}, \ldots, e^{j\frac{2\pi}{\lambda}x_{N_{t}}[n]\sin(\theta_{T}[n])} \right]^{T},
\end{equation}
\begin{equation}
    \mathbf{b}[n] = \left[ e^{j\frac{2\pi}{\lambda}y_{1}[n]\sin(\theta_{T}[n])}, \ldots, e^{j\frac{2\pi}{\lambda}y_{N_{r}}[n]\sin(\theta_{T}[n])} \right]^{T},
\end{equation}
where $N_{r}$ represents the number of receive MAs, and $y_{k}[n]$ defines the local position of the $k$-th receive MA.

\subsubsection{Dynamic BS and Dynamic Eve}

The most complex and generalized scenario involves continuous motion from both the BS (e.g., a UAV) and the Eve (e.g., a malicious tracking UAV or vehicle). This creates a highly dynamic and adversarial environment that is frequently modeled as a non-cooperative Stackelberg game. Here, the malicious UAV strategically adapts its trajectory to maximize eavesdropping capabilities, while the legitimate network orchestrates its joint beamforming, trajectory, and MA positioning to guarantee both sensing accuracy and secure communication \cite{11131476}.

Addressing this threat necessitates sophisticated multi-agent coordination, where the UAV swarm forms a comprehensive two-level mobile adaptive system for cooperative target detection and secure transmission \cite{10791445}. The associated optimization problem aims to maximize the minimum achievable secure rate among all users by jointly optimizing the continuous 3D placement of the UAV fleet, the microscopic local MA positions on each UAV, and the receive beamforming, subject to sensing constraints and maximum power budgets. Solving these multi-dimensional, non-convex problems typically requires advanced computational techniques, including multi-agent deep reinforcement learning for predictive trajectory planning and successive convex approximation for real-time resource allocation \cite{11329133}.

In summary, the intricate interplay between the mobility of the legitimate infrastructure and the adversary fundamentally dictates the design paradigms of MA-enabled secure ISAC systems. As explored in the preceding case studies, the required countermeasures evolve from single-instance deterministic spatial optimizations in purely static environments to highly complex, predictive, and multi-agent resource allocations in fully dynamic networks, all while maintaining the dual functionality of sensing and secure communication.
%To synthesize these diverse operational regimes and provide a holistic perspective, Table~\ref{tab:mobility_scenarios} summarizes the core characteristics, primary technical approaches, and representative practical applications corresponding to four distinct mobility scenarios.

\vspace{-0.5em}
\subsection{Chapter Summary}
% This section systematically explored the integration of MA technology into PLS for ISAC systems, demonstrating that spatial flexibility is the key enabler for unifying communication, sensing, and security. To begin with, we established the theoretical foundations of PLS by reviewing security metrics such as secrecy capacity and outage probability, along with traditional enabling techniques including beamforming and AN. Building upon this basis, the MA-enhanced secure ISAC framework was analyzed to demonstrate how antenna position flexibility unlocks unique spatial DoF. By jointly optimizing element positions, beamforming, and AN, MAs significantly outperform FPAs in enhancing legitimate links, shaping sensing beams, and enforcing precise null-steering against Eves. Furthermore, we evaluated MA-enabled secure ISAC across diverse mobility scenarios, ranging from static to fully dynamic environments where all nodes are mobile. The case studies revealed that MAs serve as an adaptive security mechanism, capable of maintaining robust secrecy and sensing accuracy under high channel uncertainty. In summary, the transition from FPA to MA architectures represents a cornerstone for future secure and resilient ISAC infrastructures. By offering a transformative paradigm that jointly optimizes communication, sensing, and security through spatial adaptability, MA technology paves the way toward integrated and secure next-generation wireless networks.
This section reviewed the security issues of MA-enabled ISAC systems from the perspective of PLS. We first introduced representative PLS metrics and security-enhancing techniques, and then discussed how the spatial flexibility of MA can be exploited to improve secure communication and sensing performance through joint antenna position and beamforming optimization. Representative case studies under different mobility scenarios were further examined, highlighting the potential of MA to adapt legitimate and eavesdropping channels while maintaining the desired sensing performance. Overall, MA-enabled ISAC provides an additional spatial DoF for balancing communication, sensing, and security requirements, while robustness under channel uncertainty, dynamic mobility, and practical implementation constraints remains an important direction for further research.

\section{Open Issues and Future Directions}
\subsection{Adaptive Beamforming and Antenna Positioning}
Compared with conventional FPAs, MA introduces additional spatial DoFs by allowing antenna elements to adjust their positions, providing greater flexibility for beamforming design in ISAC systems. Since antenna positions affect the array response and radiation pattern, the joint optimization of antenna positions and beamforming weights can provide additional flexibility in controlling beam directions and sidelobe levels, thereby improving communication, interference management, sensing, and parameter estimation performance. However, such joint optimization is generally highly coupled and non-convex, making efficient optimization challenging even with accurate CSI. Moreover, iterative updates of beamforming and antenna positions may incur considerable computational overhead, particularly in large-scale or dynamic scenarios. Future research should therefore investigate low-complexity joint optimization methods, efficient strategies for selecting candidate antenna positions, and learning-assisted position adaptation, while considering hardware constraints to facilitate practical deployment of MA-enabled ISAC systems.

\vspace{-0.5em}
\subsection{Low-Overhead and Scalable Parameter Estimation}
The spatial flexibility of MA introduces position-dependent channel and sensing characteristics that need to be properly captured for effective communication and sensing. In MA-enabled ISAC systems, accurate estimation of channel and sensing parameters is essential for beamforming, resource allocation, target detection, localization, and other sensing tasks. By collecting observations at multiple antenna positions, MA can provide additional spatial information and potentially improve parameter resolvability. However, practical antenna movement is subject to positioning errors, motion constraints, and energy consumption, while the coupling between antenna positions, temporal channel variations, and sensing parameters increases the estimation complexity. Existing estimation methods may also incur considerable training or computational overhead when extended to large moving regions, dense antenna configurations, or dynamic environments. Future research should therefore develop physically consistent spatio-temporal models and low-overhead, robust, and scalable estimation methods, preferably by jointly considering antenna trajectory design, measurement selection, and communication-sensing requirements.

\vspace{-0.5em}
\subsection{Security and Privacy Protection}
The spatial reconfigurability and mobility of MA introduce new security and privacy considerations for ISAC systems. Since ISAC signals support both communication and sensing, security threats may affect not only confidential information transmission but also the protection of sensing-related information. Although MA provides additional spatial DoFs that can be exploited to enhance PLS through antenna-position and beamforming optimization, its dynamic operation also introduces new security and privacy risks, particularly under channel uncertainty and adversarial mobility. Moreover, antenna-position adaptation may incur a trade-off among security performance, sensing accuracy, energy consumption, and mobility cost. Future research should therefore investigate dynamic security models and low-overhead privacy-preserving mechanisms, together with the joint design of antenna positioning, beamforming, sensing, and security under practical mobility and channel constraints.

\vspace{-0.5em}
\subsection{Synergistic Integration of MA and RIS}
The integration of MA and RIS provides additional spatial DoFs and channel reconfigurability for ISAC systems. While MA enables flexible antenna-position adaptation, RIS can reshape the propagation environment through programmable reflection. Their joint deployment therefore offers complementary means for adapting both the transmitter and propagation environment, which may be particularly beneficial in complex or obstructed wireless environments. Although initial studies have explored joint MA-RIS design for communication and sensing, the potential of their cooperation remains largely unexplored. Moreover, jointly optimizing MA positions and RIS configurations results in a highly coupled and high-dimensional optimization problem, especially for large-scale systems and dynamic environments. Future research should investigate low-complexity joint optimization and real-time control methods, while accounting for hardware constraints, mobility, and signaling overhead in practical MA-RIS-enabled ISAC systems.

\section{Conclusion}
\input{Conclusion}
\balance
\bibliographystyle{IEEEtran}
\bibliography{reference}

\end{document}

%% file: Conclusion.tex
This tutorial has provided a comprehensive overview of MA-enabled ISAC systems, with a particular focus on their fundamentals, parameter estimation, and security issues. First, we reviewed the hardware architectures, spatial DoFs, and key characteristics of MA, together with representative techniques for antenna positioning, beamforming, and signal and channel modeling. Representative application scenarios were also discussed to illustrate the potential of MA-enabled ISAC. We then reviewed parameter estimation methods for communication and sensing, including LS, subspace-based, CS-based, and tensor decomposition-based approaches, and compared their estimation accuracy and computational complexity. Furthermore, we investigated security issues from the perspective of PLS and discussed how the spatial flexibility of MA can be exploited to enhance secure communication and sensing through joint antenna-position and beamforming optimization. Representative case studies under different mobility scenarios were also presented to demonstrate the benefits of MA-enabled secure ISAC. Several open issues and future research directions remain to be explored. Overall, MA introduces an additional spatial DoF that provides a useful design dimension for jointly addressing communication, sensing, and security in ISAC systems. We hope that this tutorial provides a unified perspective on MA-enabled ISAC and serves as a useful reference for further research in this emerging area.